\documentclass[11pt,a4paper]{article}
\usepackage{cite}
\usepackage{graphicx}
\usepackage{amssymb}
\usepackage{amsmath}
\usepackage{amsfonts}
\usepackage{dsfont}
\usepackage{mathtools}
\usepackage{array}
\usepackage{rotating}
\usepackage{bbold,amsfonts}

\usepackage[utf8]{inputenc}
\usepackage{bm}
\usepackage{xcolor}
\usepackage{float}
\usepackage{braket}

\usepackage[height=8.8in,width=6.45in]{geometry}
\usepackage[font=small,labelfont=bf]{caption}
\usepackage[hidelinks]{hyperref}
\usepackage{booktabs,float,slashed}
\usepackage{standalone}
\usepackage{caption}
\usepackage{subcaption}
\numberwithin{equation}{section}

\newcommand{\beq}{\begin{equation}}
\newcommand{\eeq}{\end{equation}}

\newcommand{\overbar}[1]{\mkern 1.5mu\overline{\mkern-1.5mu#1\mkern-1.5mu}\mkern 1.5mu}

\newcommand{\vvev}[1]{\langle #1 \rangle}

\DeclareMathOperator{\tr}{tr}

\newcommand{\ii}{\mathrm{i}}

\makeatletter
\newcommand*{\letterdef@}{}
\newcommand*{\letterdef}[3]{%
	\def\letterdef@##1{\expandafter\newcommand\csname #1\endcsname{#2{##1}}}%
	\@tfor\@tempa :=#3\do{\expandafter\letterdef@\expandafter{\@tempa}}}
\makeatother
\letterdef{c#1} {\mathcal}{ABCDEFGHIJKLMNOPQRSTUVWXYZ} 
\letterdef{rm#1}{\mathrm} {dDeimM} 

\newdimen\tableauside\tableauside=1.0ex
\newdimen\tableaurule\tableaurule=0.4pt
\newdimen\tableaustep
\def\phantomhrule#1{\hbox{\vbox to0pt{\hrule height\tableaurule
			width#1\vss}}}
\def\phantomvrule#1{\vbox{\hbox to0pt{\vrule width\tableaurule
			height#1\hss}}}
\def\sqr{\vbox{%
		\phantomhrule\tableaustep
		\hbox{\phantomvrule\tableaustep\kern\tableaustep\phantomvrule\tableaustep}%
		\hbox{\vbox{\phantomhrule\tableauside}\kern-\tableaurule}}}
\def\squares#1{\hbox{\count0=#1\noindent\loop\sqr
		\advance\count0 by-1 \ifnum\count0>0\repeat}}
\def\tableau#1{\vcenter{\offinterlineskip
		\tableaustep=\tableauside\advance\tableaustep by-\tableaurule
		\kern\normallineskip\hbox
		{\kern\normallineskip\vbox
			{\gettableau#1 0 }%
			\kern\normallineskip\kern\tableaurule}%
		\kern\normallineskip\kern\tableaurule}}
\def\gettableau#1 {\ifnum#1=0\let\next=\null\else
	\squares{#1}\let\next=\gettableau\fi\next}
\allowdisplaybreaks

\begin{document}
\begin{titlepage}
\vspace*{10mm}
\begin{center}
{\LARGE \bf 
Localization vs holography in \texorpdfstring{$4d~\mathcal{N}=2$}{}
quiver theories 
}

\vspace*{15mm}

{\Large M. Bill\`o${}^{\,a,c}$, M. Frau${}^{\,a,c}$, A. Lerda${}^{\,b,c}$, A. Pini${}^{\,c}$, P. Vallarino${}^{\,a,c}$}

\vspace*{8mm}
	
${}^a$ Universit\`a di Torino, Dipartimento di Fisica,\\
			Via P. Giuria 1, I-10125 Torino, Italy
			\vskip 0.3cm

${}^b$  Universit\`a del Piemonte Orientale,\\
			Dipartimento di Scienze e Innovazione Tecnologica\\
			Viale T. Michel 11, I-15121 Alessandria, Italy
			\vskip 0.3cm
			
${}^c$   I.N.F.N. - sezione di Torino,\\
			Via P. Giuria 1, I-10125 Torino, Italy

\vskip 0.8cm
	{\small
		E-mail:
		\texttt{billo,frau,lerda,apini,vallarin@to.infn.it}
	}
\vspace*{0.8cm}
\end{center}

\begin{abstract}
We study 4-dimensional $\mathcal{N}=2$ superconformal quiver gauge theories obtained with an orbifold projection from $\mathcal{N}=4$ SYM, and compute the 2- and 3-point correlation functions among chiral/anti-chiral single-trace scalar operators and the corresponding structure constants.
Exploiting localization, we map the computation to an interacting matrix model and obtain expressions for the correlators and the structure constants that are valid for any value of the 't Hooft coupling in the planar limit of the theory. At strong coupling, these expressions simplify and allow us to extract the leading behavior in an analytic way. Finally, using the AdS/CFT correspondence, we compute the structure constants from the dual supergravity theory and obtain results that perfectly match the strong-coupling predictions from localization.
\end{abstract}
\end{titlepage}
\setcounter{tocdepth}{2}
\tableofcontents
\vspace{1cm}
\section{Introduction and summary of results}
\label{sec:intro}
The study of gauge theories with extended supersymmetry, like $\mathcal{N}=4$ 
and $\mathcal{N}=2$ Super Yang-Mills (SYM) theories, offers many insightful points of view
and provides powerful techniques for the non-perturbative analysis of quantum field theories in general. In the present paper, which is an expanded version of a recent short
letter \cite{Billo:2022gmq}, we exploit two of these techniques that are complementary to each other, localization and holography, 
to study a class of $\mathcal{N}=2$ superconformal quiver gauge theories in four dimensions.

\begin{figure}[ht]
	\center{\includegraphics[scale=0.6]{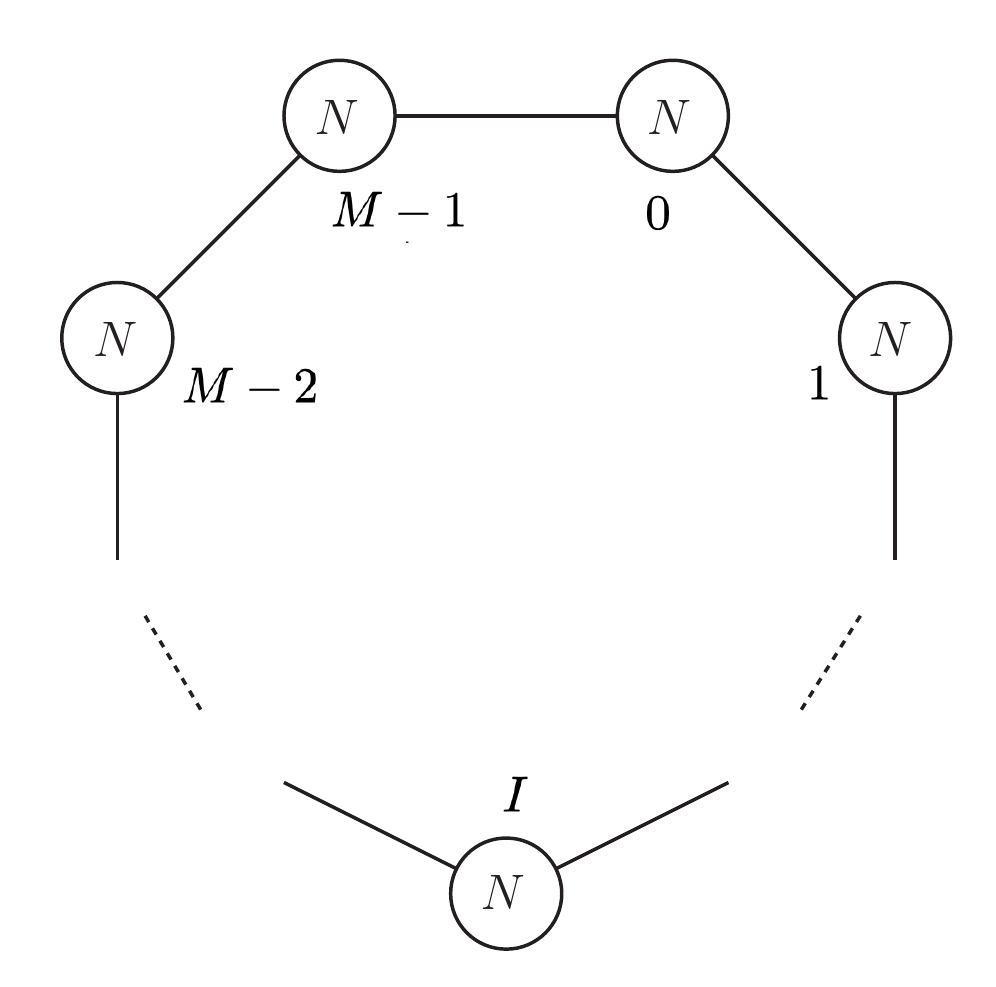}
		\caption{A graphical representation of the quiver theory with $M$ nodes. Each node, labeled by an index $I=0,1,\ldots,M-1$, stands for a SU($N$) factor with its adjoint vector multiplet. The lines connecting two neighboring nodes represent bi-fundamental matter hypermultiplets. 
		\label{fig:1_quiver}} 
	}  
\end{figure}

These quiver theories, which are obtained with a $\mathbb{Z}_M$ orbifold projection from $\mathcal{N}=4$ SYM, can be represented as in Fig.~\ref{fig:1_quiver}, where a 
$\mathrm{SU}(N)$ group factor is associated to each node and bi-fundamental matter hypermultiplets are in correspondence to the links.
They can be realized in string theory as the low-energy theory of the massless sector of open strings attached to fractional D3 branes on a $\mathbb{C}^2/\mathbb{Z}_M$ orbifold background \cite{Douglas:1996sw}, and possess a holographic dual given by Type II B string theory on the $\mathrm{AdS}_5\times S^5/\mathbb{Z}_M$ space \cite{Kachru:1998ys,Gukov:1998kk}. 
Because of these features, they represent one of the simplest set-ups to investigate 
the strong-coupling regime and to explore the holographic correspondence when supersymmetry
is not maximal. Beside being very interesting in their own, these quiver theories give rise, upon further orbifold or orientifold projections, to other superconformal $\mathcal{N}=2$ models, like for instance the so-called $\mathbf{E}$-theory with gauge group SU$(N)$ and matter in the antisymmetric plus the symmetric representation, recently discussed in \cite{Beccaria:2021hvt,Billo:2022xas}. Furthermore, they are also interesting from the integrability point of view (see \cite{Beisert:2010jr} and references therein) and can be used to analyze the tensionless limit of string theory in the 
$\mathrm{AdS}_5\times S^5/\mathbb{Z}_M$ background and to probe the dual free 
world-sheet description in a context with $\mathcal{N}=2$ supersymmetry \cite{Gaberdiel:2022iot}.

In the quiver theories there are sectors of BPS protected observables that can be profitably studied with supersymmetric localization techniques \cite{Pestun:2007rz,Pestun:2016jze}. Examples of such observables
are the partition function and the expectation value of circular Wilson loops \cite{Rey:2010ry,Mitev:2015oty,Zarembo:2020tpf,Fiol:2020ojn,Ouyang:2020hwd,Beccaria:2021ksw,Beccaria:2021vuc,Galvagno:2021bbj,Beccaria:2021ism}. 
In this paper we focus on another class of observables, namely the 2- and 3-point functions of a set of single-trace scalar operators. These correlation functions have been considered in the 
literature from many points of view and for various $\mathcal{N}=2$ superconformal theories
(see for example \cite{Baggio:2014sna,Baggio:2015vxa,Gerchkovitz:2016gxx,Baggio:2016skg,Rodriguez-Gomez:2016ijh,Rodriguez-Gomez:2016cem,Pini:2017ouj,Billo:2017glv,Bourget:2018obm,Beccaria:2018xxl,Billo:2019fbi,Beccaria:2020azj,Beccaria:2020hgy,Niarchos:2020nxk,Galvagno:2020cgq,Beccaria:2021hvt,Fiol:2021icm,Billo:2021rdb,Billo:2022xas}). Here we will study such correlators for the $\mathbb{Z}_M$ quiver theories in the 't Hooft limit of a large number of colors and derive from them the normalized structure constants, which are part of the
intrinsic conformal field theory data. In general, these structure constants are non-trivial functions of the 't Hooft coupling $\lambda$, and an efficient way to get them is to use localization, which maps the computation to an interacting matrix model. By exploiting the recursive matrix-model techniques introduced in \cite{Billo:2017glv} and named ``full Lie algebra approach'' in \cite{Fiol:2018yuc}, it is possible to generate explicit expressions for the coefficients of the 2- and 3-point functions and the corresponding structure constants, both at finite $N$ and in the large-$N$ limit. This expressions are typically power series in $\lambda$ and are therefore valid only in the weak-coupling regime when $\lambda\to 0$. However, for the leading terms at large $N$ we manage to resum these perturbative expansions into functions that are valid for \emph{all} values of $\lambda$. From these resummed expressions, we then extract the leading behavior for $\lambda\to\infty$ .

On the other hand, the strong-coupling regime of these quiver theories can be accessed by means of the AdS/CFT correspondence \cite{Maldacena:1997re}. Applying the AdS/CFT dictionary 
we identify the supergravity modes that are dual to the operators of the quiver theory and from their effective action in the AdS space we obtain the 2- and 3-point
functions. The corresponding structure constants computed with these holographic methods perfectly
match those obtained from localization at strong coupling.
This agreement can be seen either as a validation of the strong-coupling extrapolation of the localization results or, alternatively, as an explicit check of the AdS/CFT correspondence for a non-maximally supersymmetric theory in four dimensions. To our knowledge, this is the first example where an analytic interpolation between weak and strong coupling is presented
together with an independent holographic calculation for a four-dimensional $\mathcal{N}=2$ superconformal theory. A summary of these
results for the quiver theory with two nodes has been recently published in \cite{Billo:2022gmq} where only the main ideas have been described, omitting demonstrations and technical details. In this paper, instead, we consider the general $M$-node quiver theories and discuss in detail all aspects of the calculations.

\subsection*{Summary of results}

We have studied the 2- and 3-point functions of a set of gauge-invariant
operators built with the scalar fields of the adjoint vector multiplets of the
various nodes of the quiver. As in \cite{Billo:2021rdb}, we have actually introduced specific combinations of these operators denoted as
$U_k$ and $T_{\alpha,k}$ which are called, respectively, untwisted and twisted and are in one-to-one correspondence with the conjugacy classes of $\mathbb{Z}_M$. These operators are chiral, have a protected conformal dimension
\begin{equation}
\Delta_{U_k}=\Delta_{T_{\alpha,k}}=k
\label{confdim}
\end{equation}
and carry a U(1)$_R$ charge which, in our conventions, equals to $k$. By replacing the adjoint scalar fields with their complex conjugates, one obtains the anti-chiral operators $\overbar{U}_k$ and $\overbar{T}_{\alpha,k}$
which have the same conformal dimension as the chiral ones, but
opposite charge. The detailed expressions of these operators
can be found in Section~\ref{secn:quiver}.

The correlators that involve only untwisted operators are planar equivalent to those of the $\cN=4$ SYM theory and in the holographic correspondence they are associated to Kaluza-Klein modes of the metric and of 
the 4-form of the Ramond/Ramond sector, exactly as in \cite{Lee:1998bxa}. 
The twisted operators, instead, are dual to Kaluza-Klein modes 
of the scalar fields that arise from wrapping
the Neveu-Schwarz/Neveu-Schwarz and Ramond/Ramond 2-forms
around the exceptional cycles of the orbifold resolution \cite{Gukov:1998kk}. From the string point of view, these are modes of twisted closed strings, and their correlators receive corrections at all orders in perturbation theory even in the large-$N$ limit.

In the first part of the paper (Sections~\ref{secn:matrix} and \ref{secn:strong}) we exploit matrix-model techniques
to determine the leading contribution at large $N$ to the 2-point and 3-point correlators of untwisted and twisted operators.
As already discussed in \cite{Billo:2021rdb}, the untwisted 2-point functions $\big\langle U_k\, \overbar U_k\big\rangle$ do not depend on $\lambda$, while the twisted ones, 
$\big\langle T_{\alpha,k}\, \overbar T_{\alpha,k}\big\rangle$ do and are
proportional to $1/\lambda$ at strong coupling.
The 3-point functions can be of various types depending on how many twisted operators they involve. If there are only untwisted operators, they are of the type $\big\langle U_k\,U_\ell\,\overbar{U}_p
\big\rangle$ (or with chiral and anti-chiral operators exchanged). These correlators are non zero 
only if $p=k+\ell$ because of charge conservation and are independent of $\lambda$ in the planar approximation. When there are two twisted operators, the correlators can be of the type $\big\langle
U_k\, T_{\alpha,\ell}\, \overbar T_{\alpha,p}\big\rangle$ or
$\big\langle T_{\alpha,k}\, T_{M-\alpha,\ell}\, \overbar U_{p} \big\rangle$
(or their conjugates) and are analogous to
the ones computed in \cite{Billo:2022xas} for the so-called $\mathbf{E}$-theory. In this paper we evaluate them by means of a more compact and general method based on the properties of an infinite matrix 
$\mathsf{X}$, firstly introduced in \cite{Beccaria:2020hgy} and later
extended in \cite{Billo:2021rdb}, that encodes all interactions of 
the matrix model in a convolution of Bessel functions and allows one to obtain results that are valid at all values of the 't Hooft coupling. This method can be nicely rephrased with a diagrammatic formalism which helps
in organizing the various contributions in a neat way. Extrapolating these results at strong coupling, we find that the correlators involving two twisted operators of the type mentioned above scale as $1/\lambda$ when $\lambda\to\infty$. These same methods allow us to compute also 
the correlators with three twisted fields, namely $\big\langle
T_{\alpha,k}\, T_{\beta,\ell}\, \overbar T_{\gamma,p}\big\rangle$, 
which are present for $\mathbb{Z}_M$ quivers with $M\geq 3$. They are
proportional to $\delta_{\alpha+\beta,\gamma}$ and behave as $1/\lambda^{3/2}$ at strong coupling. 

Of course, both the 2- and the 3-point functions depend on the way the operators have been normalized. To overcome this ambiguity we compute the normalization-independent structure constants, namely the 3-point functions of the normalized operators. In the planar limit and at strong coupling, we find that these structure constants are simple functions of the conformal dimensions (\ref{confdim}) given by
\begin{subequations}
\begin{align}
		C_{U_k,U_\ell,\overbar{U}_p} &= \frac{1}{\sqrt{M}\,N}\,
		\sqrt{k\,\ell\,p}~,\label{CUUU}\\
	C_{U_k,T_{\alpha,\ell},\overbar{T}_{\alpha,p}}	&= \frac{1}{\sqrt{M}\,N}\,
	\sqrt{k\,(\ell-1)\,(p-1)}~,\label{CUTT}\\
C_{T_{\alpha,k},T_{M-\alpha,\ell},\overbar{U}_{p}}	&= \frac{1}{\sqrt{M}\,N}\,
	\sqrt{(k-1)\,(\ell-1)\,p}~,\label{CTTU}\\
	C_{T_{\alpha,k},T_{\beta,\ell},\overbar{T}_{\gamma,p}}	
	&= \frac{1}{\sqrt{M}\,N}\,
	\sqrt{(k-1)\,(\ell-1)\,(p-1)}\,\,\delta_{\alpha+\beta,\gamma}~.
	\label{CTTT}
\end{align}
\label{Cintro}%
\end{subequations}
All these formulas, in which we have understood the factor $\delta_{k+\ell-p,0}$ for charge conservation, are strictly valid in the large-$N$ and large-$\lambda$ limits. We notice that the untwisted structure constants (\ref{CUUU}) are actually $\lambda$-independent and, apart from the
factor of $\sqrt{M}$ due to the $\mathbb{Z}_M$ orbifold, they coincide
with those of the $\mathcal{N}=4$ SYM theory \cite{Lee:1998bxa}. The
other structure constants are, instead, a new strong-coupling result.

In the second part of the paper (Sections~\ref{secn:holo} and \ref{secn:effective}), we study the strong coupling regime of the quiver gauge theories by means of the AdS/CFT correspondence.
As mentioned above, the holographic dual description of the $\mathbb{Z}_M$ quivers is given in terms of Type II B strings propagating in an orbifold geometry of the form AdS$_5\times S^5/\mathbb{Z}_M$
\cite{Kachru:1998ys,Gukov:1998kk}. In the untwisted sector of this orbifold we identify the scalar modes $s_k$ that are dual to the untwisted operators $U_k$ and compute their 2- and 3-point functions in a holographic manner from their effective supergravity action. 
Just like in the $\mathcal{N}=4$ case \cite{Lee:1998bxa}, also for the quiver theories these untwisted correlators do not depend on $\lambda$. Extending this analysis to the twisted sectors of AdS$_5\times S^5/\mathbb{Z}_M$, we identify the scalar modes $\eta_{\alpha,k}$ that are dual to the twisted operators $T_{\alpha,k}$. These scalar modes \cite{Gukov:1998kk} are localized at the orbifold fixed-locus and their effective action can be obtained by dimensionally reducing the Type II B supergravity action to AdS$_5\times S^1$, where $S^1 \subset S^5$ is the circle fixed by the orbifold projection. In this way we can obtain the 2-point functions 
$\big\langle T_{\alpha,k}\, \overbar T_{\alpha,k}\big\rangle$ and the 3-point functions $\big\langle U_k\,T_{\alpha,\ell}\, \overbar T_{\alpha,p}\big\rangle$, respectively, from the quadratic and cubic parts of the twisted effective action.  Of course the normalization chosen in this supergravity calculations is different from the one of the localization approach, but it is interesting to see that 
these holographic correlators scale as $1/\lambda$, exactly as their matrix-model counterparts at strong coupling. To obtain a more convincing check, we compute the normalization-independent structure constants $C_{U_k,T_{\alpha,\ell},\overbar{T}_{\alpha,p}}$ and find exactly the same expression as in (\ref{CUTT}), showing a perfect agreement between localization and
holography. However, we note that the effective twisted action that is derived
from Type II B supergravity is quadratic in the twisted modes and thus cannot yield the 3-point functions with three twisted fields which instead can be easily found in the gauge theory. Presumably, to obtain these correlators from the AdS/CFT correspondence 
one has to consider higher-derivative string corrections to the effective action of the twisted modes. We leave this issue to future investigations.

This paper contains also four appendices. In the first one we provide details on the asymptotic
behavior at strong coupling of the building blocks that are used to write the 2- and 3-point functions in the localization approach. In the second appendix we show that the typical geometric structure of the resolved $\mathbb{Z}_M$ orbifold used in the holographic approach actually emerges from the localization results once they are extrapolated at strong coupling. Finally, the last two appendices contain details and identities that are useful for the calculations done with the AdS/CFT correspondence.

\section{The \texorpdfstring{$\mathcal{N}=2$}{} quiver theory}
\label{secn:quiver}
The 4-dimensional $\mathcal{N}=2$ quiver theory represented by the necklace diagram in Fig.~\ref{fig:1_quiver}
has a gauge group that is a product of $M$ factors of SU($N$), labeled with an index $I=0,1,\ldots,M-1$ (defined modulo $M$), and has matter fields that form $M$ bi-fundamental hypermultiplets. Given this field content, there are eight conserved supercharges and in each node the $\beta$-function vanishes. This means that we have $\mathcal{N}=2$ superconformal invariance at the quantum level. While in general it would be possible to assign different gauge couplings to the different nodes, we take the most symmetric attitude and assume that all Yang-Mills couplings are equal to $g$. Since we will be interested in the planar limit, we introduce the 't Hooft coupling
\begin{equation}
\lambda=N g^2
\label{lambda}
\end{equation}
which is kept fixed when $N\to\infty$. In this limit, instanton contributions are exponentially suppressed and will not be considered.

It is interesting to note that this quiver theory can be obtained as a 
$\mathbb{Z}_M$ orbifold projection from a parent $\mathcal{N}=4$ SYM theory 
with gauge group SU($MN$). Under this orbifold projection the gauge group is broken to $\mathrm{SU}(N)^{M}$ and
the $R$-symmetry group SU(4)$_R$ is broken to its SU(2)$_R \times$ 
U(1)$_R$ subgroup. Thus, only half of the original sixteen supersymmetries survive
and the resulting theory has $\mathcal{N}=2$ supersymmetry.
Moreover, the gauge field $\mathcal{A}$ 
and one of the three complex scalars of the original $\mathcal{N}=4$ SYM, which we call $\phi$, 
are reduced to $M$ block-diagonal components of size $N\times N$ denoted, respectively, $\mathcal{A}_I$ and $\phi_I$. These fields, together with their fermionic super-partners, form $M$ vector multiplets in the adjoint representation 
of SU($N$) associated to the $M$ nodes of the quiver. The other two complex scalars of the parent 
$\mathcal{N}=4$ SYM are instead reduced to $N\times N$ off-diagonal 
blocks which, together with their fermionic super-partners, give rise to $M$ bi-fundamental hypermultiplets represented by the lines connecting two neighboring nodes of the quiver.
This orbifold construction has a direct realization in Type II B string theory by means of fractional D3-branes placed on a $\mathbb{C}^2/\mathbb{Z}_M$ orbifold singularity \cite{Douglas:1996sw,Bertolini:2000dk,Billo:2001vg}. This fact will be exploited when we will discuss the holographic calculations in Section~\ref{secn:holo}.

Our main interest is to study this theory in the large-$N$ limit and in the regime where the 't Hooft coupling $\lambda$ tends to infinity. To this aim, we will consider the correlation functions of a special class of protected gauge-invariant local operators defined as
\begin{subequations}
\begin{align}
U_k(x)&=\frac{1}{\sqrt{M}}\,\Big[\tr \phi_0^k(x)+\tr \phi_1^k(x)+\ldots+\tr \phi_{M-1}^k(x)\Big]~,\label{defUk}\\
T_{\alpha,k}(x)&=\frac{1}{\sqrt{M}}\,\sum_{I=0}^{M-1}\rho^{-\alpha\,I}\,\tr \phi_I^k(x)~,
\label{defTk}
\end{align}
\label{operators}%
\end{subequations}
where $k=2,3,\ldots$, $\alpha=1,\ldots,M-1$ and $\rho$ is the $M$-th root of unity:
\begin{equation}
\rho=\mathrm{e}^{\frac{2\pi\ii}{M}}~.
\label{rho}
\end{equation}
The operators $U_k$ are called untwisted, while the operators $T_{\alpha,k}$ are called twisted.
The reason for this terminology is two-fold. Firstly, under the cyclic permutation $\phi_I\to\phi_{I+1}$ that generates $\mathbb{Z}_M$, the operators $U_k$ and $T_{\alpha,k}$ transform as
\begin{equation}
U_k(x)\to U_k(x)\quad\mbox{and}\quad
T_{\alpha,k}(x)\to\rho^\alpha\,T_{\alpha,k}(x)~,
\label{TtoT}
\end{equation}
and are thus associated to the untwisted and twisted sectors of $\mathbb{Z}_M$. Secondly, as shown in
\cite{Gukov:1998kk,Billo:2021rdb}, in the holographic correspondence
the operators $U_k$ are dual to supergravity modes that belong to the untwisted sector of $\mathbb{C}^2/\mathbb{Z}_M$, while the operators $T_{\alpha,k}$ are dual to string excitations that belong to the $\alpha$-twisted sector of the orbifold. 

The operators (\ref{operators}) are chiral primary operators of conformal dimension $k$ and charge $k$\,%
\footnote{In our conventions the chiral fields $\phi_I$ have charge $+1$ and the anti-chiral fields $\overbar{\phi}_I$ have charge $-1$.}. If we replace the chiral fields $\phi_I$ with their complex conjugates $\overbar{\phi}_I$, we obtain the anti-chiral
operators $\overbar{U}_k(x)$ and $\overbar{T}_{\alpha,k}(k)$, which are primary operators of dimension $k$ and charge $-k$. Using these definitions, it is easy to see that
\begin{equation}
\overbar{U}_k(x)=\big[U_k(x)\big]^*
\quad\mbox{and}\quad
\overbar{T}_{\alpha,k}(x)=\big[T_{M-\alpha,k}(x)\big]^*
~.
\label{UUbarTTbar}
\end{equation}

Charge conservation, conformal invariance and symmetry under $\mathbb{Z}_M$ fix the form of the 2-point functions of the above operators to be
\begin{align}
\big\langle U_k(x)\,\overbar{U}_k(y)\big\rangle=\,
\frac{G_{U_k}}{|x-y|^{2k}}\quad\mbox{and}\quad
\big\langle T_{\alpha,k}(x)\,\overbar{T}_{\alpha,k}(y)\big\rangle=\,
\frac{G_{T_{\alpha,k}}}{|x-y|^{2k}}~,
\label{GUUGTT}
\end{align}
where the coefficients $G_{U_k}$ and $G_{T_{\alpha,k}}$ are functions of $N$ and $\lambda$.
Also the 3-point functions are constrained by the symmetries of the quiver theory. Because of charge
conservation, two of the operators must be chiral and the third one must be anti-chiral  (or viceversa) to soak up the charge. 
In addition, because of the $\mathbb{Z}_M$ symmetry the total twist of the three operators must be 0 modulo $M$. Taking these constraints into account we have the following possibilities:
\begin{itemize}
\item a 3-point function with all untwisted operators
\begin{align}
\big\langle U_k(x)\,U_\ell(y)\,\overbar{U}_p(z)\big\rangle =
\frac{G_{U_k,U_\ell,\overbar{U}_p}}{|x-z|^{2k}\,|y-z|^{2\ell}}~;
\label{GUUUbar}
\end{align}
\item a 3-point function with chiral and anti-chiral twisted operators
\begin{align}
\big\langle U_k(x)\,T_{\alpha,\ell}(y)\,\overbar{T}_{\alpha,p}(z)\big\rangle =
\frac{G_{U_k,T_{\alpha,\ell},\overbar{T}_{\alpha,p}}}{|x-z|^{2k}\,|y-z|^{2\ell}}~;
\label{GUTTbar}
\end{align}
\item a 3-point function with two chiral twisted operators belonging to conjugate sectors
\begin{align}
\big\langle T_{\alpha,k}(x)\,T_{M-\alpha,\ell}(y)\,\overbar{U}_{p}(z)\big\rangle =
\frac{G_{T_{\alpha,k},T_{M-\alpha,\ell},\overbar{U}_{p}}}{|x-z|^{2k}\,|y-z|^{2\ell}}~;
\label{GTTUbar}
\end{align}
\item a 3-point function with all twisted operators
\begin{align}
\big\langle T_{\alpha,k}(x)\,T_{\beta,\ell}(y)\,\overbar{T}_{\alpha+\beta,p}(z)\big\rangle =
\frac{G_{T_{\alpha,k},T_{\beta,\ell},\overbar{T}_{\alpha+\beta,p}}}{|x-z|^{2k}\,|y-z|^{2\ell}}~,
\label{GTTTbar}
\end{align}
which is possible if $M\not=2$.
\end{itemize}
In all above correlators we have understood the $\delta$-function $\delta_{k+\ell-p,0}$ enforcing charge conservation, and determined the space dependent part using conformal invariance.
All coefficients in the numerators of (\ref{GUUUbar})--(\ref{GTTTbar}) are functions of
$N$ and $\lambda$. Besides these 3-point functions, we have of course also the conjugate ones in which chiral and anti-chiral operators are exchanged.

Both the 2-point functions (\ref{GUUGTT}) and the 3-point functions (\ref{GUUUbar})--(\ref{GTTTbar})  are sensitive to the normalization of the operators. To remove such dependence, one introduces the structure constants:
\begin{equation}
\begin{aligned}
C_{U_k,U_\ell,\overbar{U}_p}&=\frac{G_{U_k,U_\ell,\overbar{U}_p}}{\sqrt{G_{U_k}\,G_{U_\ell}\,G_{U_p}}}
~,\qquad
C_{U_k,T_{\alpha,\ell},\overbar{T}_{\alpha,p}}=
\frac{G_{U_k,T_{\alpha,\ell},\overbar{T}_{\alpha,p}}}
{\sqrt{G_{U_k}\,G_{T_{\alpha,\ell}}\,G_{T_{\alpha,p}}}}
~,\\
C_{T_{\alpha,k},T_{M-\alpha,\ell},\overbar{U}_{p}}&=
\frac{G_{T_{\alpha,k},T_{M-\alpha,\ell},\overbar{U}_{p}}}
{\sqrt{G_{T_{\alpha,k}}\,G_{T_{M-\alpha,\ell}}\,G_{U_{p}}}}
~,\qquad
C_{T_{\alpha,k},T_{\beta,\ell},\overbar{T}_{\gamma,p}}=
\frac{G_{T_{\alpha,k},T_{\beta,\ell},\overbar{T}_{\gamma,p}}}
{\sqrt{G_{T_{\alpha,k}}\,G_{T_{\beta,\ell}}\,G_{T_{\gamma,p}}}}
~,\label{structure}
\end{aligned} 
\end{equation}
which are part of the intrinsic CFT data of the theory.

In the following we will study the 3-point correlators (\ref{GUUUbar})--(\ref{GTTTbar}) 
with particular emphasis on the
strong-coupling behavior of the coefficients $G$ in the large-$N$ limit and on the structure constants (\ref{structure}), generalizing what was done in \cite{Billo:2021rdb} for the 2-point functions. We will do this by employing two methods:
\begin{itemize}
\item[I)] supersymmetric localization, which will be discussed in the next two sections;
\item[II)] holography within the AdS/CFT correspondence, as we will see in the second part of the paper.
\end{itemize}

\part{Localization}
\label{part:1}
The computation of the 2- and 3-point functions introduced in Section~\ref{secn:quiver} can be efficiently performed using 
supersymmetric localization \cite{Pestun:2016jze}. With this technique 
one maps a $\mathcal{N}=2$ superconformal theory in
$\mathbb{R}^4$ to an interacting matrix model defined on a 4-sphere
${S}^4$ \cite{Pestun:2007rz}, so that the computation of correlation functions is reduced to the evaluation of finite dimensional matrix integrals. In the following we will employ the ``full Lie algebra" approach to localization introduced in \cite{Billo:2017glv}, which permits to obtain explicit expressions for the matrix integrals both at finite $N$ and in the large $N$-limit by exploiting recursion relations.

\section{The matrix-model description}
\label{secn:matrix}

Once the quiver theory is placed on a sphere ${S}^4$ with unit radius, its partition function 
$\mathcal{Z}$ localizes and can be written as an integral over a set of $M$
traceless Hermitian $N \times N$ matrices $a_I$:
\begin{align}
\label{Zpartition}
\mathcal{Z} = \int \prod_{I=0}^{M-1}\left(da_I \ \textrm{e}^{-\textrm{tr}a_I^2}\right)|Z_{\textrm{1-loop}}\,Z_{\textrm{inst}}|^2 ~.
\end{align}
Here, $Z_{\textrm{1-loop}}$ encodes the 1-loop determinants of the fluctuations around the fixed points, while $Z_{\textrm{inst}}$ accounts for the non-perturbative instanton corrections. However, since in the large $N$-limit instantons are exponentially suppressed, henceforth we will set $Z_{\textrm{inst}}=1$. In the ``full Lie algebra" approach the integrations in (\ref{Zpartition}) are
performed over all matrix elements. If we write $a_I$ as a linear combination of the $\mathfrak{su}(N)$ generators $T_b$ in the fundamental representation, namely
\begin{align}
a_I = a_I^b\,T_b\quad\mbox{where}\quad \tr T_b\,T_c = \frac{1}{2}\,\delta_{b,c}
\quad\mbox{with}\quad b,c=1,\cdots ,N^2-1~,
\end{align} 
then the integration measure in (\ref{Zpartition}) is
\begin{align}
\label{daI}
da_I \equiv \prod_{b=1}^{N^2-1}\frac{da_I^b}{\sqrt{2\pi}}~,
\end{align}
where the normalization is fixed by requiring that the Gaussian integration for each $a_I$ gives 1. 
The 1-loop contribution $Z_{\textrm{1-loop}}$
can be written in terms of an interaction action $S_{\textrm{int}}$
as follows
\begin{align}
|Z_{\textrm{1-loop}}|^2 = \textrm{e}^{-S_{\textrm{int}}} ~,
\end{align}
with\,%
\footnote{Recall that $a_M\equiv a_0$.}
\begin{align}
\label{Sintstart}
S_{\textrm{int}}= \sum_{I=0}^{M-1}\bigg[\sum_{m=2}^{\infty}\sum_{k=2}^{2m}(-1)^{m+k}\Big(\frac{\lambda}{8\pi^2N}\Big)^{\!m}
\binom{2m}{k}\,
\frac{\zeta_{2m-1}}{2m}
\big(\tr a_I^{2m-k}-\tr a_{I+1}^{2m-k}\big)
\big(\tr a_I^{k}-\tr a_{I+1}^{k}\big)\bigg]
\end{align}
where $\lambda$ is the 't Hooft coupling (\ref{lambda}) and $\zeta_{2m-1}$ is 
the Riemann $\zeta$-value $\zeta(2m-1)$.

In this set-up, given a generic function $f$ of the $a_I$'s, its expectation value reads
\begin{align}
\big\langle f \big\rangle = \frac{1}{\mathcal{Z}}\int \!\bigg(\prod_{I=0}^{M-1} da_I  \ \textrm{e}^{-\tr a_I^2}\bigg) \, 
f \ \mathrm{e}^{-S_{\textrm{int}}} = \frac{\big\langle f\ \mathrm{e}^{-S_{\textrm{int}}} \big\rangle_0}{\big\langle \mathrm{e}^{-S_{\textrm{int}}} \big\rangle_{0}}
\label{vevf}
\end{align}
where $\big\langle \cdot \big\rangle_0$ denotes the vacuum expectation value in the Gaussian matrix model. 
These Gaussian expectation values can be efficiently evaluated exploiting 
the recursion relations obeyed by the set of the following quantities
\begin{align}
	\label{tks}
		t_{n_1 n_2 \ldots n_p} = \big\langle
		\tr a^{n_1}\, \tr a^{n_2}\, \ldots \tr a^{n_p}\big\rangle_0~,  
\end{align}
where $a$ stands for any of the matrices $a_I$, which were derived in 
\cite{Billo:2017glv} starting from the fusion/fission identities of $\mathrm{SU}(N)$.

\subsection{Untwisted and twisted operators}
In analogy with what one does in the quiver gauge theory, also in the matrix model it's natural to introduce the untwisted and twisted linear combinations
\begin{subequations}
\begin{align}
& A_k = 
\frac{1}{\sqrt{M}}\,\Big[\tr a_0^k+\tr a_1^k+\ldots+\tr a_{M-1}^k\Big]
-\sqrt{M}\, t_k~,
\label{Au} \\
& A_{\alpha,k} = \frac{1}{\sqrt{M}}\sum_{I=0}^{M-1}\rho^{-\alpha I}\,\tr a_{I}^k ~,  \label{At}
\end{align}
\label{Aut}%
\end{subequations}
with $k=2,3,4,\ldots$. Here the index $\alpha=1,\ldots,M-1$ labels the twisted sector and
$\rho$ is the $M$-th root of unity (\ref{rho})\,%
\footnote{Notice that because of the term proportional to $t_k$ that appears in the definition (\ref{Au}), the untwisted operators $A_k$ have vanishing vacuum expectation value: $\langle A_k\rangle_0=0$. The same is true for the twisted operators: $\langle A_{\alpha,k}
\rangle_0=0$.}.
Actually, we can combine the two definitions (\ref{Aut}) and write more compactly
\begin{equation}
A_{\widehat{\alpha},k}=\frac{1}{\sqrt{M}}\sum_{I=0}^{M-1}\rho^{-\widehat{\alpha} I}\,\tr a_{I}^k-\sqrt{M}\, t_k\,\delta_{\widehat{\alpha},0}
\label{Ahat}
\end{equation}
where the new index $\widehat{\alpha}=0,1,\ldots,M-1$ is defined modulo $M$. In particular, the value $\widehat{\alpha}=0$ is associated to the untwisted sector, {\it{i.e.}} $A_{0,k}=A_k$, while 
the non-zero values of $\widehat{\alpha}$ correspond to the twisted sectors. Using these definitions, we easily realize that
\begin{align}
A_{\widehat{\alpha},k}^{\dagger} = A_{M-\widehat{\alpha},k}
\end{align}
for any $\widehat{\alpha}$.
It will be useful in the following to collect the single trace operators in each sector $\widehat{\alpha}$
into an infinite column vector $\mathbf{A}_{\widehat{\alpha}}$
defined as
\begin{align}
\mathbf{A}_{\widehat{\alpha}}=
\begin{pmatrix}
A_{\widehat\alpha,2}\\
A_{\widehat\alpha,3}\\
\vdots
\end{pmatrix}~.
\label{Avec}
\end{align}

Even if the operators (\ref{Ahat}) have a structure resembling that of the primary operators (\ref{operators}), they cannot properly represent the latter
\cite{Gerchkovitz:2016gxx}.
Indeed, differently from the operators of the quiver gauge theory,
the matrix-model operators (\ref{Ahat}) mix with those of lower dimension. To disentangle this mixing, it is necessary to 
introduce a normal-ordered version of $A_{\widehat{\alpha},k}$ which can
be obtained by applying the Gram-Schmidt orthogonalization procedure
in each sector $\widehat{\alpha}$. 
As shown in \cite{Baggio:2016skg,Billo:2022xas}, in planar limit it is 
enough to consider the mixing of the single-trace operators among themselves, and thus we define
\begin{equation}
P_{\widehat{\alpha},k}(\lambda)=A_{\widehat{\alpha},k}
-\sum_{\ell<k}\mathsf{C}^{(\widehat{\alpha})}_{k,\ell}(\lambda)\,P_{\widehat{\alpha},\ell}(\lambda)
\label{Pk}
\end{equation}
where the mixing coefficients $\mathsf{C}^{(\widehat{\alpha})}_{k,\ell}(\lambda)$ 
form a lower triangular matrix $\mathsf{C}^{(\widehat{\alpha})}(\lambda)$ and are determined by demanding that $P_{\widehat{\alpha},k}(\lambda)$ be orthogonal to all
lower dimensional operators, namely
\begin{equation}
\big\langle\, P_{\widehat{\alpha},k}^{\phantom{\dagger}}(\lambda)\,\,P_{\widehat{\alpha},\ell}^\dagger(\lambda)
\,\big\rangle=0\quad\mbox{for all}\quad \ell<k~.
\label{ortho0}
\end{equation}
Imposing this requirement leads to
\begin{equation}
\mathsf{C}^{(\widehat{\alpha})}_{k,\ell}(\lambda)=\frac{\big\langle  A_{\widehat{\alpha},k}^{\phantom{\dagger}}\,\,P_{\widehat{\alpha},\ell}^\dagger(\lambda)
\big\rangle\phantom{\Big|}}{\big\langle P_{\widehat{\alpha},\ell}^{\phantom{\dagger}}(\lambda) \,P_{\widehat{\alpha},\ell}^\dagger(\lambda)
\big\rangle\phantom{\Big|}}=\frac{\big\langle  A_{\widehat{\alpha},k}^{\phantom{\dagger}}\,\,P_{\widehat{\alpha},\ell}^\dagger(\lambda)
\big\rangle\phantom{\Big|}}{\big\langle A_{\widehat{\alpha},\ell}^{\phantom{\dagger}} \,P_{\widehat{\alpha},\ell}^\dagger(\lambda)
\big\rangle\phantom{\Big|}}
\end{equation}
where $k+\ell$ is even with $\ell<k$.
Notice that we can also rewrite (\ref{Pk}) to express $P_{\widehat{\alpha},k}(\lambda)$ as a combination
of the original operators $A_{\widehat{\alpha},k}$. 
Defining the infinite column vector 
$\mathbf{P}_{\widehat{\alpha}}(\lambda)$ in analogy with (\ref{Avec}), we have
\begin{equation}
\mathbf{P}_{\widehat{\alpha}}(\lambda) =
\mathsf{M}^{(\widehat{\alpha})}(\lambda)\, \mathbf{A}_{\widehat{\alpha}}
\label{PversusA}
\end{equation}
where $\mathsf{M}^{(\widehat{\alpha})}(\lambda)=
\big[\mathbb{1}+\mathsf{C}^{(\widehat{\alpha})}(\lambda)\big]^{-1}$.

As mentioned above, the operators $P_{\widehat{\alpha},k}(\lambda)$ properly represent in the
matrix model the primary operators (\ref{operators}) of the gauge theory; more precisely the
correspondence is 
\begin{equation}
U_k(x)~\longleftrightarrow~P_{0,k}(\lambda)\quad\mbox{and}\quad
T_{\alpha,k}(x)~\longleftrightarrow~P_{\alpha,k}(\lambda)~,
\label{UTP}
\end{equation} 
with similar relations for the conjugate operators. Therefore,
using this map, the 2- and 3-point functions of the primary operators defined in
Section~\ref{secn:quiver} are computed in the matrix model by the 2- and 3-point correlators of the
normal-ordered operators $P_{\widehat{\alpha},k}(\lambda)$. For example we have\,%
\footnote{Here and in the following, we use the symbol $\simeq$ to denote the leading term in the large-$N$ limit.}
\begin{equation}
G_{T_{\alpha,k}} \,\simeq\,\big\langle P_{\alpha,k}^{\phantom{\dagger}}(\lambda)\,P_{\alpha,k}^\dagger(\lambda)\big\rangle\quad
\mbox{and}\quad
G_{U_k,T_{\alpha,\ell},\overbar{T}_{\alpha,p}}\,\simeq\,
\big\langle P_{0,k}^{\phantom{\dagger}}(\lambda)\,P_{\alpha,\ell}^{\phantom{\dagger}}(\lambda)\,P_{\alpha,p}^\dagger(\lambda)\big\rangle~,
\label{examples}
\end{equation}
with analogous expressions for the other cases.

\subsection{Free correlators in the large-\texorpdfstring{$N$}{} limit}
The matrix-model correlators like those in (\ref{examples}) can be evaluated using the definition (\ref{vevf}), even if it is difficult to write the result in a compact way for the general case. 
However, in the large-$N$ limit remarkable simplifications occur. 

As shown in \cite{Billo:2021rdb}, the key to compute correlators at large 
$N$ is to first analyze the scaling of correlators with $N$ in the free model. Using the recursion relations satisfied by the multi-traces (\ref{tks}), for the 2-point functions
we find that 
\begin{align}
	\label{AA}
		\big\langle A_{\widehat{\alpha},k}^{\phantom{\dagger}}
		\, A_{\widehat{\beta},\ell}^\dagger \big\rangle_0 \,\propto\, 
		N^{\frac{k+\ell}{2}} \,\delta_{\widehat\alpha,\widehat{\beta}}
\end{align}
if $k+\ell$ is even, otherwise the correlator vanishes. For the 3-point functions, instead, we have  
\begin{align}
	\label{AAA}
	\big\langle A_{\widehat{\alpha},k}^{\phantom{\dagger}}
	\, A_{\widehat{\beta},\ell}^{\phantom{\dagger}}\, A_{\widehat{\gamma},p}^\dagger\big\rangle_0 
	\,\propto\,  N^{\frac{k+\ell+p}{2}-1}\, \delta_{\widehat\alpha + \widehat{\beta} ,\widehat{\gamma}}
\end{align}
if $k+\ell+p$ is even, otherwise the correlator vanishes.  In higher correlators with an even number of $A$'s, the leading contribution at large $N$
can be computed by factorizing them \`a la Wick in terms of 2-point functions. Indeed, it follows from (\ref{AA}) and (\ref{AAA}) that decompositions which include two 3-point functions are suppressed by 
$1/N^2$ and thus are sub-leading. On the other hand, for correlators containing an odd number of $A$'s, the leading large-$N$ terms arise from  factorizations \`a la Wick into exactly one 3-point function and as many 2-point functions as needed.

To export these large-$N$ simplifications in the computation of the gauge theory correlators, we have to change basis and consider the normal-ordered operators
(\ref{PversusA}).
At $\lambda=0$ further simplifications occur. Indeed, in the free theory
the coefficients of the change of basis at large $N$ are related to the power expansion of Chebyshev polynomials \cite{Rodriguez-Gomez:2016cem} and in all sectors read
\begin{align}
\label{M0}
\lim_{\lambda\to 0} M_{k,\ell}^{(\widehat{\alpha})}(\lambda) ~\simeq~ 
\frac{k}{\ell}\, \binom{\frac{k+\ell-2}{2}}{\frac{k-\ell}{2}}\,\Big(\!-\frac{N}{2}\Big)^{\frac{k-\ell}{2}}
\,\equiv\,\mathsf{M}_{k,\ell} 
\end{align}
when $k+\ell$ is even with $\ell\leq k$, and vanish in all other cases.
Therefore, at $\lambda=0$ we simply have
\begin{equation}
\mathbf{P}_{\widehat{\alpha}}(0) \simeq\, \mathsf{M}\, \mathbf{A}_{\widehat{\alpha}}~.
\label{P0MA}
\end{equation}
The lower triangular matrix $\mathsf{M}$ can be easily inverted and the elements of its inverse read
\begin{align}
\label{M0inverse}
(\mathsf{M}^{-1})_{k,\ell} = \Big(\frac{N}{2}\Big)^{\frac{k-\ell}{2}}
\,\binom{k}{\frac{k-\ell}{2}}
\end{align}
where again $\ell\leq k$, with $k+\ell$ even.

Combining (\ref{AA}) and (\ref{AAA}) with (\ref{P0MA}), we obtain the 2- and 3-point correlators
of the normal-ordered operators in the free theory. For the 2-point functions we find
\begin{equation}
\big\langle P_{\widehat{\alpha},k}^{\phantom{\dagger}}(0)\,P_{\widehat{\beta},\ell}^\dagger(0)
\big\rangle_0~\simeq~ \mathcal{G}_k \,\delta_{\widehat{\alpha},\widehat{\beta}}\,\delta_{k,\ell} \quad
\mbox{with}\quad \mathcal{G}_k=k\,\Big(\frac{N}{2}\Big)^k~.
\label{2ptP}
\end{equation}
This result naturally suggests to introduce the normalized operators
\begin{equation}
	\mathcal{P}_{\widehat{\alpha},k}\,\equiv\, \frac{1}{\sqrt{\mathcal{G}_k}}\,P_{\widehat{\alpha},k}(0)
	\label{calP}
\end{equation}
which obviously have a canonical ``propagator''
\begin{equation}
	\big\langle\, \mathcal{P}_{\widehat{\alpha},k}^{\phantom{\dagger}}
	\,\mathcal{P}_{\widehat{\beta},\ell}^\dagger\,\big\rangle_0~\simeq~  \delta_{\widehat{\alpha},\widehat{\beta}}\,\delta_{k,\ell}
	~\equiv ~\parbox[c]{.20\textwidth}{\includegraphics[width = .20\textwidth]{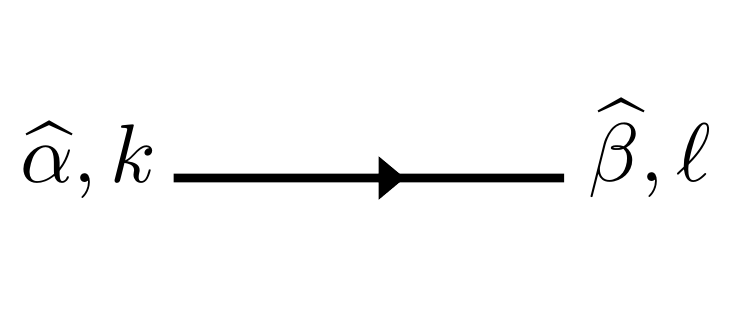}}
	\label{cPcP}
\end{equation}
At $\lambda=0$ it is easy to compute also the 3-point correlators. Indeed one finds
\begin{equation}
	\big\langle \,\mathcal{P}_{\widehat{\alpha},k}^{\phantom{\dagger}}
	\,\mathcal{P}_{\widehat{\beta},\ell}^{\phantom{\dagger}}
	\,\mathcal{P}_{\widehat{\gamma},p}^\dagger\,\big\rangle_0~\simeq~C_{k,\ell,p} \,
	\delta_{\widehat{\alpha}+\widehat{\beta}\,,\,\widehat{\gamma}}
	\label{cPcPcP}
\end{equation}
for $k+\ell+p$ even, where
\begin{equation}
C_{k,\ell,p}=
\frac{1}{\sqrt{M}\,N}\,\sqrt{k\,\ell\,p}~.
\label{Cklp}
\end{equation}
Note that the factor of $1/N$ with respect to the 2-point function (\ref{cPcP}) agrees with the large-$N$ behavior of the correlators of the operators $A_{\widehat{\alpha},k}$ given in (\ref{AA}) and (\ref{AAA}).
The cubic correlator (\ref{cPcPcP}) can be interpreted as a Feynman diagram built with three external free propagators (\ref{cPcP}) attached to a cubic vertex as follows:
\begin{align}
	\label{cubvertex}
		\parbox[c]{.20\textwidth}{\includegraphics[width = .20\textwidth]{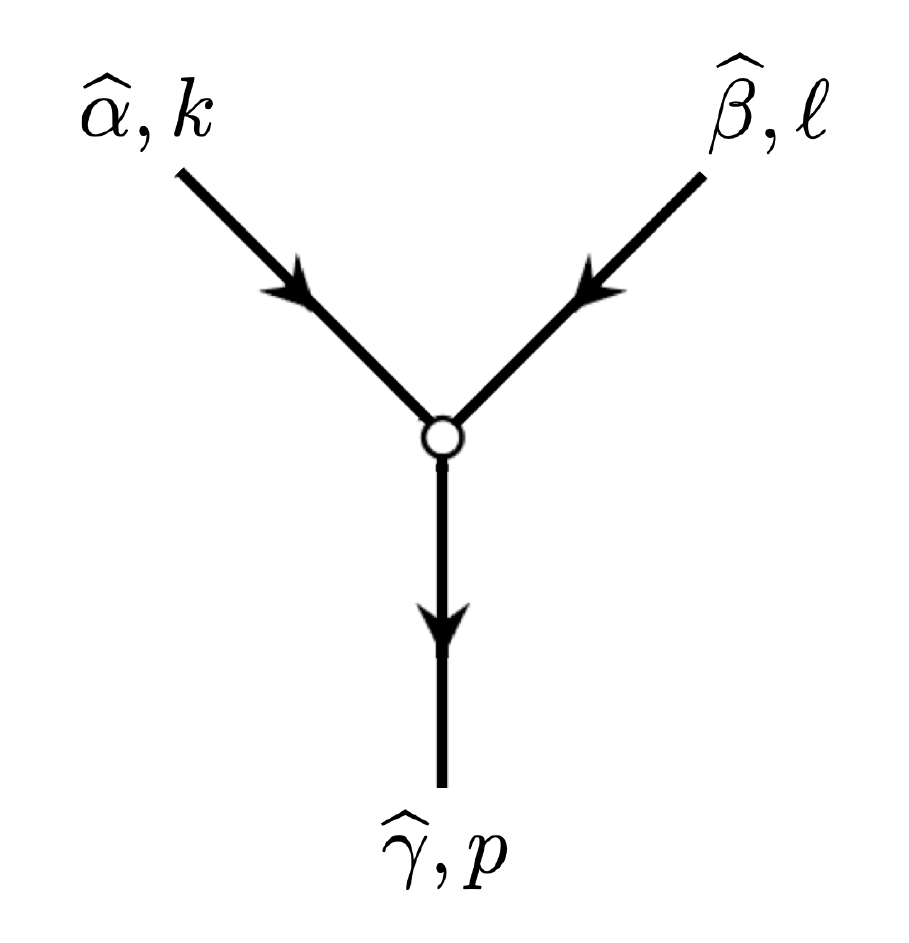}}
		\equiv\, C_{k,\ell,p}\,\, \delta_{\widehat{\alpha}+\widehat{\beta}\,,\,\widehat{\gamma}}~.
\end{align}	
In terms of the unnormalized operators, the 3-point functions (\ref{cPcPcP})
become
\begin{equation}
\big\langle\, P_{\widehat{\alpha},k}^{\phantom{\dagger}}(0)
\,P_{\widehat{\beta},\ell}^{\phantom{\dagger}}(0)\,
P_{\widehat{\gamma},p}^\dagger(0)\,\big\rangle_0~\simeq~\mathcal{G}_{k,\ell,p} \,\,
\delta_{\widehat{\alpha}+\widehat{\beta},\widehat{\gamma}}
\label{PPP}
\end{equation}
where
\begin{equation}
\mathcal{G}_{k,\ell,p}=\sqrt{\mathcal{G}_k\,\mathcal{G}_\ell\,\mathcal{G}_p}\,\,C_{k,\ell,p}=
\frac{k\,\ell\,p}{2\sqrt{M}}\,\Big(\frac{N}{2}\Big)^{\frac{k+\ell+p}{2}-1}~.
\label{calGklp}
\end{equation}
The quantities $\mathcal{G}_k$ and $\mathcal{G}_{k,\ell,p}$ defined in (\ref{2ptP}) and (\ref{calGklp}) are precisely the coefficients that appear in the numerator of the 2- and 3-point functions of the gauge theory operators at $\lambda=0$, while the quantities
$C_{k,\ell,p}$ in (\ref{Cklp}) are the normalized structure constants of the free theory. Notice that, apart from
the factors of $\sqrt{M}$ due to the quiver construction, these coefficients match those of the
$\mathcal{N}=4$ SYM theory in the planar limit.

\subsection{The interaction action at large \texorpdfstring{$N$}{}}

The crucial observation is that the 2- and 3-point functions of the normalized operators given
in (\ref{cPcP}) and (\ref{cPcPcP}) can be computed in full generality also in the interacting theory
when $\lambda\neq 0$. 
The reason for this is that the interaction action
(\ref{Sintstart}) can be written in terms of the normalized operators
$\mathcal{P}_{\widehat{\alpha},k}$ and their conjugates in a very simple form. 
To see this, following \cite{Billo:2021rdb},
let us first rewrite $S_{\mathrm{int}}$ in the basis of the operators (\ref{operators}). This yields
\begin{align}
	\label{Sint}
	S_{\mathrm{int}} = \sum_{\widehat{\alpha}=0}^{M-1}\bigg[
	4\,s_{\widehat{\alpha}}\sum_{m=2}^{\infty}\sum_{k=2}^{2m}(-1)^{m+k}
	\Big(\frac{\lambda}{8\pi^2N}\Big)^{\!m}\,\binom{2m}{k}\,
	\frac{\zeta_{2m-1}}{2m}\,A^{\dagger}_{\widehat{\alpha},2m-k}\,
	A_{\widehat{\alpha},k}^{\phantom{\dagger}}\bigg] ~,
\end{align}
where
\begin{align}
	s_{\widehat{\alpha}} = \textrm{sin}^2\Big(\frac{\pi\widehat{\alpha}}{M}\Big) ~.
	\label{salpha}
\end{align}
Note that $s_0=0$, so that the sum over $\widehat{\alpha}$ effectively reduces to a sum over only the twisted sectors. However, we find convenient to keep the extended sum over $\widehat{\alpha}$.
Using the inverse matrix \eqref{M0inverse}, we can rewrite the operators $A_{\widehat{\alpha},k}$ in terms of the normalized operators $\mathcal{P}_{\widehat{\alpha},k}$ and put 
the interaction action \eqref{Sint} in the form
\begin{align}
	\label{Sintshort}
	S_{\mathrm{int}} = -\frac{1}{2}\sum_{\widehat{\alpha}=0}^{M-1}
	s_{\widehat{\alpha}}\,\boldsymbol{\mathcal{P}}_{\widehat{\alpha}}^{\dagger}\,\,\mathsf{X}\,\,\boldsymbol{\mathcal{P}}_{\widehat{\alpha}} ~.
\end{align}
Here $\boldsymbol{\mathcal{P}}_{\widehat{\alpha}}$ is an infinite vector with
components $\mathcal{P}_{\widehat{\alpha},k}$ and $\mathsf{X}$ is an infinite symmetric matrix in which the entries with opposite parity vanish, namely
\begin{align}
	\mathsf{X}_{2n,2m+1} = 0~,
\end{align}
while the entries with the same parity are non-trivial functions of $\lambda$ which
can be expressed in terms of a convolution of Bessel functions of the first kind \cite{Billo:2021rdb} as follows
\begin{align}
	&  \mathsf{X}_{k,\ell}
	= -8(-1)^{\frac{k+\ell+2k\,\ell}{2}}\,\sqrt{k\,\ell}\int_0^{\infty}\!\frac{dt}{t}
	\,\frac{\mathrm{e}^t}{(\mathrm{e}^t-1)^2}\,J_{k}\Big(\frac{t\sqrt{\lambda}}{2\pi}\Big)
	\,J_{\ell}\Big(\frac{t\sqrt{\lambda}}{2\pi}\Big) \label{Xkl} 
\end{align}
with $k,\ell\geq 2$.
Note that when we write the interaction action in the form (\ref{Sintshort}) all dependence on $\lambda$ is inside the matrix $\mathsf{X}$ and, differently from what we had in the original expression (\ref{Sintstart}), this is {\emph{not}} provided through a
weak-coupling expansion but through the Bessel functions. In other words, in writing the
interaction action as in (\ref{Sintshort})
we managed to effectively resum the perturbative expansion and obtain the exact dependence
on $\lambda$ through the matrix $\mathsf{X}$. If we Taylor expand the Bessel functions
in \eqref{Xkl} and then analytically perform the integration over $t$,
we recover the perturbative expansion \eqref{Sint} in the weak-coupling regime. On the other
hand, using the inverse Mellin transform of the product of two Bessel functions, \eqref{Xkl} can 
be expanded asymptotically for large $\lambda$, providing in this way information about the strong-coupling regime of the theory.

\subsection{Correlators in the interacting theory}
Since the normalized operators $\mathcal{P}_{\widehat{\alpha},k}$ defined in (\ref{calP}) are linear combinations of the operators $A_{\widehat{\alpha},k}$, they inherit from the latter the large-$N$ factorization properties discussed after (\ref{AAA}). Thus, the free correlator of an even number of $\mathcal{P}$'s can be simply obtained, at large $N$, by making all possible contractions using the propagator (\ref{cPcP}); indeed, for every insertion of the cubic vertex
(\ref{cubvertex}) we would get an extra factor of $1/N$, as shown in (\ref{Cklp}). Furthermore, 
since the free propagator only connects conjugated twisted sectors, the correlator factorizes with respect to the indices $\widehat{\alpha}$. 

\subsubsection{The propagator}
\label{subsubsec:tpf}
As already discussed in \cite{Billo:2021rdb}, the interaction action (\ref{Sint})
modifies the propagator. At the first order in $S_{\mathrm{int}}$, one has 
\begin{align}
	\label{foprop}
		\big\langle \mathcal{P}_{\widehat{\alpha},k}^{\phantom{\dagger}}\,\mathcal{P}_{\widehat{\beta},\ell}^\dagger\big\rangle
		& = \big\langle \mathcal{P}_{\widehat{\alpha},k}^{\phantom{\dagger}}\,\mathcal{P}_{\widehat{\beta},\ell}^\dagger\big\rangle_0 
		- \big\langle \mathcal{P}_{\widehat{\alpha},k}^{\phantom{\dagger}}\,\mathcal{P}_{\widehat{\beta},\ell}^\dagger\,S_{\mathrm{int}}\big\rangle_0
		+ \big\langle \mathcal{P}_{\widehat{\alpha},k}^{\phantom{\dagger}}\,\mathcal{P}_{\widehat{\beta},\ell}^\dagger\big\rangle_0 \, \big\langle S_{\rm int}\big\rangle_0 + \ldots \notag\\[2mm]
		& = \delta_{\widehat{\alpha},\widehat{\beta}} \,\delta_{k,\ell} + 
		\frac{1}{2} \sum_{\widehat{\gamma}} 
		\sum_{m,n} \,s_{\widehat{\gamma}}\,\,
		\big\langle \mathcal{P}_{\widehat{\alpha},k}^{\phantom{\dagger}}\,\mathcal{P}_{\widehat{\beta},\ell}^\dagger\,\big(
		\mathcal{P}_{\widehat{\gamma},m}^\dagger\,\mathsf{X}_{m,n}\,\mathcal{P}_{\widehat{\gamma},n}^{\phantom{\dagger}}
		\big)\big\rangle_0\,{\Big|}_{\mathrm{conn}}
	 + \ldots~, 
\end{align}		
where by the notation $|_{\mathrm{conn}}$ we mean the connected part of the correlator in which
the two operators coming from $S_{\mathrm{int}}$ are not contracted with each other. Performing the allowed Wick contractions with the free propagator, we get
\begin{align}
	\label{foprop2}
	\big\langle \mathcal{P}_{\widehat{\alpha},k}^{\phantom{\dagger}}\,\mathcal{P}_{\widehat{\beta},\ell}^\dagger\big\rangle\,
	\simeq \,\delta_{\widehat{\alpha},\widehat{\beta}} \,\big( 
	 \mathbb{1}+ s_{\widehat{\alpha}}\,\mathsf{X} \big)_{k,\ell} + \ldots~.
\end{align}
Thus, diagrammatically, the insertion of $S_{\mathrm{int}}$ corresponds
to a quadratic vertex:
\begin{align}
	\label{fox}
		\parbox[c]{.22\textwidth}{\includegraphics[width = .22\textwidth]{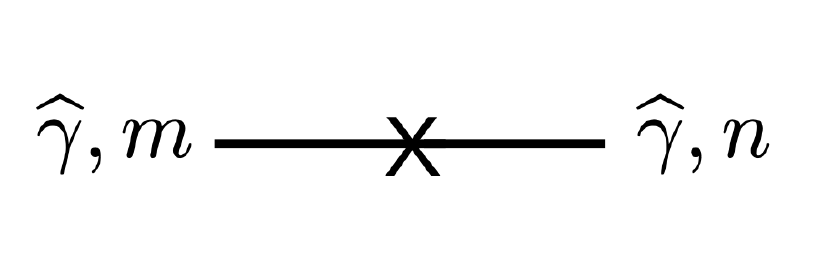}}	
		\equiv \,s_{\widehat{\gamma}}\, \mathsf{X}_{m,n}~.
\end{align}
Taking into account the subsequent orders in $S_{\mathrm{int}}$ leads simply to
\begin{align}
	\label{foprop3}
		\big\langle \mathcal{P}_{\widehat{\alpha},k}^{\phantom{\dagger}}\,\mathcal{P}_{\widehat{\beta},\ell}^\dagger\big\rangle
		& \simeq \parbox[c]{.75\textwidth}{\includegraphics[width = .75\textwidth]{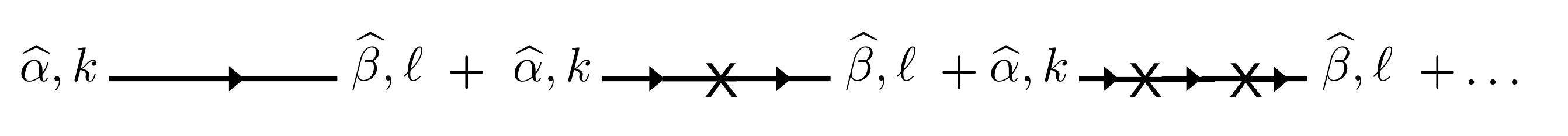}} \notag\\
		& \simeq\, \delta_{\widehat{\alpha},\widehat{\beta}} \,\big( \mathbb{1} + s_{\widehat\alpha}\, \mathsf{X} + 
		 s^2_{\widehat\alpha} \,\mathsf{X}^2 + \ldots\big)_{k,\ell}~.
\end{align}
Resumming the formal geometric series, we find
\begin{align}
	\label{intprop}
		\big\langle \mathcal{P}_{\widehat{\alpha},k}^{\phantom{\dagger}}\,\mathcal{P}_{\widehat{\beta},\ell}^\dagger\big\rangle
		\,\simeq\,
		\delta_{\widehat{\alpha},\widehat{\beta}} \,\Big(\frac{1}{\mathbb{1} - s_{\widehat\alpha}\, \mathsf{X}}\Big)_{k,\ell}
		\equiv \,\delta_{\widehat{\alpha},\widehat{\beta}} \,\mathsf{D}^{(\widehat{\alpha})}_{k,\ell}
		\,\equiv\,\parbox[c]{.22\textwidth}{\includegraphics[width = .22\textwidth]{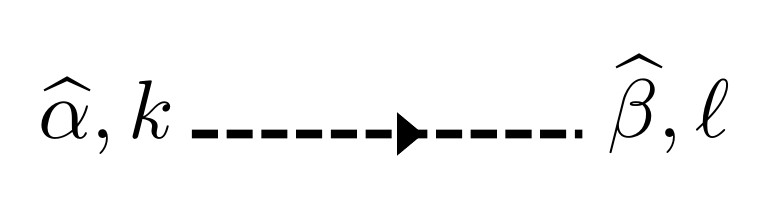}}
		~.  
\end{align} 
We stress that this formula, which vanishes if $k$ and $\ell$ have different parity, is exact in $\lambda$ and that 
the entire dependence on the coupling is encoded in the Bessel functions inside $\mathsf{X}$ and 
hence inside $\mathsf{D}^{(\widehat{\alpha})}$.
Note that in the untwisted sector $\widehat{\alpha}=0$, we have 
$s_0 = 0$ and thus the propagator remains the free one, namely
\begin{equation}
\big\langle \mathcal{P}_{0,k}^{\phantom{\dagger}}\,\mathcal{P}_{0,\ell}^\dagger\big\rangle
		\,\simeq\,
\mathsf{D}^{(0)}_{k,\ell} \,=\, \delta_{k,\ell}~.
\label{D0}
\end{equation}
We can therefore conclude that in the interacting theory 
the large-$N$ leading contribution to correlators of an even number 
of $\mathcal{P}$'s are given by all possible 
Wick contractions as in the free theory, but performed using the interacting 
propagator (\ref{intprop}).

\subsubsection{The 2-point functions}
The operators $\mathcal{P}_{\widehat{\alpha},k}$, which are orthonormal at
$\lambda=0$, are no longer so when $\lambda\not=0$ but can be easily normal-ordered 
by applying the Gram-Schmidt procedure to their scalar product (\ref{intprop}). Taking into account the normalization factor
$\mathcal{G}_k$ appearing in (\ref{calP}), we can thus express the 
operators $P_{\widehat{\alpha},k}(\lambda)$, which represent the gauge theory operators according to (\ref{UTP}), in terms of the $\mathcal{P}$'s and write
\begin{equation}
	P_{\widehat{\alpha},k}(\lambda)= \sqrt{\mathcal{G}_k} \,\Bigl(\mathcal{P}_{\widehat{\alpha},k}
	-\sum_{\ell<k} \mathsf{Q}^{(\widehat{\alpha})}_{k,\ell}(\lambda)\,\mathcal{P}_{\widehat{\alpha},\ell}
	\Bigr)~,
	\label{PktocPk}
\end{equation}
where the mixing coefficients $\mathsf{Q}^{(\widehat{\alpha})}_{k,\ell}(\lambda)$ 
are determined by demanding that $P_{\widehat{\alpha},k}(\lambda)$ be orthogonal to all
lower dimensional operators. 
This process is trivial in the untwisted case since
\begin{equation}
P_{0,k}(\lambda)\,\simeq\,\sqrt{\mathcal{G}_k}\,\mathcal{P}_{0,k}
\label{P0calP0}
\end{equation}
for any $k$, while it can be carried out iteratively to the desired level, independently in each twisted sector $\alpha$. For the first few values of $k$, we have
\begin{equation}
\begin{aligned}
P_{\alpha,2}(\lambda) &\,\simeq\, \sqrt{\mathcal{G}_2}\,\mathcal{P}_{\alpha,2}~,\qquad\qquad\qquad\qquad~\,\,
P_{\alpha,3}(\lambda) \,\simeq\,  \sqrt{\mathcal{G}_3}\,\mathcal{P}_{\alpha,3}~,
\\[2mm]
P_{\alpha,4}(\lambda) &\,\simeq\,   \sqrt{\mathcal{G}_4}\,
\Big(\mathcal{P}_{\alpha,4} - 
\frac{\mathsf{D}^{(\alpha)}_{4,2}}{\mathsf{D}^{(\alpha)}_{2,2}}\,
\mathcal{P}_{\alpha,2}\Big)~,\qquad
P_{\alpha,5}(\lambda) \,\simeq\,   \sqrt{\mathcal{G}_5}\,\Big(\mathcal{P}_{\alpha,5} 
- 	\frac{\mathsf{D}^{(\alpha)}_{5,3}}{\mathsf{D}^{(\alpha)}_{3,3}} \,
\mathcal{P}_{\alpha,3} \Big)~,
\end{aligned}
\label{P2345calP2345}
\end{equation}
and so on. 

Using these formulas, the correspondence (\ref{examples}) and the interacting propagator (\ref{intprop}), it is straightforward to obtain the coefficients $G_{U_{k}}$  and $G_{T_{\alpha,k}}$ appearing in the 
twisted 2-point functions. Indeed, in the untwisted sector we have
\begin{equation}
G_{U_{k}} \,\simeq\, \big\langle P_{0,k}^{\phantom{\dagger}}(\lambda)\,
P_{0,k}^\dagger(\lambda)\big\rangle
		\,\simeq\,  \mathcal{G}_k\, \big\langle \mathcal{P}_{0,k}^{\phantom{\dagger}}\,\mathcal{P}_{0,k}^\dagger\big\rangle  \,\simeq\,\mathcal{G}_k
\end{equation}
for all $k$, while in each twisted sector $\alpha$ we obtain
\begin{align}
	\label{lowG}
		G_{T_{\alpha,2}} & \,\simeq\, \big\langle P_{\alpha,2}^{\phantom{\dagger}}(\lambda)\,P_{\alpha,2}^\dagger(\lambda)\big\rangle
		\,\simeq\,  \mathcal{G}_2\, \big\langle \mathcal{P}_{\alpha,2}^{\phantom{\dagger}}\,\mathcal{P}_{\alpha,2}^\dagger\big\rangle
		 \,\simeq\, \mathcal{G}_2\, \mathsf{D}^{(\alpha)}_{2,2}~,\notag\\[2mm]
		G_{T_{\alpha,3}} & \,\simeq\, \big\langle P_{\alpha,3}^{\phantom{\dagger}}(\lambda)\,P_{\alpha,3}^\dagger(\lambda)\big\rangle
		 \,\simeq\, \mathcal{G}_3\, \big\langle \mathcal{P}_{\alpha,3}^{\phantom{\dagger}}\,\mathcal{P}_{\alpha,3}^\dagger\big\rangle
		\,\simeq\,  \mathcal{G}_3\, \mathsf{D}^{(\alpha)}_{3,3}~,\notag\\
		G_{T_{\alpha,4}} & \,\simeq\, \big\langle P_{\alpha,4}^{\phantom{\dagger}}(\lambda)\,P_{\alpha,4}^\dagger(\lambda)\big\rangle
		\,\simeq\,  \sqrt{\mathcal{G}_4}\, \big\langle {P}_{\alpha,4}^{\phantom{\dagger}}(\lambda)\,\mathcal{P}_{\alpha,4}^\dagger\big\rangle
		\,\simeq\,  \mathcal{G}_4\, 
\big\langle \Big(\mathcal{P}_{\alpha,4} - 
\frac{\mathsf{D}^{(\alpha)}_{4,2}}{\mathsf{D}^{(\alpha)}_{2,2}}\,
\mathcal{P}_{\alpha,2}\Big) 	\mathcal{P}_{\alpha,4}^\dagger\big\rangle	
		\notag\\
		&  \,\simeq\, \mathcal{G}_4\, \frac{\mathsf{D}^{(\alpha)}_{4,4}\, \mathsf{D}^{(\alpha)}_{2,2} - \mathsf{D}^{(\alpha)}_{4,2}\,\mathsf{D}^{(\alpha)}_{2,4}}{\mathsf{D}^{(\alpha)}_{2,2}}~,\\
		G_{T_{\alpha,5}} & \,\simeq\, \big\langle P_{\alpha,5}^{\phantom{\dagger}}(\lambda)\,P_{\alpha,5}^\dagger(\lambda)\big\rangle
		\,\simeq\,  \sqrt{\mathcal{G}_5}\, \big\langle {P}_{\alpha,5}^{\phantom{\dagger}}(\lambda)\,\mathcal{P}_{\alpha,5}^\dagger\big\rangle
		\,\simeq\,  \mathcal{G}_5\, 
\big\langle \Big(\mathcal{P}_{\alpha,5} - 
\frac{\mathsf{D}^{(\alpha)}_{5,3}}{\mathsf{D}^{(\alpha)}_{3,3}}\,
\mathcal{P}_{\alpha,3}\Big) 	\mathcal{P}_{\alpha,5}^\dagger\big\rangle	
		\notag\\
		&  \,\simeq\, \mathcal{G}_5\, \frac{\mathsf{D}^{(\alpha)}_{5,5}\, \mathsf{D}^{(\alpha)}_{3,3} - \mathsf{D}^{(\alpha)}_{5,3}\,\mathsf{D}^{(\alpha)}_{3,5}}{\mathsf{D}^{(\alpha)}_{3,3}}~.\notag
\end{align}
Similar formulas can be easily worked out for other values of $k$.
In fact, as shown in \cite{Billo:2021rdb}, it is possible to write a compact expression 
for $G_{T_{\alpha,k}}$ at a generic $k$. If we 
introduce the matrices $\mathsf{D}^{\mathrm{even},(\alpha)}$ and 
$\mathsf{D}^{\mathrm{odd},(\alpha)}$ defined as
\begin{equation}
\big(\mathsf{D}^{\mathrm{even},(\alpha)}\big)_{n,m}=
\mathsf{D}^{(\alpha)}_{2n,2m}\quad\mbox{and}\quad
\big(\mathsf{D}^{\mathrm{odd},(\alpha)}\big)_{n,m}=
\mathsf{D}^{(\alpha)}_{2n+1,2m+1}~,
\end{equation}
then we have
\begin{align}
	\label{Gtgenk}
		G_{T_{\alpha,2n}}\, \simeq\, \mathcal{G}_{2n}\, 
		\frac{\det \big[\mathsf{D}^{\mathrm{even},(\alpha)}_{(n)}\big]}
		{\det\big[\mathsf{D}^{\mathrm{even},(\alpha)}_{(n-1)}\big]}
		\quad\mbox{and}\quad
		G_{T_{\alpha,2n+1}} \, \simeq\,  \mathcal{G}_{2n+1}\, 
		\frac{\det \big[\mathsf{D}^{\mathrm{odd},(\alpha)}_{(n)}\big]}
		{\det\big[\mathsf{D}^{\mathrm{odd},(\alpha)}_{(n-1)}\big]}
\end{align}	
where the subscript $_{(n)}$
indicates the upper-left $n\times n$ block of the matrix, with the convention that $\mathsf{D}^{\mathrm{even},(\alpha)}_{(0)} = \mathsf{D}^{\mathrm{odd},(\alpha)}_{(0)}=1$.

Such expressions can be used to extract very efficiently the perturbative expansion in $\lambda$. To do so, one simply has to expand the propagators $\mathsf{D}^{(\alpha)}$ as geometric series in $\mathsf{X}$ and then exploit the weak coupling expansion of the latter
that is obtained by expanding the Bessel functions. However, using the properties of these
functions when the argument tends to infinity, one can also obtain the asymptotic expansion of the
2-point functions in the strong-coupling regime, as we will see in Section~\ref{secn:strong}.
 
\subsubsection{The 3-point functions} 
\label{subsubsec:3pf}
We now extend the above arguments to the calculation of the 3-point functions at large $N$. At $\lambda=0$ they are given in (\ref{cPcPcP}) and (\ref{Cklp}) and have the diagrammatic interpretation shown in (\ref{cubvertex}).
When the interaction is turned on, at order $n$ in $S_{\mathrm{int}}$ we have to evaluate correlators involving $(3+2n)$ operators $\mathcal{P}$ in the free theory. As said before, the operators $\mathcal{P}$ inherit the large-$N$ factorization properties of the operators $A$; in particular, as described after (\ref{AAA}), the leading term of the correlator of an odd number of $\mathcal{P}$'s is given by diagrams constructed out of just one cubic vertex (\ref{cubvertex}) plus the needed contractions with the free propagators. Thus, for a correlator with $(3+2n)$ operators arising at order $n$ in $S_{\mathrm{int}}$, we employ one cubic vertex and $n$ free propagators. 
For instance, at the first order in $S_{\mathrm{int}}$ we have 
\begin{align}
	\label{focub}
	 \big\langle
		\mathcal{P}_{\widehat{\alpha},k}^{\phantom{\dagger}}
		\,\mathcal{P}_{\widehat{\beta},\ell} &\,\mathcal{P}_{\widehat{\gamma},p}^\dagger \big\rangle
		= \big\langle
		\mathcal{P}_{\widehat{\alpha},k}^{\phantom{\dagger}}
		\,\mathcal{P}_{\widehat{\beta},\ell} \,\mathcal{P}_{\widehat{\gamma},p}^\dagger \big\rangle_0 
		-\big\langle
		\mathcal{P}_{\widehat{\alpha},k}^{\phantom{\dagger}}
		\,\mathcal{P}_{\widehat{\beta},\ell} \,\mathcal{P}_{\widehat{\gamma},p}^\dagger\,S_{\mathrm{int}} \big\rangle_0\Big|_{\mathrm{conn}}+\ldots
\notag\\[2mm]
		= & ~  \delta_{\widehat{\alpha}+\widehat{\beta},\widehat{\gamma}}
		\, C_{k,\ell,p} + 
		\frac 12 \sum_{\widehat{\delta}}\sum_{m,n}  s_{\widehat{\delta}}\,\,
		\big\langle
		\mathcal{P}_{\widehat{\alpha},k}^{\phantom{\dagger}}
		\,\mathcal{P}_{\widehat{\beta},\ell} \,\mathcal{P}_{\widehat{\gamma},p}^\dagger\,
		\mathcal{P}_{\widehat{\delta},m}^\dagger\,
		\mathsf{X}_{m,n}\,\mathcal{P}_{\widehat{\delta},n}
		\big\rangle_0\Big|_{\mathrm{conn}}+\ldots
~. 
\end{align}		
Factorizing the second term into a cubic vertex times a Wick contraction 
in all possible ways, we get  
\begin{align}
	\label{focub1}
		\big\langle
		\mathcal{P}_{\widehat{\alpha},k}^{\phantom{\dagger}}
		\,\mathcal{P}_{\widehat{\beta},\ell} \,\mathcal{P}_{\widehat{\gamma},p}^\dagger \big\rangle \,\simeq\,  \delta_{\widehat{\alpha}+\widehat{\beta},\widehat{\gamma}}\,\Big(
		 C_{k,\ell,p} + \!\sum_{k^\prime}\! s_{\widehat{\alpha}}\,
		 \mathsf{X}_{kk^\prime} \,
		 C_{k^\prime,\ell,p}
		+\! \sum_{\ell^\prime}\! s_{\widehat{\beta}}
		\,\mathsf{X}_{\ell\ell^\prime} \,
		 C_{k,\ell^\prime,p}
		+\! \sum_{p^\prime}\! s_{\widehat{\gamma}}\,\mathsf{X}_{pp^\prime} \,
		C_{k,\ell,p^\prime} \Big) + \ldots~.
\end{align}
 Diagrammatically, this reads
 \begin{align}
 	\label{focubdia}
 		\big\langle
		\mathcal{P}_{\widehat{\alpha},k}^{\phantom{\dagger}}
		\,\mathcal{P}_{\widehat{\beta},\ell} \,\mathcal{P}_{\widehat{\gamma},p}^\dagger \big\rangle
		\,\simeq \parbox[c]{.75\textwidth}{\includegraphics[width = .75\textwidth]{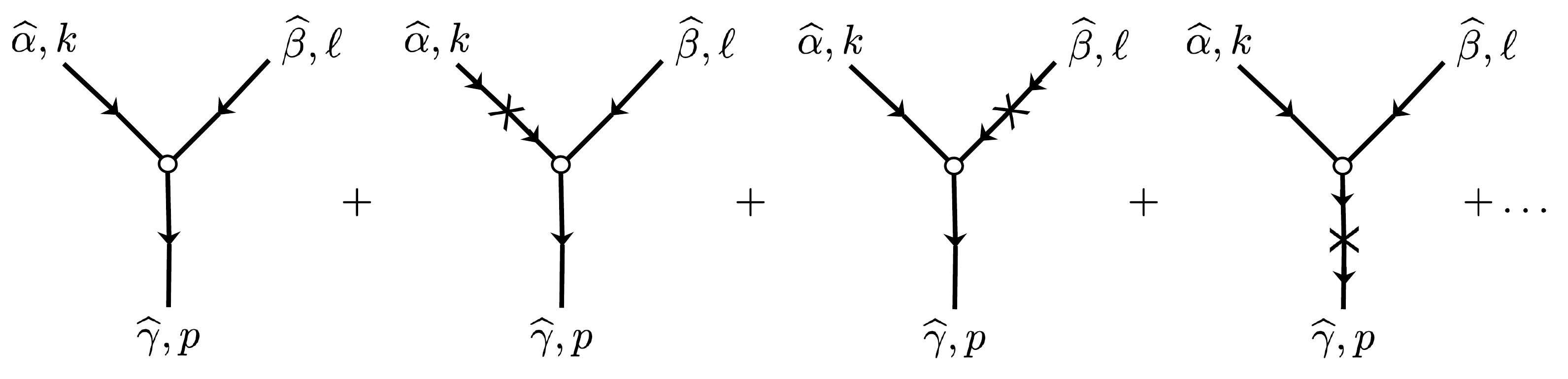}} ~.
 \end{align}
It is easy to see that the inclusion of all subsequent orders in $S_{\mathrm{int}}$ has the effect of promoting the external lines to the fully interacting propagators (\ref{intprop}). The result, at the leading order 
in $N$ but exact in $\lambda$, is thus
\begin{align}
	\label{cubres}
		\big\langle
		\mathcal{P}_{\widehat{\alpha},k}^{\phantom{\dagger}}
		\,\mathcal{P}_{\widehat{\beta},\ell} \,\mathcal{P}_{\widehat{\gamma},p}^\dagger \big\rangle
		\,\simeq  \parbox[c]{.20\textwidth}{\includegraphics[width = .20\textwidth]{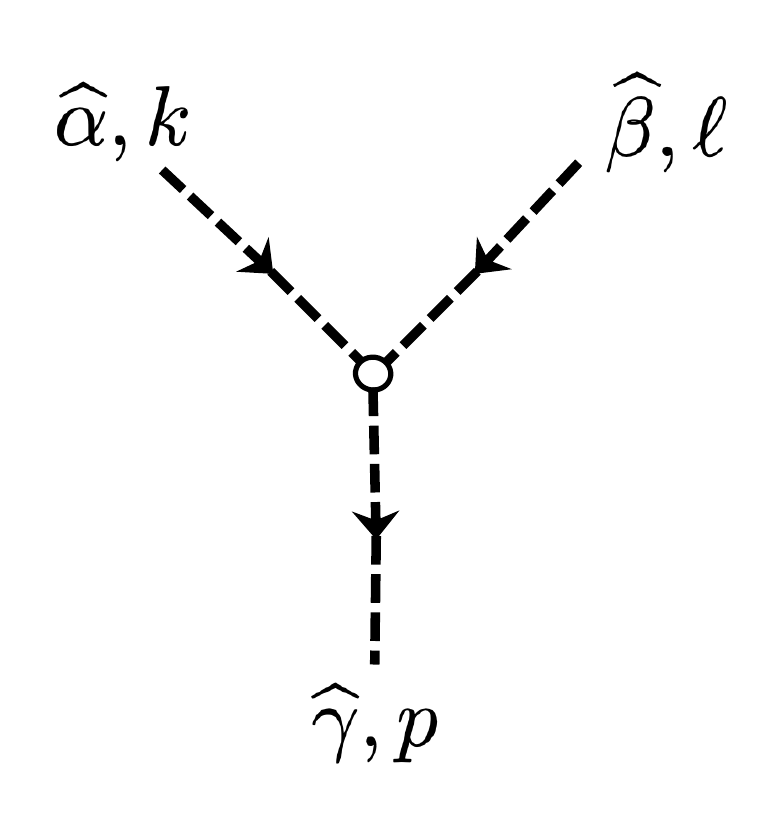}}
		\,\equiv~ \delta_{\widehat{\alpha}+\widehat{\beta},\widehat{\gamma}}\sum_{k^\prime, \ell^\prime, p^\prime} \mathsf{D}^{(\widehat{\alpha})}_{k k^\prime}\, \mathsf{D}^{(\widehat{\beta})}_{\ell \ell^\prime}\,
		\mathsf{D}^{(\widehat{\gamma})}_{p p^\prime}\, C_{k^\prime,\ell^\prime,p^\prime}~. 
 \end{align}
 Since the structure constants $C_{k^\prime,\ell^\prime,p^\prime}$
 are factorized (see (\ref{Cklp})), the above expression can be written as
\begin{align}
	\label{cubres2}
		\big\langle
		\mathcal{P}_{\widehat{\alpha},k}^{\phantom{\dagger}}
		\,\mathcal{P}_{\widehat{\beta},\ell} \,
		\mathcal{P}_{\widehat{\gamma},p}^\dagger \big\rangle
		\,\simeq \,\frac{\delta_{\widehat{\alpha}+\widehat{\beta},\widehat{\gamma}}}
		{\sqrt{M}\,N}\,\,
		\mathsf{d}_k^{(\widehat\alpha)}\, 
		\mathsf{d}_\ell^{(\widehat\beta)}\, 
		\mathsf{d}_p^{(\widehat\gamma)} ~,
\end{align}
where
\begin{align}
	\label{cadef}
		\mathsf{d}^{(\widehat\alpha)}_k = \sum_{k^\prime} 
		\mathsf{D}^{(\widehat{\alpha})}_{k k^\prime} \,\sqrt{k^\prime}~.
\end{align}
In the untwisted case we simply have
\begin{align}
	\label{cuntw}
	\mathsf{d}^{(0)}_k = \sqrt{k} ~,
\end{align}
and the correlator of three untwisted operators retains its free value also for $\lambda\not= 0$:
\begin{align}
	\label{3cP0}
		\big\langle
		\mathcal{P}_{0,k}^{\phantom{\dagger}}
		\,\mathcal{P}_{0,\ell} \,
		\mathcal{P}_{0,p}^\dagger \big\rangle
		\,\simeq \,C_{k,\ell, p} = \frac{1}{\sqrt{M} \,N}\,\sqrt{k\,\ell\, p}
\end{align}
provided $k+\ell+p$ is even. 
Instead, in the twisted case the quantities in (\ref{cadef})
are non-trivial functions of $\lambda$. At weak coupling it is quite straightforward 
to obtain their perturbative expressions. For example, 
for $k=2,3$ we have
\begin{align}
	\label{c2is}
		\mathsf{d}_2^{(\alpha)} 
		&= \sqrt{2} \,\bigg[1 - 6 s_\alpha\, \zeta_3 \,\widehat{\lambda}^2
		+ 50 \, s_\alpha \,\zeta_5\, \widehat{\lambda}^3
		- \Big(\frac{735}{2}\,s_\alpha\, \zeta_7 - 36\, s_\alpha^2 \,
		\zeta_3^2\Big) \widehat{\lambda}^4
		+ \Big(2646 \,s_\alpha\, \zeta_9
		 - 540\, s_\alpha^2\,\zeta_3\,\zeta_5\Big) 
		 \widehat{\lambda}^5\notag\\
		&\qquad- \Big(\frac{38115}{2}\,s_\alpha\,\zeta_{11}- 
		3675 \,s_\alpha^2\,\zeta_3\,\zeta_7 - 2050\, s_\alpha^2\,\zeta_5^2
		+ 216 \,s_\alpha^3\, \zeta_3^3\Big) \widehat{\lambda}^6 + \ldots
		\bigg]~, 
\end{align}
and
\begin{align}
	\label{c3is}	
		\mathsf{d}_3^{(\alpha)} & = \sqrt{3} \,\bigg[
		1 - 10\, s_\alpha \,\zeta_5\,  \widehat{\lambda}^3
		+\frac{ 245}{2}\, s_\alpha\, \zeta_7\,  \widehat{\lambda}^4 - 1134\, s_\alpha \,\zeta_9\,  \widehat{\lambda}^5
		+ \Big(\frac{38115}{4}\, s_\alpha \,\zeta_{11} + 100 \,
		s_\alpha^2\, \zeta_5^2\Big)\widehat{\lambda}^6 + \ldots
		\bigg] ~,
\end{align}	
where for compactness of notation we introduced the rescaled coupling $\widehat{\lambda}=\lambda/(8\pi^2)$.
Similar perturbative expansions can be easily derived for higher values of $k$.

Using (\ref{cubres2}) and the relation (\ref{PktocPk}), we can obtain the 3-point 
functions of the normal-ordered operators $P_{\widehat{\alpha},k}
(\lambda)$ and thus find 
the 3-point coefficients of the quiver gauge theory at generic value of $\lambda$. 
If all three operators are untwisted, the 3-point function is unmodified with respect to the free theory expression (\ref{PPP}):
\begin{align}
	\label{GUUUres}
	G_{U_k,U_\ell,\overbar{U}_p} & \,\simeq\, 
	\big\langle P_{0,k}^{\phantom{\dagger}}(\lambda)\,P_{0,\ell}^{\phantom{\dagger}}(\lambda) \,P_{0,p}^\dagger(\lambda)\big\rangle
	\,\simeq\, \sqrt{\mathcal{G}_k\,\mathcal{G}_\ell\,\mathcal{G}_p}\,\,
	\big\langle
		\mathcal{P}_{0,k}^{\phantom{\dagger}}
		\,\mathcal{P}_{0,\ell}^{\phantom{\dagger}} \,
		\mathcal{P}_{0,p}^\dagger \big\rangle
	\, \simeq\, \sqrt{\mathcal{G}_k\,\mathcal{G}_\ell\,\mathcal{G}_p}\,\, C_{k,\ell, p} = \mathcal{G}_{k,\ell, p}
\end{align} 
where the $\delta$ function imposing charge conservation is understood. 

When there are twisted operators, the correlators deviate from those of the
free theory. To give some explicit expressions, let us fix for simplicity $M=3$. In the corresponding $\mathbb{Z}_3$ quiver theory there are two
twisted sectors, $\alpha=1$ and $\alpha=2$, that are conjugated to each other, and the coefficients $s_{\widehat{\alpha}}$ take the values:
$s_0 = 0$, $s_1 = s_2 = 3/4$. A typical 3-point twisted correlator in this theory is, for example, $G_{U_k,T_{\alpha,\ell},\overbar{T}_{\alpha,p}}$
with $p=k+\ell$ for charge conservation. If for simplicity we take $k=2$,
$\ell=3$, $p=5$ and $\alpha=1$, using (\ref{P0calP0}) and (\ref{P2345calP2345}) we have
\begin{align}
	\label{G235}
		G_{U_2,T_{1,3},\overbar{T}_{1,5}} & \,\simeq\, \big\langle P_{0,2}^{\phantom{\dagger}}(\lambda)\,P_{1,3}^{\phantom{\dagger}}(\lambda) \,P_{1,5}^\dagger(\lambda)\big\rangle
		\,\simeq\, \sqrt{\mathcal{G}_2\,\mathcal{G}_3\,\mathcal{G}_5}\,\,
		\big\langle 
		\mathcal{P}_{0,2}^{\phantom{\dagger}}
		\,\mathcal{P}_{1,3}^{\phantom{\dagger}} 
\Big(\mathcal{P}_{1,5}^\dagger- \frac{\mathsf{D}^{(1)}_{5,3}}{\mathsf{D}^{(1)}_{3,3}}\,\mathcal{P}_{1,3}^\dagger\Big)\big\rangle
		\notag\\
		& \,\simeq\, \frac{\sqrt{\mathcal{G}_2\,\mathcal{G}_3\,\mathcal{G}_5}}{\sqrt{3}\,N}\, \sqrt{2}\,\, \mathsf{d}^{(1)}_3
		\Big(\mathsf{d}^{(1)}_5 - \frac{\mathsf{D}^{(1)}_{5,3}}{\mathsf{D}^{(1)}_{3,3}} \,\mathsf{d}^{(1)}_3 \Big)
\end{align} 
where the last step follows from (\ref{cubres2}) and (\ref{cuntw}). Expanding at weak coupling, we obtain
\begin{align}
	\label{G235pert}
		G_{U_2,T_{1,3},\overbar{T}_{1,5}}  \, \simeq\, \mathcal{G}_{2,3,5} \,
		\bigg[1 - \frac{15}{2}\, \zeta_5\, \widehat{\lambda}^3 
		+ \frac{735}{8} \,\zeta_7\, \widehat{\lambda}^4
		- \frac{6993}{8}\, \zeta_9\, \widehat{\lambda}^5
		+ \Big(7623\, \zeta_{11} + \frac{225}{4}\, \zeta_5^2\Big)  \widehat{\lambda}^6 + \ldots \bigg]~,
\end{align} 
where $\mathcal{G}_{2,3,5} = 5\sqrt{3} (N/2)^4$
in accordance with (\ref{calGklp}). Other correlators involving operators with different dimensions can be obtained in a similar way.

In the $\mathbb{Z}_3$ quiver theory, we can have also a 3-point function with three twisted operators. An example is given by the following correlator
\begin{align}
	\label{G235t}
	G_{T_{1,2},T_{1,3},\overbar{T}_{2,5}} & \,\simeq\,
	\big\langle P_{1,2}^{\phantom{\dagger}}(\lambda)\,
	P_{1,3}^{\phantom{\dagger}}(\lambda) \,P_{2,5}^\dagger(\lambda)
	\big\rangle
\,\simeq\,\sqrt{\mathcal{G}_2\,\mathcal{G}_3\,\mathcal{G}_5}\,\,
		\big\langle 
		\mathcal{P}_{1,2}^{\phantom{\dagger}}
		\,\mathcal{P}_{1,3}^{\phantom{\dagger}} 
\Big(\mathcal{P}_{2,5}^\dagger- \frac{\mathsf{D}^{(2)}_{5,3}}{\mathsf{D}^{(2)}_{3,3}}\,\mathcal{P}_{2,3}^\dagger\Big)\big\rangle	
	\notag\\
	&\, \simeq \,\frac{\sqrt{\mathcal{G}_2\,\mathcal{G}_3\,\mathcal{G}_5}}{\sqrt{3}\,N}\, \mathsf{d}^{(1)}_2\, \mathsf{d}^{(1)}_3
		\Big(\mathsf{d}^{(2)}_5 - \frac{\mathsf{D}^{(2)}_{5,3}}{\mathsf{D}^{(2)}_{3,3}} \,\mathsf{d}^{(2)}_3 \Big)
\end{align} 
whose perturbative expansion is
\begin{align}
	\label{G235pertT}
	G_{T_{1,2},T_{1,3},\overbar{T}_{2,5}} & \simeq\,
	\mathcal{G}_{2,3,5}
\,		\bigg[1 - \frac{9}{2}\, \zeta_3\, \widehat{\lambda}^2 + 30\,
\zeta_5\, \widehat{\lambda}^3
	- \Big(\frac{735}{4}\, \zeta_7 - \frac{81}{4}\,\zeta_3^2\Big) \widehat{\lambda}^4
	+ \Big(\frac{8883}{8}\, \zeta_9 - 270\,\zeta_3\,\zeta_5\Big) \widehat{\lambda}^5
	\notag\\
	& \qquad
	- \Big(\frac{53361}{8}\, \zeta_{11} - \frac{6615}{4} \,\zeta_3\,\zeta_7
	- \frac{7425}{8}\,\zeta_5^2
	+ \frac{729}{8}\,\zeta_3^3\Big)  \widehat{\lambda}^6 + \ldots \bigg]~.
\end{align}
These prototypical examples, which can be easily generalized in many ways, show that
the 3-point correlators of the quiver theory can be written algebraically in terms of the quantities $\mathsf{D}^{(\widehat{\alpha})}_{k,\ell}$ and $\mathsf{d}^{(\widehat{\alpha})}_k$ that contain the exact dependence on $\lambda$. By expanding these quantities at weak coupling, we easily generate the perturbative series and, more importantly, by studying their
behavior for large values of $\lambda$ we can access the strong-coupling regime as we are going to do in the next section.

\section{Strong-coupling results from localization}
\label{secn:strong}
The expressions for the correlators we have derived in the previous section for a generic value of $\lambda$ remarkably simplify at strong coupling. Indeed, as discussed in \cite{Billo:2021rdb}, using the properties of the Bessel functions
one can show that the matrix $\mathsf{X}$ behaves for $\lambda\to \infty$ as
\begin{align}
	\label{XtoS}
		\mathsf{X} \underset{\lambda\to\infty}{\sim} - \frac{\lambda}{2\pi^2} \,\mathsf{S}
\end{align}	
where $\mathsf{S}$ is a three-diagonal matrix with elements
\begin{align}
	\label{Sres}
		\mathsf{S}_{k,\ell} = \sqrt{\frac{\ell}{k}}\, \Big(- \frac{\delta_{k-2,\ell}}{2(k-2)(k-1)}
		+ \frac{\delta_{k,\ell}}{(k-1)(k+1)} - \frac{\delta_{k+2,\ell}}{2(k+1)(k+2)} \Big)
\end{align}
for $k+\ell$ even and zero otherwise.
Using this result in (\ref{intprop}) yields
\begin{align}
	\label{Dllres}
		\mathsf{D}^{(\alpha)}_{k,\ell} ~\underset{\lambda\to\infty}{\sim}~ \frac{2\pi^2}{s_\alpha\,\lambda}\,
		\big( \mathsf{S}^{-1}\big)_{k,\ell}=
		\frac{\pi^2}{s_\alpha\,\lambda} \,\Big[
		\sqrt{k\,\ell}\big(\min(k^2,\ell^2) - \delta_{k\!\!\!\!\!_{\mod\!2},1}\big)\Big]~.
\end{align}
Writing out explicitly the first entries, we have\,%
\footnote{Recall that $k$ and $\ell$ are $\geq 2$.}
\begin{align}
	\label{Dllrese}
		\mathsf{D}^{(\alpha)}_{k,\ell} \underset{\lambda\to\infty}{\sim}  \  \frac{8\pi^2}{s_\alpha\,\lambda}\,
		\begin{pmatrix}
		1 & 0 & \sqrt{2} & 0 & \sqrt{3} & 0 & \cdots\\
			0 & 3 & 0 & \sqrt{15} & 0 & \sqrt{21} & \cdots\\
			\sqrt{2} & 0 & 8 & 0 & 4 \sqrt{6} & 0 &\cdots \\
			0 & \sqrt{15} & 0 & 15 & 0 & 3 \sqrt{35}& \cdots \\
			\sqrt{3} & 0 & 4 \sqrt{6} & 0 & 27 & 0 & \cdots\\
			0 & \sqrt{21} & 0 & 3 \sqrt{35} & 0 & 42 &\cdots \\
			\vdots &   \vdots   & \vdots    &   \vdots    &  \vdots  & \vdots    & \ddots
		\end{pmatrix}~.
\end{align}

\subsection{The 2- and 3-point functions}
Through (\ref{intprop}), the matrix (\ref{Dllres}) encodes the 2-point functions of the operators 
$\mathcal{P}_{\alpha,k}$ in the large-$N$ and large-$\lambda$ regime. 
Ultimately, however, we are interested in the correlators of the 
normal-ordered operators $P_{\widehat{\alpha},k}(\lambda)$ determined by the Gram-Schmidt procedure. Quite remarkably,
at large $\lambda$, this procedure takes a very simple form. Indeed, the 2-point functions can be diagonalized with the following change of basis\,%
\footnote{From now on, the symbol $\simeq$ denotes the leading term both at large $N$ and at large $\lambda$.}
\begin{align}
	P_{0,k}(\infty) &\,\simeq\, \sqrt{\mathcal{G}_k} \,\mathcal{P}_{0,k}
	~,\qquad
	P_{\alpha,2}(\infty)\,\simeq\,
	 \sqrt{\mathcal{G}_2} \,\mathcal{P}_{\alpha,2}~,\notag\\
		P_{\alpha,k}(\infty)& \,\simeq\,
		\sqrt{\mathcal{G}_k}\, \Big(\mathcal{P}_{\alpha,k} - \sqrt{\frac{k}{k-2}}\,\mathcal{P}_{\alpha,k-2}\Big)\quad\mbox{for}\ k>2~.
		\label{inftyno}
\end{align}
If we express $\mathcal{P}_{\alpha,k}$ in the basis of the operators $A_{\alpha,k}$ through (\ref{calP}) and (\ref{P0MA}), we obtain
\begin{equation}
\mathbf{P}_{\alpha}(\infty)\,\simeq\,\mathsf{M}^{(\infty)}
\mathbf{A}_{\alpha}
\label{Mstrong}
\end{equation}
where 
\begin{align}
\mathsf{M}^{(\infty)}_{k,\ell}=\frac{2(k-1)}{k+\ell-2}\,\mathsf{M}_{k,\ell}
\label{Mstrongkl}
\end{align}
with $\mathsf{M}_{k,\ell}$ defined in (\ref{M0})\,%
\footnote{It is interesting to notice that the operators $P_{\alpha,k}(\infty)$ as given in (\ref{Mstrong}) are related to the Gegenbauer polynomials 
$C_k^{(w)}$ of order $k$ with weight parameter $w=-1$. 
More precisely, we have $P_{\alpha,k}(\infty)=
(\frac{N}{2})^{\frac{k}{2}}\,\frac{k!}{(w)_k}\,\big[C_k^{(w)}(x/\sqrt{2N})-C_k^{(w)}(0)\big]
\big|_{w=-1}$, where $(w)_k$ is the Pochammer symbol and $x^n$ has to be interpreted as $A_{\alpha,n}$. This fact is the strong-coupling counterpart of the relation of the operators
$P_{\widehat{\alpha},k}(0)$ with the Chebyshev polynomials observed at $\lambda=0$
\cite{Rodriguez-Gomez:2016cem}.}.

Then, using the correspondence (\ref{UTP}), it is immediate to find that
the coefficients in the 2-point functions of the quiver operators at strong coupling are
\begin{align}
	\label{PPin}
	G_{U_k}&\,\simeq\,\big\langle P_{0,k}^{\phantom{\dagger}}(\infty)\,
	P_{0,k}^\dagger(\infty)\big\rangle
	\simeq \,\mathcal{G}_k~,\notag\\
	G_{T_{\alpha,k}}&\,\simeq\,
		\big\langle P_{\alpha,k}^{\phantom{\dagger}}(\infty)\,
		P_{\alpha,k}^\dagger(\infty)\big\rangle
	\simeq \,\mathcal{G}_k\,
		\frac{4\pi^2 \,k\,(k-1)}{s_\alpha\,\lambda}~,
\end{align}
confirming the results of \cite{Billo:2021rdb}.

To find the strong-coupling behavior of the 3-point functions we need a new ingredient, namely
the strong-coupling limit of the quantities $\mathsf{d}^{(\widehat{\alpha})}_k$ defined in (\ref{cadef}). Of course, in the untwisted case $\widehat{\alpha}=0$, we already know that $\mathsf{d}^{(0)}_k=\sqrt{k}$ (see 
(\ref{cuntw})), but in the twisted sectors we have to work out how these coefficients behave
when $\lambda\to \infty$. Since the strong-coupling expansions are asymptotic series, we cannot simply plug-in the large-$\lambda$ limit in each term of the sum that defines 
$\mathsf{d}^{(\widehat{\alpha})}_k$, but we have to perform a new independent analysis. This is discussed
in detail in Appendix~\ref{app:ck} using two different methods that lead to the following very
compact result
\begin{align}
	\label{call}
		\mathsf{d}^{(\alpha)}_k &\underset{\lambda\to\infty}{\sim} 
		\frac{\pi}{\sqrt{s_\alpha\,\lambda}} \,\Big[
		\frac{\sqrt{k}}{2} \big(k^2 - \delta_{k\!\!\!\!\!\mod 2,1}\big)\Big]
		~.
\end{align}

Through the change of basis (\ref{inftyno}), the correlators
of three normal-ordered operators $P_{\widehat{\alpha},k}(\infty)$
are reduced to linear combinations of the 3-point functions of the
$\mathcal{P}_{\widehat{\alpha},k}$ operators given in (\ref{cubres2}). Since the latter factorize in terms of the $\mathsf{d}^{(\widehat{\alpha})}_k$ coefficients, it is sufficient to introduce the quantities 
\begin{align}
	\label{ctilde}
	\widetilde{\mathsf{d}}^{(0)}_k=\mathsf{d}^{(0)}_k\quad\mbox{and}\quad
	\widetilde{\mathsf{d}}^{(\alpha)}_k=\mathsf{d}^{(\alpha)}_k-
	\sqrt{\frac{k}{k-2}}\,\,\mathsf{d}^{(\alpha)}_{k-2}~,
\end{align}
and obtain
\begin{equation}
\big\langle P_{\widehat{\alpha},k}^{\phantom{\dagger}}(\infty)\,P_{\widehat{\beta},\ell}^{\phantom{\dagger}}(\infty)
\,P_{\widehat{\gamma},p}^\dagger(\infty)\big\rangle
\simeq\,\frac{\delta_{\widehat{\alpha}+\widehat{\beta},\widehat{\gamma}}}{
\sqrt{M}\,N}\, \sqrt{\mathcal{G}_k\,\mathcal{G}_\ell\,\mathcal{G}_p}\,\,\,
\widetilde{\mathsf{d}}^{(\widehat{\alpha})}_k\,
\widetilde{\mathsf{d}}^{(\widehat{\beta})}_\ell\,
\widetilde{\mathsf{d}}^{(\widehat{\gamma})}_p
\label{PPPin}
\end{equation}
where, as usual, we have omitted the $\delta$-function imposing charge conservation.
At strong coupling, we have
\begin{align}
	\label{ctll}
	\widetilde{\mathsf{d}}^{(0)}_k=\sqrt{k}\quad\mbox{and}\quad
	\widetilde{\mathsf{d}}^{(\alpha)}_k \underset{\lambda\to\infty}{\sim} 
	\frac{2\pi}{\sqrt{s_\alpha\,\lambda}}\,\sqrt{k}\,(k-1)~,
\end{align}
which follows upon inserting (\ref{call}) into (\ref{ctilde}). With all these ingredients it is now straightforward to obtain the expression of the various 3-point correlators at strong-coupling. When all operators are untwisted we simply have
\begin{align}
	\label{GUUUstrong}
	G_{U_k,U_\ell,\overbar{U}_p} & \,\simeq\,
	\big\langle P_{0,k}^{\phantom{\dagger}}(\infty)\,P_{0,\ell}^{\phantom{\dagger}}(\infty) \,P_{0,p}^\dagger(\infty)\big\rangle
	\, \simeq\, \mathcal{G}_{k,\ell, p}
\end{align}
as in the free theory (see (\ref{GUUUres})), whereas when there are some twisted operators we have new results. In particular we find
\begin{align}
	\label{GUTTstrong}
		G_{U_k,T_{\alpha,\ell},\overline{T}_{\alpha,p}} & \,\simeq\,
		\big\langle P_{0,k}^{\phantom{\dagger}}(\infty)\,P_{\alpha,\ell}^{\phantom{\dagger}}(\infty) \,P_{\alpha,p}^\dagger(\infty)\big\rangle
	\, \simeq\,\frac{1}{\sqrt{M}\,N}\,\sqrt{\mathcal{G}_k\,\mathcal{G}_\ell\,\mathcal{G}_p}\,\,
		\sqrt{k}\, \,\widetilde{\mathsf{d}}^{(\alpha)}_\ell\,
\widetilde{\mathsf{d}}^{(\alpha)}_p
		\notag\\
		& \,\simeq \,\mathcal{G}_{k,\ell,p} \,\Big[\frac{2\pi}{\sqrt{s_\alpha\,
		\lambda}}\,(\ell-1)\Big]
		\Big[\frac{2\pi}{\sqrt{s_\alpha\,\lambda}}\,(p-1)\Big]~, 
\end{align}
and
\begin{align}
	\label{GTTUstrong}
		G_{T_{\alpha,k},T_{M-\alpha,\ell},\overline{U}_{p}} & \,\simeq\,
		\big\langle P_{\alpha,k}^{\phantom{\dagger}}(\infty)\,
		P_{M-\alpha,\ell}^{\phantom{\dagger}}(\infty) \,
		P_{0,p}^\dagger(\infty)\big\rangle
	\, \simeq\,\frac{1}{\sqrt{M}\,N}\,\sqrt{\mathcal{G}_k\,\mathcal{G}_\ell\,\mathcal{G}_p}\,\,
		\widetilde{\mathsf{d}}^{(\alpha)}_k\,
		\widetilde{\mathsf{d}}^{(M-\alpha)}_\ell \sqrt{p}
		\notag\\
		& \,\simeq \,\mathcal{G}_{k,\ell,p} \,\Big[\frac{2\pi}{\sqrt{s_\alpha\,
		\lambda}}\,(k-1)\Big]
		\Big[\frac{2\pi}{\sqrt{s_{M-\alpha}\,\lambda}}\,(\ell-1)\Big]~.
\end{align}
Finally, if all three operators are twisted we have
\begin{align}
	\label{GTTTstrong}
		G_{T_{\alpha,k},T_{\beta,\ell},\overline{T}_{\gamma,p}} & \,\simeq\, 
		\big\langle P_{\alpha,k}^{\phantom{\dagger}}(\infty)\,
		P_{\beta,\ell}^{\phantom{\dagger}}(\infty) \,
		P_{\gamma,p}^\dagger(\infty)\big\rangle	\, \simeq\,
		 \frac{\delta_{\alpha+\beta,\gamma}}{\sqrt{M}\,N} \,
		 \sqrt{\mathcal{G}_k\,\mathcal{G}_\ell\,\mathcal{G}_p}\,\,
		 \widetilde{\mathsf{d}}^{(\alpha)}_k
		 \widetilde{\mathsf{d}}^{(\beta)}_\ell
		 \widetilde{\mathsf{d}}^{(\gamma)}_p
		\notag\\
		& \,\simeq\, \mathcal{G}_{k,\ell,p} \,\Big[\frac{2\pi}{\sqrt{s_\alpha\,\lambda}}\,(k-1)\Big]
		\Big[\frac{2\pi}{\sqrt{s_\beta\,\lambda}}\,(\ell-1)\Big]
		\Big[\frac{2\pi}{\sqrt{s_\gamma\,\lambda}}\,(p-1)\Big]
		\delta_{\alpha+\beta,\gamma}~. 
\end{align}
Of course this last possibility exists only in quivers with more than two nodes.

\subsection{The structure constants at strong coupling}
Combining our strong-coupling results on the 2- and 3-point correlators, we can obtain the structure constants which do not depend on the normalization of the operators and are part of the intrinsic data of the conformal field theory. For the various cases we have considered, these structure constants, defined in (\ref{structure}), at strong coupling read
\begin{subequations}
\begin{align}
C_{U_k,U_\ell,\overbar{U}_p}&
\,\simeq\,
\frac{1}{\sqrt{M}\,N}\,\sqrt{k\,\ell\,p}~,\label{CUUUloc}\\
C_{U_k,T_{\alpha,\ell},\overbar{T}_{\alpha,p}}&
\,\simeq\,\frac{1}{\sqrt{M}\,N}\,\sqrt{k\,(\ell-1)\,(p-1)}~,\label{CUTTloc}\\
C_{T_{\alpha,k},T_{M-\alpha,\ell},\overbar{U}_{p}}&
\,\simeq\,\frac{1}{\sqrt{M}\,N}\,\sqrt{(k-1)\,(\ell-1)\,p}~,\label{CTTUloc}\\
C_{T_{\alpha,k},T_{\beta,\ell},\overbar{T}_{\gamma,p}}&
\,\simeq\,\frac{1}{\sqrt{M}\,N}\,\sqrt{(k-1)\,(\ell-1)\,(p-1)}\,\,\delta_{\alpha+\beta,\gamma}
~.\label{CTTTloc}
\end{align}
\label{Cloc}%
\end{subequations}
In all these expressions we have understood the factor $\delta_{k+\ell-p,0}$ which enforces the charge conservation.

\part{Holography}
\label{part:2}
In this part of the paper we derive the 2- and 3-point functions of the scalar operators at strong coupling using the AdS/CFT correspondence. Since we are interested in the planar limit, we can work at the level of supergravity. However, we find more convenient to start from a more general string theory set-up.

\section{The holographic description}
\label{secn:holo}

The quiver theory under consideration can be obtained from a parent $\cN=4$ SYM theory with gauge group SU($MN$) engineered with $MN$ regular D3-branes of Type II B string theory placed on a $\mathbb{C}^2/\mathbb{Z}_M$ orbifold singularity
(see for instance \cite{Douglas:1996sw,Kachru:1998ys}). 
Denoting by $z_2$ and $z_3$ the complex coordinates of $\mathbb{C}^2$, the action
of $\mathbb{Z}_M$ is simply
\begin{equation}
z_2~\to~ \rho\,z_2\quad\mbox{and}\quad
z_3~\to~ \rho^{-1}\,z_3~,
\label{orbifold}
\end{equation}
where $\rho=\mathrm{e}^{\frac{2\pi\ii}{M}}$.
By breaking the pile of $MN$ D3-branes 
into $M$ stacks of $N$ fractional D3-branes located at 
the orbifold fixed-locus $z_2=z_3=0$,
we obtain the quiver theory of Fig.\,\ref{fig:1_quiver}, in which 
each node corresponds to one of the $M$ stacks. The latter can therefore be labeled by the same index $I$ used for the quiver nodes.
In the field-theory limit, the massless excitations of the open strings starting and ending
on the $I$-th branes give rise to the adjoint vector multiplet of the
$I$-th node of the quiver, while the massless excitations of the open strings stretching 
between the $I$-th branes and the $(I\pm1)$-th branes yield the bi-fundamental matter hypermultiplets. The massless excitations associated to open strings stretching between 
the $I$-th branes and the $(I\pm k)$-th branes with $k\geq 2$ are instead removed by the orbifold projection and this explains why there are no links between non-adjacent nodes of the
quiver. 

From a geometrical point of view, the fractional D3-branes can be interpreted as D5-branes wrapped around the exceptional 2-cycles $e_i$ (with $i=1,\ldots,M-1$) of the $\mathbb{Z}_M$
orbifold singularity. These 2-cycles are associated to anti-self dual 2-forms 
$\omega^i$ such that
\begin{equation}
\int_{e_i}\omega^j=\delta_i^{\,j}~,
\label{eomega}
\end{equation}
which are normalized as follows
\begin{equation}
\int_{\mathcal{M}}\omega^i\wedge\omega^j=-\big(C^{-1}\big)^{ij}~.
\label{Cij}
\end{equation}
Here $\mathcal{M}$ is the ALE space obtained by resolving the $\mathbb{C}^2/\mathbb{Z}_M$
orbifold singularity and $C$ is the Cartan matrix of SU($M$), namely
\begin{equation}
C=\begin{pmatrix}
2&-1&0&0&0&\ldots\\
-1&2&-1&0&0&\ldots\\
0&-1&2&-1&0&\ldots\\
\vdots&\vdots&\vdots&\vdots&\vdots&\ddots
\end{pmatrix}~.
\label{Cartanmatrix}
\end{equation}
Note that there are $M$ types of fractional branes but there are only $(M-1)$ exceptional 
2-cycles $e_i$. 
In fact, the fractional branes corresponding to the trivial representation, {\it{i.e.}} those with $I=0$ in our conventions, are D5-branes wrapped around 
the 2-cycle $e_0=-\sum_i e_i$ (in presence
of an additional magnetic background flux on the world-volume). 

From the closed string point of view, the fractional D3-branes can be seen
as soliton configurations that emit the metric, a 4-form $C_4$ with a self-dual field strength, and the scalars corresponding to 
the wrapping of the 2-forms $B_{2}$ and $C_{2}$ around the 2-cycles $e_i$ of the orbifold\,%
\footnote{See for instance \cite{Bertolini:2000dk,Polchinski:2000mx,Billo:2001vg} where
it is also shown that the axio-dilaton of fractional D3-branes is constant.}, namely
\begin{equation}
\hat{b}_i=\frac{1}{2\pi\alpha^\prime}\int_{e_i}B_{2}\quad\mbox{and}\quad
\hat{c}_i=\frac{1}{2\pi\alpha^\prime}\int_{e_i}C_{2}~,
\label{bIcI}
\end{equation}
where $\alpha^\prime$ is the square of the string length. The factors of $1/(2\pi\alpha^\prime)$ have been inserted in order to make $\hat{b}_i$ and $\hat{c}_i$
dimensionless, just like their parent 2-forms $B_2$ and $C_2$.

The low-energy effective dynamics of the metric and the 4-form $C_4$ is captured by the following 10-dimensional action
\begin{equation}
S_{10}=\frac{1}{2\kappa_{10}^2} 
\int\! \!d^{10}x\,\sqrt{G}\, \Big(R+\frac{1}{4\cdot 5!}\,\big(d C_{4}\big)^{2}\Big)
\label{S10}
\end{equation}
with the understanding that the self-duality condition on the field strength $dC_{4}$ has to be imposed on the field equations.
Here $G$ is the determinant of the metric, $R$ is the scalar curvature
and $2\kappa_{10}^2$ is the gravitational constant in ten dimensions:
\begin{equation}
2\kappa_{10}^2=(2\pi)^7\,g_s^2\,\alpha^{\prime\,4}
\label{kappa10}
\end{equation}
where $g_s$ is the string coupling.

To describe the dynamics of the scalars $\hat{b}_i$ and $\hat{c}_i$,
we have to consider the part of the low-energy effective action of Type II B string theory that depends
on $B_2$ and $C_2$, which is given by\,%
\footnote{Here we have used the fact that in the presence of fractional D3-branes, the axio-dilaton is constant. Therefore, the only dilaton dependence in $S_{10}^\prime$ 
is through $g_s$ which is the exponential of the vacuum expectation value of the dilaton. Moreover, without any loss of generality, we have set the axion to zero, so that the 
field strength of $C_2$ is just its exterior derivative.}
 \begin{equation}
S_{10}^\prime
=\frac{1}{2\kappa_{10}^2}\bigg[ \int d^{10}x\,\sqrt{G}~ \Big(\frac{1}{12}
\big(dB_{2})^2+\frac{1}{12}\big(dC_{2})^2\Big)-4\!
\int \!C_{4}\wedge dB_{2} \wedge dC_{2}\bigg]~.
\label{S10bis}
\end{equation}
From (\ref{bIcI}) we see that $B_2=(2\pi\alpha^\prime)\sum_i \hat{b}_i\,\omega^i$ and
$C_2=(2\pi\alpha^\prime)\sum_i \hat{c}_i\,\omega^i$. Inserting these expansions in the above action and using (\ref{Cij}), we obtain
the following six-dimensional action
\begin{equation}
S_6=\frac{(2\pi\alpha^\prime)^2}{2\kappa_{10}^2} \sum_{i,j=1}^{M-1}\bigg[
\int d^{6}x\,\sqrt{G^\prime}~ \Big(\frac{1}{2}\,
\nabla \hat{b}_i \!\cdot\!\nabla \hat{b}_j+\frac{1}{2}\,\nabla \hat{c}_i \!\cdot\!\nabla \hat{c}_j\Big)+
4\!\int \!C_{4}\wedge d\hat{b}_{i} \wedge d\hat{c}_{j}
\bigg]\big(C^{-1}\big)^{ij}
\label{S6}
\end{equation}
where $G^\prime$ is the determinant of the metric in the space transverse to the
$\mathbb{C}^2/\mathbb{Z}_M$ singularity. 

We now perform a change of basis and rewrite everything in terms of the fields 
associated to the twisted sectors of the orbifold 
\cite{Billo:2001vg,Ashok:2020jgb,Billo:2021rdb}.
To do so, we first introduce two additional scalars\,%
\footnote{Here $B_{2}^\prime=B_{2}+2\pi\alpha^\prime\mathcal{F}$ where $\mathcal{F}$ is a constant background representing a unit magnetic flux.}
\begin{equation}
\hat{b}_0=\frac{1}{2\pi\alpha^\prime}\int_{e_0}B_{2}^\prime
\quad\mbox{and}\quad
\hat{c}_0=\frac{1}{2\pi\alpha^\prime}\int_{e_0}C_{2}~.
\label{b0c0}
\end{equation}
Since $e_0=-\sum_i e_i$, these fields are not independent; indeed one has
\begin{equation}
\hat{b}_0=1-\sum_{i=1}^{M-1}\hat{b}_i\quad\mbox{and}\quad
\hat{c}_0=-\sum_{i=1}^{M-1}\hat{c}_i~.
\label{relations}
\end{equation}
Nevertheless, it is useful to use them in order to define the following combinations
\begin{subequations}
\begin{align}
b_0&=\frac{1}{2}\sum_{I=0}^{M-1} \hat{b}_{I}~,\qquad
b_{\alpha}=\frac{1}{2}\sum_{I=0}^{M-1} \rho^{-\alpha I}\,\hat{b}_{I}~,\\
c_0&=\frac{1}{2}\sum_{I=0}^{M-1} \hat{c}_{I}~,\qquad
c_{\alpha}=\frac{1}{2}\sum_{I=0}^{M-1} \rho^{-\alpha I}\,\hat{c}_{I}~,
\end{align}
\label{bcIbcalpha}%
\end{subequations}
where $\alpha=1,\ldots,M-1$.
Note that, in view of (\ref{relations}), $b_0$ is constant and $c_0$ vanishes.
Therefore, these scalars do not have any dynamical role.
The other scalars $b_\alpha$ and $c_\alpha$ (with $\alpha\not=0$)
are called twisted since they are associated to the $(M-1)$ twisted
sectors of the $\mathbb{Z}_M$ orbifold. They are complex fields and
satisfy the the following conjugation rules
\begin{equation}
b_\alpha^{*}=b_{M-\alpha}\quad\mbox{and}\quad
c_\alpha^{*}=c_{M-\alpha}~.
\label{conjugatebc}
\end{equation}
Now we rewrite $\hat{b}_i$ and $\hat{c}_i$ in terms of $b_\alpha$ and $c_\alpha$ using the
inverse of (\ref{bcIbcalpha}) and, after some simple algebra, we find
that (\ref{S6}) becomes
\begin{equation}
S_6=\frac{(2\pi\alpha^\prime)^2}{2\kappa_{10}^2 M}\,
\sum_{\alpha=1}^{M-1}\frac{1}{2s_\alpha}\bigg[
\int \!d^{6}x\,\sqrt{G^\prime}~ \Big(
\nabla b_{\alpha}^{*} \!\cdot\!\nabla b_{\alpha}+
\nabla c_{\alpha}^{*} \!\cdot\!\nabla c_{\alpha}\Big)+8\!
\int \!C_{4}\wedge db_{\alpha}^{*} \wedge\, dc_{\alpha}~,
\bigg]
\label{S6bis}
\end{equation}
where
\begin{equation}
s_\alpha=\sin^2\big(\frac{\pi\alpha}{M}\big)~,
\label{salpha1}
\end{equation}
which is just eq. (\ref{salpha}) for $\widehat{\alpha}=\alpha$.
The actions (\ref{S10}) and (\ref{S6bis}) are the starting point for our holographic 
computations.

\subsection{The near-horizon limit and Kaluza-Klein expansions}
The next step is to consider the near-horizon limit of the fractional D3-brane geometry
\cite{Gukov:1998kk}. This means that the 10-dimensional space is taken to be of the
form
\begin{equation}
\mathrm{AdS}_5\times S^5/\mathbb{Z}_M
\label{AdS5S5}
\end{equation}
and the field-stregth $F_5=dC_4$ to be proportional to the volume form of the AdS$_5$ space. The 6-dimensional space transverse to the orbifold fixed locus where the 
scalars $b_\alpha$ and $c_\alpha$ propagate is instead taken to be of the form
\begin{equation}
\mathrm{AdS}_5\times S^1~.
\label{AdSS1}
\end{equation}
In Appendix\,\ref{app:harmonics} we provide an explicit parametrization of $S^5$ and of
the $\mathbb{Z}_M$ action on its angular coordinates, from which one can easily see that the fixed
locus is indeed a circle $S^1$ inside $S^5$.

The analysis of Type II B supergravity in the AdS$_5\times S^5$ space was performed long ago in
\cite{Kim:1985ez} where the full spectrum of excitations around that background was derived from the equations obeyed by all supergravity fields. 
For our purposes, here it is enough to consider the fluctuations of the metric and the 
4-form which are the fields appearing in the action (\ref{S10}). Therefore, we write\,%
\footnote{Our conventions are the following: 10-dimensional indices are denoted by Latin letters $m,n,\ldots$; 5-dimensional indices in the AdS$_5$ space are denoted by Greek letters from the middle part of the alphabet $\mu,\nu,\ldots$, whereas 5-dimensional indices along 
the 5-sphere are denoted by Greek letters from the beginning part of the alphabet $\alpha,\beta,\ldots$ The notation $(mn)$ means that this pair of indices is symmetrized with strength one and with the trace removed.}
\begin{equation}
G_{mn}=g_{mn}+h_{mn}\quad\mbox{and}\quad
{C_4}_{\,m_1\ldots m_4}={c}_{m_1\ldots m_4}+{a}_{m_1\ldots m_4}~,
\label{back1}
\end{equation}
where $g_{mn}$ and ${c}_{m_1\ldots m_4}$ are the background fields, while 
the fluctuations are as in \cite{Kim:1985ez,Lee:1998bxa}, namely\,%
\footnote{In writing these expressions we have already implemented several consistency conditions and constraint equations. For details we refer to the original paper \cite{Kim:1985ez}. Moreover, we have written only those fluctuations which contain terms that are scalars with respect to $S^5$.}
\begin{equation}
\begin{aligned}
&h_{\mu\nu}=h_{(\mu\nu)}^\prime-\frac{3}{25}\,h_2\,g_{\mu\nu}~~~\mbox{with}~~~
g^{\mu\nu}\,h_{(\mu\nu)}^\prime=0~,\\
&h_{\alpha\beta}=h_{(\alpha\beta)}^\prime+\frac{1}{5}\,h_2\,g_{\alpha\beta}~~~~\mbox{with}~~~
g^{\alpha\beta}\,h_{(\alpha\beta)}^\prime=0~,\\
&a_{\mu_1\mu_2\mu_3\mu_4}=-\,\epsilon_{\mu_1\mu_2\mu_3\mu_4\nu}\,\partial^\nu a~,
\\[1mm]
&a_{\alpha_1\alpha_2\alpha_3\alpha_4}=\,\epsilon_{\alpha_1\alpha_2\alpha_3\alpha_4\beta}\,\partial^\beta a~.
\end{aligned}
\label{fluctuations}
\end{equation}
We then expand these fluctuations in the spherical harmonics of $S^5$ to
find the Kaluza-Klein (KK) modes that propagate in the AdS$_5$ space. Since we are interested in those KK modes that are scalars in $S^5$ and that are dual to the untwisted operators of the quiver theory, among all possible scalar harmonics we consider only the following ones:
\begin{equation}
Y^{\pm n}=\frac{1}{2^{\frac{n}{2}}}\,\cos^n\!\phi\,\mathrm{e}^{\pm\,\ii\,n\,\theta}~,
\label{y+-n}
\end{equation}
where we have used the parametrization of $S^5$ discussed in Appendix\,\ref{app:harmonics}. In
this parametrization, the condition $\phi=0$ defines the fixed locus of the $\mathbb{Z}_M$ orbifold which is a circle $S^1$ parametrized by $\theta\in[0,2\pi]$. Notice that these harmonics
are normalized in the standard way (see Appendix\,\ref{subapp:ZM}) and remain non-trivial at the
orbifold fixed locus. The relevant expansions are then\,%
\footnote{We do not consider the symmetric traceless fluctuation $h_{(\alpha\beta)}$ since it
does not yield KK modes that are scalars in $S^5$.}
\begin{equation}
h_2=\sum_{k\in\mathbb{Z}} h_{2,k}\, Y^k~,\quad
a=\sum_{k\in\mathbb{Z}} a_{k}\, Y^k~,\quad
h_{(\mu\nu)}^\prime=\sum_{k\in\mathbb{Z}} h_{(\mu\nu),k}^\prime\, Y^k~.
\label{expansion}
\end{equation}
Following the same steps described in \cite{Kim:1985ez,Lee:1998bxa}, one can show that the equations of motion for $h_{2,k}$ and $a_k$, which descend from (\ref{S10}), can be diagonalized by introducing the combinations
\begin{equation}
\begin{aligned}
s_k&=\frac{1}{20(k+2)}\,\big[h_{2,k}-10\,(k+4)\,a_k\big]~,\qquad\qquad\,
t_k=\frac{1}{20(k+2)}\,\big[h_{2,k}+10\,k \,a_k\big]~,\\
s_k^*&=\frac{1}{20(k+2)}\,\big[h_{2,-k}-10\,(k+4)\,a_{-k}\big]~,\quad
\qquad
t_k^*=\frac{1}{20(k+2)}\,\big[h_{2,-k}+10\,k \,a_{-k}\big]~,
\end{aligned}
\label{sk}
\end{equation}
for $k\geq 2$. Indeed, one finds
\begin{equation}
\nabla_\mu \nabla^\mu s_k=k (k-4)\,s_k\quad\mbox{and}\quad
\nabla_\mu \nabla^\mu \,t_k=(k+4)(k+8)\,t_k~,
\label{masssk}
\end{equation}
and similarly for $s_k^*$ and $t_k^*$. The relations (\ref{sk}) can be easily inverted, getting
\begin{equation}
h_{2,k}=10\,k\,s_k+10\,(k+4)\,t_k~,\qquad
a_k=-s_k+t_k~,
\label{hst}
\end{equation}
with analogous expressions for their conjugates.
Furthermore, as shown in \cite{Lee:1998bxa}, the constraint equations satisfied by $h_{(\mu\nu)}^\prime$ can be solved by requiring that
\begin{equation}
h_{(\mu\nu),k}^\prime=\frac{2\,\nabla_{(\mu}\nabla_{\nu)}\,\big(h_{2,k}-30\,a_k\big)}{5(k+1)(k+3)}
=\frac{4\,\nabla_{(\mu}\nabla_{\nu)}\,s_k}{k+1}+\frac{4\,\nabla_{(\mu}\nabla_{\nu)}\,t_k}{k+3}
\label{hprimek}
\end{equation}
for $k\geq 2$, and an analogous relation for $h_{(\mu\nu),-k}^{\prime}$ in terms of $s_k^*$
and $t_k^*$.

The KK modes $s_k$ and $s_k^*$ are dual to the untwisted primary operators $U_k(x)$ and $\overbar{U}_k(x)$ of the gauge theory and will be the focus of our attention in the following. 
The KK modes
$t_k$ and $t_k^*$ correspond instead to scalar descendants of these primary operators and will not be considered any longer. Inserting this information into (\ref{hst}) and (\ref{hprimek}), we obtain the following effective substitution rules
\begin{equation}
h_{2,k} \,\to\,10\,k\,s_k~,\quad
a_k\,\to\,-s_k~,\quad
h_{(\mu\nu),k}^\prime\,\to\,
\frac{4\,\nabla_{(\mu}\nabla_{\nu)}\,s_k}{k+1}~,
\label{ruleuntw}
\end{equation}
for $k\geq 2$, and similarly for the complex conjugate modes.

The dynamics of $s_k$ and $s_k^*$ can be obtained by using  the above harmonic expansions and the effective rules into the action (\ref{S10}). In the AdS$_5\times S^5$ case, corresponding to the $\mathcal{N}=4$ SYM theory, this is precisely what has been done in \cite{Lee:1998bxa}. We can therefore heavily rely on that analysis
and simply rephrase those findings in our notations, adapting them to the $\mathcal{N}=2$ orbifold
theory. Proceeding in this way, at the quadratic level we obtain the following effective action 
in AdS$_5$:
\begin{equation}
S_{\mathrm{untw}}^{(2)}=\frac{4(M N)^2}{(2\pi)^5}\int_{\mathrm{AdS}_5}\!\!d^5z\,\sqrt{g}
\,\sum_{k\geq2}A_k\Big(\nabla_\mu s_k^*\,\nabla^\mu s_k+k(k-4)
\,s_k^*\,s_k\Big)\,\frac{\pi^3}{M}~.
\label{Suntw}
\end{equation}
Let us comment on the various terms appearing in this expression. The prefactor is simply the rewriting of the gravitational constant using the AdS/CFT dictionary for the case at hand, namely
\begin{equation}
4\pi g_s=\frac{\lambda}{MN}\quad\mbox{and}\quad
\alpha^\prime=\frac{R^2}{\sqrt{\lambda}}
\label{dictionary}
\end{equation}
where $R$ is the radius of AdS$_5$ and of $S^5$. Therefore, in units where this radius is set to 1, we have
\begin{equation}
\frac{1}{2\kappa_{10}^2}=\frac{1}{(2\pi)^7\,g_s^2\,\alpha^{\prime\,4}}=\frac{4(M N)^2}{(2\pi)^5}
\end{equation}
which is the prefactor in (\ref{Suntw}). The normalization factor of the kinetic term is
\begin{equation}
A_k=\Big[\frac{32\,k(k-1)(k+2)}{k+1}\Big]\,\Big[\frac{1}{2^{k-1}(k+1)(k+2)}\Big]~.
\end{equation}
where the first bracket has been explicitly derived in \cite{Lee:1998bxa}, while the second
bracket comes from the overlap of the spherical harmonics (see (\ref{YYbis})).
Finally, the last factor of $\pi^3/M$ is simply the volume of $S^5/\mathbb{Z}_M$ (see (\ref{volS5M})).
Up to the $M$-dependence, which is due to the orbifold, the action (\ref{Suntw}) is the same as
the one appearing in \cite{Lee:1998bxa}, and of course yields the equations (\ref{masssk}) satisfied by the KK modes $s_k$ and $s_k^*$.

Let us now turn to the twisted sectors. We start from the action (\ref{S6bis}) and take the background geometry to be (\ref{AdSS1}) with the 5-form $F_5$ proportional to the volume form of
AdS$_5$. Then, we expand the twisted scalars in the harmonics of $S^1$, according to
\begin{equation}
b_\alpha=\sum_{k\in \mathbb{Z}}b_{\alpha,k}\,\mathrm{e}^{\ii k\theta} 
\quad\mbox{and}\quad
c_\alpha=\sum_{k\in \mathbb{Z}}c_{\alpha,k}\,\mathrm{e}^{\ii k\theta} ~.
\label{bck}
\end{equation}
Notice that as a consequence of (\ref{conjugatebc}), the KK modes $b_{\alpha,k}$ and
$c_{\alpha,k}$ satisfy the following complex conjugation rules
\begin{equation}
b_{\alpha,k}^{*}=b_{M-\alpha,-k}\quad\mbox{and}\quad
c_{\alpha,k}^{*}=c_{M-\alpha,-k}~.
\end{equation}
As shown in \cite{Gukov:1998kk,Billo:2021rdb}, the equations of motion that follow from (\ref{S6bis})
after using the above expansions can be diagonalized by introducing in each twisted sector $\alpha$
the combinations
\begin{equation}
\begin{aligned}
\eta_{\alpha,k}&=c_{\alpha,k}-\ii\,b_{\alpha,k}~,\qquad\qquad\qquad
\gamma_{\alpha,k}=c_{\alpha,k}+\ii\,b_{\alpha,k}~,\\
\eta_{\alpha,k}^*&=c_{M-\alpha,-k}+\ii\,b_{M-\alpha,-k}~,\quad\quad
\gamma_{\alpha,k}^*=c_{M-\alpha,-k}-\ii\,b_{M-\alpha,-k}~,
\end{aligned}
\label{gammak}
\end{equation}
for $k\geq 2$. In fact, one has
\begin{equation}
\nabla_\mu \nabla^\mu \eta_{\alpha,k}=k (k-4)\,\eta_{\alpha,k}\quad\mbox{and}\quad
\nabla_\mu \nabla^\mu \gamma_{\alpha,k}=k(k+4)\,\gamma_{\alpha,k}~,
\label{massgammak}
\end{equation}
and similarly for $\eta_{\alpha,k}^*$ and $\gamma_{\alpha,k}^*$.
The relations (\ref{gammak}) can be easily inverted leading to
\begin{equation}
c_{\alpha,k}=\frac{1}{2}\big(\eta_{\alpha,k}+\gamma_{\alpha,k}\big)~,\qquad
b_{\alpha,k}=\frac{\ii}{2}\big(\eta_{\alpha,k}-\gamma_{\alpha,k}\big)~,
\label{bcetagamma}
\end{equation}
with analogous expressions for the conjugate modes.

In \cite{Gukov:1998kk} it was proved that the mass spectrum (\ref{massgammak}) perfectly accounts for the scalar
operators of the quiver gauge theory. In particular, the KK modes $ \eta_{\alpha,k}$
and $\eta_{\alpha,k}^*$ are dual to the twisted primary 
operators $T_{\alpha,k}(x)$ and $\overbar{T}_{\alpha,k}(x)$
defined in (\ref{defTk}) for $k\geq 2$, and thus from now on we will focus on them and disregard
$\gamma_{\alpha,k}$ and $\gamma_{\alpha,k}^*$. This means that 
we can use the following effective replacements
\begin{equation}
c_{\alpha,k}~\to~\frac{1}{2}\,\eta_{\alpha,k}~,
\quad
b_{\alpha,k}~\to~\frac{\ii}{2}\,\eta_{\alpha,k}~,
\quad
c_{\alpha,k}^*~\to~\frac{1}{2}\,\eta_{\alpha,k}^*~,
\quad
b_{\alpha,k}^*~\to~-\frac{\ii}{2}\,\eta_{\alpha,k}^*~,
\label{ruletw}
\end{equation}
which follow from (\ref{bcetagamma}).

Inserting the above expansions into the twisted action (\ref{S6bis}), it is easy to obtain
\begin{equation}
S_{\mathrm{tw}}^{(2)}=\frac{4(M N)^2}{(2\pi)^3 M \lambda}\sum_{\alpha=1}^{M-1}
\int_{\mathrm{AdS}_5}\!\!d^5z\,\sqrt{g}\,\frac{1}{2s_\alpha}
\sum_{k\geq2}\Big(\nabla_\mu\eta_{\alpha,k}^*\,\nabla^\mu\eta_{\alpha,k}+k(k-4)\,\eta_{\alpha,k}^*\,\eta_{\alpha,k}\Big)\,2\pi~,
\label{Stw}
\end{equation}
from which the equations of motion for $\eta_{\alpha,k}$ and $\eta_{\alpha,k}^*$ given in
(\ref{massgammak}) immediately follow. Notice that
the prefactor in (\ref{Stw}) 
is just the rewriting of the effective 6-dimensional gravitational constant using the
holographic dictionary (\ref{dictionary}):
\begin{equation}
\frac{(2\pi\alpha^\prime)^2}{2\kappa_{10}^2 M} = \frac{1}{(2\pi)^5\,g_s^2\,\alpha^{\prime\,2}M}=
\frac{4(M N)^2}{(2\pi)^3 M \lambda}~,
\label{kappa6}
\end{equation}
and the last factor of $2\pi$ in (\ref{Stw}) is simply the length of the unit circle $S^1$.

The quadratic actions (\ref{Suntw}) and (\ref{Stw}) can be used to compute the 2-point functions
of the untwisted and twisted primary operators of the quiver theory in a holographic way.

\subsection{The 2-point functions}
\label{subsecn:2pointholo}
To obtain the 2-point functions we follow the procedure
described in detail in \cite{Witten:1998qj,Gubser:1998bc}
and also in \cite{Freedman:1998tz}.

\subsubsection{The untwisted sector}
We introduce a linear coupling on the boundary of the AdS$_5$ space between the KK modes $s_k$ and $s_k^*$ and the corresponding dual operators $U_k$ and $\overbar{U}_k$, which reads
\begin{equation}
S_{\mathrm{untw}}^\prime=\int_{\partial(\mathrm{AdS}_5)} ~\sum_{k\geq2} w_k
\big(s_k^*\,U_k+s_k\,\overbar{U}_k\big)~.
\label{Suntwbound}
\end{equation}
Here, as in \cite{Lee:1998bxa}, we have put an arbitrary coefficient $w_k$ to parametrize our lack of knowledge of the boundary action. Indeed, we only know that the supergravity fields that couple
to $U_k$ and $\overbar{U}_k$ are proportional to the untwisted KK modes $s_k^*$ and $s_k$. Then, adding the boundary action (\ref{Suntwbound}) to the 
bulk action (\ref{Suntw}), and using the explicit formulas of
\cite{Freedman:1998tz}\,%
\footnote{See in particular Eq.\,(17) with the correction factor in Eq.\,(95).}, we obtain
\begin{equation}
\big\langle U_k(x)\,\overbar{U}_k(y)\big\rangle=\frac{G_{U_k}}{|x-y|^{2k}}
\label{2p-untw-sugra}
\end{equation}
where
\begin{equation}
G_{U_k}=\frac{4(M N)^2}{(2\pi)^5}\,
\frac{A_k}{w_k^2}\,\Big[\frac{1}{\pi^2}\,\frac{\Gamma(k+1)}{\Gamma(k-2)}\,\frac{2(k-2)}{k}\Big]\,\frac{\pi^3}{M}~.
\label{GUksugra}
\end{equation}
Here we have kept the various terms separate so that it is easier to trace their origin. Notice that the factor $w_k^2$ in the denominator is due to the rescaling of the KK modes $s_k$ and $s_k^*$ 
into $s_k/w_k$ and $s_k^*/w_k$ which puts the boundary coupling (\ref{Suntwbound}) in
the canonical form $\big(s_k^*\,U_k+s_k\,\overbar{U}_k\big)$ but changes the normalization of the kinetic term in the bulk action from $A_k$ to
$A_k/w_k^2$. The 2-point amplitude (\ref{GUksugra}) can be simplified as
\begin{equation}
G_{U_k}=M N^2\,
\frac{k\,(k-1)^2(k-2)^2}{2^{k-4}\,\pi^4\,w_k^2\,(k+1)^2}~.
\label{GUksugra1}
\end{equation}
\subsubsection{The twisted sectors}
In the twisted sectors we proceed in a similar way. We first introduce the boundary action
\begin{equation}
S_{\mathrm{tw}}^\prime=\sum_{\alpha=1}^{M-1}
\int_{\partial(\mathrm{AdS}_5)} ~\sum_{k\geq 2}\varpi_{\alpha,k}
\big(\eta_{\alpha,k}^*\,T_{\alpha,k}+\eta_{\alpha,k}\,\overbar{T}_{\alpha,k}\big)
\label{Stwbound}
\end{equation}
where $\varpi_{\alpha,k}$ is an arbitrary coefficient. Then, using the bulk action (\ref{Stw}) and applying the formulas 
of \cite{Freedman:1998tz}, we obtain
\begin{equation}
\big\langle T_{\alpha,k}(x)\,\overbar{T}_{\alpha,k}(y)\big\rangle=\frac{G_{T_{\alpha,k}}}{|x-y|^{2k}}
\label{2p-tw-sugra}
\end{equation}
with
\begin{equation}
G_{T_{\alpha,k}}=\frac{4(M N)^2}{(2\pi)^3 M \lambda}\,
\frac{1}{2s_\alpha}\,\frac{1}{\varpi_{\alpha,k}^2}\,\Big[\frac{1}{\pi^2}\,\frac{\Gamma(k+1)}{\Gamma(k-2)}\,\frac{2(k-2)}{k}\Big]\,2\pi~.
\label{GTksugra}
\end{equation}
Again we have kept the various factors separate in order to easily trace their origin from those appearing in (\ref{Stw}). In particular we note that the original normalization factor of the kinetic term $1/(2s_\alpha)$ becomes 
$1/(2s_\alpha\varpi_{\alpha,k}^2)$ after the rescalings 
of the KK modes $\eta_{\alpha,k}$ and $\eta_{\alpha,k}^*$ which bring 
the boundary coupling to the canonical form. Simplifying (\ref{GTksugra}), we have
\begin{equation}
G_{T_{\alpha,k}}=\frac{M N^2}{s_\alpha \lambda}\,
\frac{(k-1)(k-2)^2}{\pi^4\,\varpi_{\alpha,k}^2}~.
\label{GTksugra1}
\end{equation}

\subsection{The 3-point functions}
\label{subsecn:3pointholo}
In order to compute the 3-point functions of the gauge theory operators,
we need to find the cubic couplings of their dual KK modes.

\subsubsection{The untwisted sector}
For the untwisted case, these cubic couplings have been worked out in detail
in \cite{Lee:1998bxa}. Thus, we can rely on that analysis and simply translate those results in our conventions. 
The cubic action for the untwisted KK modes $s_k$ and $s_k^*$ is
\begin{equation}
S^{(3)}_{\mathrm{untw}}=\frac{4(M N)^2}{(2\pi)^5}\int_{\mathrm{AdS}_5}\!\!d^5z\,\sqrt{g}
\,\sum_{k,\ell,p\geq2}\Big(V_{k,\ell,p}\,s_k^*\,s_\ell^*\,s_p\,\delta_{k+\ell-p,0}+~\mbox{c.c.}\Big)\,\frac{\pi^3}{M}~,
\label{S3untw}
\end{equation}
where $V_{k,\ell,p}$ is given in Eqs.\,(3.39) and (3.40) of
\cite{Lee:1998bxa}. In our notation it reads
\begin{equation}
V_{k,\ell,p}=-a_{k,\ell,p}\,\frac{128\,(k+\ell+p)
\Big[\big(\frac{k+\ell+p}{2}\big)^2-1\Big]\Big[\big(\frac{k+\ell+p}{2}\big)^2-4\Big]\,\big(\frac{k+\ell-p}{2}\big)\,\big(
\frac{\ell+p-k}{2}\big)\,\big(\frac{p+k-\ell}{2}\big)}{(k+1)\,(\ell+1)\,(p+1)}
\label{Vklp}
\end{equation}
where
\begin{equation}
a_{k,\ell,p}=\frac{1}{2^{\frac{k+\ell+p}{2}-1}\,\big(\frac{k+\ell+p}{2}+1\big)\,\big(\frac{k+\ell+p}{2}+2\big)}
\label{aklp}
\end{equation}
is the overlap coefficient of three spherical harmonics (see (\ref{YYYbis})).

Notice that $V_{k,\ell,p}$ vanishes if one uses the $\delta$-function imposing
charge conservation. This is a well-known feature \cite{DHoker:1999jke,Rastelli:2017udc} of all coupling coefficients that are related to extremal
correlators, like the 3-point functions we are interested in, and is not in contradiction with the fact that the final holographic correlators are non-zero. Indeed, as we are going to see, the zero in the coupling is compensated by a pole in the Witten diagram that yields the 3-point function so that the final result is finite and well-defined. This can be clearly seen if one uses the $\delta$-function of charge conservation only at the end \cite{Lee:1998bxa,DHoker:1999jke,Rastelli:2017udc},
which is what we are going to do\,%
\footnote{If one enforces the charge conservation from the very beginning, one needs to carefully evaluate the contributions to the correlators coming from boundary terms which, as shown in \cite{DHoker:1999jke}, lead to the same final result in the 3-point correlators as the other approach.}.

Using the cubic action (\ref{S3untw}) together with the boundary action (\ref{Suntwbound}) and applying the formula in Eq.\,(25) of \cite{Freedman:1998tz}, we find that the coefficient of the correlator of three untwisted operators is
\begin{align}
G_{U_k,U_\ell,\overbar{U}_p}&=-\frac{4(M N)^2}{(2\pi)^5}
\,\frac{2 V_{k,\ell,p}}{w_k w_\ell w_p}\,\bigg[\frac{\Gamma\big(\frac{k+\ell-p}{2}\big)\,\Gamma\big(\frac{k+p-\ell}{2}\big)\,\Gamma\big(\frac{\ell+p-k}{2}\big)\,\Gamma\big(\frac{k+\ell+p}{2}-2\big)}
{2 \pi^4\,\Gamma(k-2)\,\Gamma(\ell-2)\,\Gamma(p-2)}\bigg]\,\frac{\pi^3}{M}
\label{GUklp}
\end{align}
where we have understood the $\delta$-function of charge conservation and have again kept separate all terms in order to easily trace their origin\,%
\footnote{The factor of 2 in front of $V_{k\ell p}$ is a multiplicity factor due to the symmetry of the correlator in $k$ and $\ell$.}.
{From} this expression, we clearly see that the vanishing factor $(k+\ell-p)$ in $V_{k,\ell,p}$
is compensated by the pole in the first $\Gamma$-function of the numerator inside the square
bracket that comes from the cubic Witten diagram; thus as mentioned above, the product is 
well-defined and non-zero when the charge conservation is imposed. 
We can drastically simplify the right hand side of
(\ref{GUklp}) and get a completely factorized expression
\begin{align}
G_{U_k,U_\ell,\overbar{U}_p}=M N^2\,  \Big(
\frac{k\,(k-1)\,(k-2)}{2^{\frac{k}{2}-2}\,\pi^2\,w_k\,(k+1)}\Big)\Big(\frac{\ell\,(\ell-1)\,(\ell-2)}{2^{\frac{\ell}{2}-2}\,\pi^2\,w_\ell\,(\ell+1)}\Big)\Big(
\frac{p\,(p-1)\,(p-2)}{2^{\frac{p}{2}-2}\,\pi^2\,w_p\,(p+1)}\Big)
\label{GUklp1}
\end{align}
where again the $\delta$-function of charge conservation is understood.

\subsubsection{The twisted sectors}
In this case we have to work out the cubic couplings involving one untwisted and two twisted modes. To do so, we start from the 6-dimensional action (\ref{S6bis})
and expand it around the AdS$_5\times S^1$ background keeping all contributions up to the third order in the fluctuations.
We treat the two parts of (\ref{S6bis}) separately. For the first term that depends on the 6-dimensional metric, we need the explicit expressions
of the latter and its fluctuations. Using (\ref{back1}) and (\ref{fluctuations}), we have
\begin{equation}
\begin{aligned}
\sqrt{G^\prime}&=\sqrt{g}-\frac{1}{5}\sqrt{g}\,h_2~,\\
G^{\prime}_{\mu\nu}&=g_{\mu\nu}-\frac{3}{25}\,h_2\,g_{\mu\nu}
+
h^{\prime}_{(\mu\nu)}~,\quad
G^{\prime}_{\mu\theta}=0~,\quad
G^{\prime}_{\theta\theta}=1+\frac{1}{5}\,h_2~,\\
G^{\prime\,\mu\nu}&=g^{\mu\nu}+\frac{3}{25}\,h_2\,g^{\mu\nu}-g^{\mu\rho}\,
h^{\prime}_{(\rho\sigma)}\,g^{\sigma\nu}~,\quad
G^{\prime\,\mu\theta}=0~,\quad
G^{\prime\,\theta\theta}=1-\frac{1}{5}\,h_2~.
\end{aligned}
\end{equation}
Inserting these expressions into the ``metric'' term of (\ref{S6bis}), we easily see that
\begin{align}
\sum_{\alpha=1}^{M-1}\frac{1}{2s_\alpha}\sqrt{G^\prime}\, \Big(
\nabla b_{\alpha}^{*} \!\cdot\! \nabla b_{\alpha}+
\nabla c_{\alpha}^{*} \!\cdot\!\nabla c_{\alpha}\Big)
&=\sum_{\alpha=1}^{M-1}\frac{1}{2s_\alpha}
\Big(\mathcal{L}^{(2)}_\alpha+\mathcal{L}^{(3)}_\alpha+\ldots\Big)
\label{bbcc}
\end{align}
where
\begin{align}
\mathcal{L}^{(2)}_\alpha=\sqrt{g}\,\Big(
\nabla_\mu b_{\alpha}^{*} \,\nabla^\mu b_{\alpha}+
\nabla_\mu c_{\alpha}^{*}\,\nabla^\mu c_{\alpha}+
\partial_{\theta\,} b_{\alpha}^{*} \,\partial_{\theta\,} b_{\alpha}+
\partial_{\theta\,} c_{\alpha}^{*} \,\partial_{\theta\,} c_{\alpha}\Big)
\label{L2tw}
\end{align}
is the Lagrangian that contributes to the quadratic action (\ref{Stw}), while 
\begin{align}
\mathcal{L}^{(3)}_\alpha&=-\frac{2}{25}\,\sqrt{g}\,h_2\Big(
\nabla_\mu b_{\alpha}^{*} \,\nabla^\mu b_{\alpha}+
\nabla_\mu c_{\alpha}^{*} \,\nabla^\mu c_{\alpha}\Big)\notag\\
&\quad-\sqrt{g}\,h^{\prime}_{(\mu\nu)}\Big(
\nabla^\mu b_{\alpha}^{*} \,\nabla^\nu b_{\alpha}+
\nabla^\mu c_{\alpha}^{*} \,\nabla^\nu c_{\alpha}\Big)
\label{L3tw}
\\
&\quad-\frac{2}{5}\,\sqrt{g}\,h_2\Big(\partial_{\theta\,} b_{\alpha}^{*} \,
\partial_{\theta\,} b_{\alpha}+
\partial_{\theta\,} c_{\alpha}^{*} \,\partial_{\theta\,} c_{\alpha}\Big)
\notag
\end{align}
is the cubic Lagrangian which is of interest for us. Finally, the ellipses in (\ref{bbcc}) stand for higher order terms in the fluctuations which we will not consider.

We now elaborate on the cubic terms (\ref{L3tw}) and expand all fields in spherical harmonics. For the twisted ones, these expansions are given in
(\ref{bck}), while for the untwisted ones, $h_2$ and $h^\prime_{(\mu\nu)}$, they are given in (\ref{expansion}) but with the spherical harmonics
evaluated at the orbifold fixed locus, namely
\begin{equation}
Y^k=\frac{1}{2^{\frac{|k|}{2}}}\,\mathrm{e}^{\ii\,k\,\theta}
\label{Yk0}
\end{equation} 
(see also (\ref{Y+-n0})).
Inserting these expansions in (\ref{L3tw}), we obtain
\begin{align}
\sum_{\alpha=1}^{M-1}\frac{1}{2s_\alpha}
\mathcal{L}^{(3)}_\alpha&=
\sum_{\alpha=1}^{M-1}\sqrt{g}\,\frac{1}{2s_\alpha}
\sum_{k,\ell,p\geq 2}
\frac{1}{2^{\frac{k}{2}}}\,\bigg[\!
-\frac{4}{25}\,h_{2,k}^*\Big(
\nabla_\mu b_{\alpha,\ell}^{*} \,\nabla^\mu b_{\alpha,p}+
\nabla_\mu c_{\alpha,\ell}^{*} \,\nabla^\mu c_{\alpha,p}\Big)
\notag\\
&\qquad\qquad\qquad\qquad
-2\,h^{\prime\,*}_{(\mu\nu),k}\Big(
\nabla^\mu b_{\alpha,\ell}^{*} \,\nabla^\nu b_{\alpha,p}+
\nabla^\mu c_{\alpha,\ell}^{*} \,\nabla^\nu c_{\alpha,p}\Big)\label{cubic1}\\
&\qquad\qquad\qquad\qquad
-\frac{4}{5}\,h_{2,k}^*\Big(\ell\,p\,b_{\alpha,\ell}^{*} \,
b_{\alpha,p}+
\ell\,p\,c_{\alpha,\ell}^{*} \,c_{\alpha,p}\Big)\bigg]\,\mathrm{e}^{-\ii\,(k+\ell-p)\theta}+\mbox{c.c.}+\ldots~.
\notag
\end{align}
Here we have explicitly exhibited only the terms with $k$, $\ell$, $p\geq2$, 
since these are those which will be relevant for the
calculation of the 3-point functions we are interested in. All other structures
are understood in the ellipses. Thanks to the effective rules (\ref{ruleuntw})
and (\ref{ruletw}), we can rewrite (\ref{cubic1}) as
\begin{align}
\sum_{\alpha=1}^{M-1}\frac{1}{2s_\alpha}
\mathcal{L}^{(3)}_\alpha&=\sum_{\alpha=1}^{M-1}\sqrt{g}\,
\frac{1}{2s_\alpha}
\sum_{k,\ell,p\geq 2}
\frac{1}{2^{\frac{k}{2}}}\,\bigg(\!
-\frac{4}{5}\,k\,s_k^* \nabla_\mu \eta_{\alpha,\ell}^{*} \,\nabla^\mu \eta_{\alpha,p}
-\frac{4}{k+1}\,\nabla_{(\mu}\nabla_{\nu)}s_k^*
\nabla^\mu \eta_{\alpha,\ell}^{*} \,\nabla^\nu \eta_{\alpha,p}
\notag\\
&\qquad\qquad\qquad\qquad
-4\,k\,\ell\,p\,s_k^* \,\eta_{\alpha,\ell}^{*} \, \eta_{\alpha,p}
\bigg)\,\mathrm{e}^{-\ii\,(k+\ell-p)\theta}+\mbox{c.c.}+\ldots~.
\label{cubic2}
\end{align} 
Then, exploiting the identities proven in Appendix~\ref{app:identities}, up to a total derivative we can recast (\ref{cubic2})
in the following form
\begin{align}
\sum_{\alpha=1}^{M-1}\frac{1}{2s_\alpha}
\mathcal{L}^{(3)}_\alpha&=\sum_{\alpha=1}^{M-1}\sqrt{g}\,
\frac{1}{2s_\alpha}
\sum_{k,\ell,p\geq2}
\Big( L_{k,\ell,p}\,s_k^*\,\eta_{\alpha,\ell}^*\,\eta_{\alpha,p}\,\mathrm{e}^{-\ii\,(k+\ell-p)\theta}+\mbox{c.c.}\Big)+\ldots
\label{cubicL}
\end{align}
where the coupling coefficients are
\begin{align}
L_{k,\ell,p}&=\frac{1}{2^{\frac{k}{2}}}\bigg\{\!-\frac{2\,k}{5}\,\big[k(k-4)-\ell(\ell-4)-p(p-4)\big]
\notag\\
&\qquad
+\frac{1}{(k+1)}\Big[\ell(\ell-4)\big(\ell(\ell-4)-k(k-4)-p(p-4)\big)\notag\\
&\qquad\qquad\qquad+p(p-4)\big(p(p-4)-k(k-4)-\ell(\ell-4)\big)\notag\\
&\qquad\qquad\qquad-\frac{3}{5}\,k(k-4)\big(k(k-4)-\ell(\ell-4)-p(p-4)\big)\Big]-4\,k\,\ell\,p\bigg\}~.
\label{Lklp}
\end{align}

Let us now consider the topological term of the action (\ref{S6bis}). The corresponding 
Lagrangian can be written as a sum of two pieces: one which is quadratic
in the fluctuations and contributes to the quadratic action (\ref{Stw}) of the
twisted fields, and one which is cubic in the fluctuations. The latter is explicitly given by
\begin{align}
\sum_{\alpha=1}^{M-1}\frac{1}{2s_\alpha}
\widetilde{\mathcal{L}}^{(3)}_\alpha&=\sum_{\alpha=1}^{M-1}\frac{1}{2s_\alpha}\Big(\frac{8}{4!}\,\epsilon^{m_1\ldots m_6}\,a_{m_1\ldots m_4}
\nabla_{m_5}b_\alpha^*\,\nabla_{m_6}c_{\alpha}\Big)
\notag\\
&=-\sum_{\alpha=1}^{M-1}\sqrt{g}\,\frac{1}{2s_\alpha}
8 \,\nabla^\mu a \Big(
\nabla_{\mu}b_\alpha^*\,\partial_\theta c_{\alpha}-
\partial_\theta b_\alpha^*\, \nabla_{\mu}c_{\alpha}
\Big)
\label{top3}
\end{align}
where the last expression follows upon using 
the fluctuation of the 4-form given in the third line of (\ref{fluctuations}).
We now repeat the same steps described in detail for the ``metric'' terms, namely we expand the fields in (\ref{top3}) in the spherical harmonics and focus only on those terms which are relevant for the calculations of the 3-point functions we are interested in. Then, using the effective rules
(\ref{ruleuntw}) and (\ref{ruletw}) and the identities in Appendix~\ref{app:identities}, up to
a total derivative we find an expression similar to (\ref{cubicL}):
\begin{align}
\sum_{\alpha=1}^{M-1}\frac{1}{2s_\alpha}
\widetilde{\mathcal{L}}^{(3)}_\alpha&=\sum_{\alpha=1}^{M-1}\sqrt{g}\,
\frac{1}{2s_\alpha}
\sum_{k,\ell,p\geq2}
\Big( \widetilde{L}_{k,\ell,p}\,s_k^*\,\eta_{\alpha,\ell}^*\,\eta_{\alpha,p}
\,\mathrm{e}^{-\ii\,(k+\ell-p)\theta}+\mbox{c.c.}\Big)+\ldots
\label{cubicLtilde}
\end{align}
where
\begin{align}
\widetilde{L}_{k,\ell,p}&=\frac{1}{2^{\frac{k}{2}}}\Big\{2\,p\,\big[p(p-4)-k(k-4)-\ell(\ell-4)\big]+2\,\ell\,
\big[\ell(\ell-4)-k(k-4)-p(p-4)\big]\Big\}~.
\label{Ltildeklp}
\end{align}
Putting everything together and integrating over AdS$_5\times S^1$,
we finally obtain the cubic action involving one untwisted and two twisted KK modes: 
\begin{equation}
S^{(3)}_{\mathrm{tw}}=
\frac{4(M N)^2}{(2\pi)^3 M \lambda}\sum_{\alpha=1}^{M-1}
\int_{\mathrm{AdS}_5}\!\!d^5z\,\sqrt{g}
\,\frac{1}{2s_\alpha}\sum_{k,\ell,p\geq2}\Big(W_{k,\ell,p}\,s_k^*\,\eta_{\alpha,\ell}^*\,\eta_{\alpha,p}\,\delta_{k+\ell-p,0}+~\mbox{c.c.}\Big)\,2\pi~.
\label{Scubictw}
\end{equation}
Here the prefactor is the same as in the quadratic action
(\ref{Stw}), the last factor of $2\pi$ arises from the integration over
$S^1$, which also produces the $\delta$-functions imposing charge conservation, and
\begin{align}
W_{k,\ell,p}&=L_{k,\ell,p}+\widetilde{L}_{k,\ell,p}=
-\frac{(k +\ell - p) (k + p - \ell) (k +\ell + p - 2) (k + \ell + p - 4)}
{2^{\frac{k}{2}} \,(k+1)}~.
\label{Wklp}
\end{align}
Notice that, differently from the partial couplings $L_{k,\ell,p}$ and
$\widetilde{L}_{k,\ell,p}$ given in (\ref{Lklp}) and (\ref{Ltildeklp}), the total
coupling $W_{k,\ell,p}$ can be written as a product of simple factors. One of these factors
vanishes if the $\delta$-function of charge conservation is used, but again this zero will be
compensated by a pole in the Witten diagram that computes the 3-point function, 
so that the final result is finite and
well-defined. Indeed, using the bulk action we just derived together with the boundary action
(\ref{Stwbound}) and applying the formulas of \cite{Freedman:1998tz}, we find that the coefficient in the 3-point function of one untwisted and two twisted operators is
\begin{align}
G_{U_k,T_{\alpha,\ell},\overbar{T}_{\alpha,p}}&=-\frac{4(M N)^2}{(2\pi)^3 M \lambda}\,\frac{1}{2 s_\alpha}
\,\frac{W_{k,\ell,p}}{w_k\, \varpi_{\alpha,\ell} \,\varpi_{\alpha,p}} 
\bigg[\frac{\Gamma\big(\frac{k+\ell-p}{2}\big)\,\Gamma\big(\frac{k+p-\ell}{2}\big)\,\Gamma\big(\frac{\ell+p-k}{2}\big)\,\Gamma\big(\frac{k+\ell+p}{2}-2\big)}
{2 \pi^4\,\Gamma(k-2)\,\Gamma(\ell-2)\,\Gamma(p-2)}\bigg] 2\pi~.
\label{GUTTklp}
\end{align}
Again we have kept separate the various factors so that it is easier to trace their origin.
Using the explicit form (\ref{Wklp}) of the coupling coefficients, the expression can be simplified and the final result is
\begin{align}
G_{U_k,T_{\alpha,\ell},\overbar{T}_{\alpha,p}}&=\frac{M N^2}{s_\alpha \lambda}\,
\Big(
\frac{k\,(k-1)\,(k-2)}{2^{\frac{k}{2}-2}\,\pi^2\,w_k\,(k+1)}\Big)
\Big(
\frac{(\ell-1)\,(\ell-2)}{\pi^2\,\varpi_{\alpha,\ell}}\Big)
\Big(
\frac{(p-1)\,(p-2)}{\pi^2\,\varpi_{\alpha,p}}\Big)
\label{GUTTklp1}
\end{align}
where the charge-conservation $\delta$-function is understood.

\section{ Structure constants and effective Witten-like diagrams}
\label{secn:effective}
The 2-point coefficients (\ref{GUksugra1}) and (\ref{GTksugra1}), and the
3-point coefficients (\ref{GUklp1}) and (\ref{GUTTklp1}) are the main results of the previous section. They represent the coefficients in the 2- and 3-point correlators of the gauge theory operators in the planar limit and at strong
coupling predicted by the AdS/CFT correspondence, and should be compared with those obtained from supersymmetric localization at strong coupling that we derived in Section~\ref{secn:strong}.

A feature of note is that these coefficients have the same dependence on the
't Hooft coupling $\lambda$ in the two approaches. In particular, 
the coefficients in the untwisted correlators are independent of $\lambda$, while those in the correlators with twisted operators behave as $1/\lambda$. 
In the localization approach this dependence arises from the strong-coupling behavior of the quantities $\mathsf{D}^{(\alpha)}$ and 
$\mathsf{d}^{(\alpha)}$, while in the holographic
approach the $1/\lambda$ comes from the gravitational constant of the
6-dimensional theory defined at the orbifold fixed point (see (\ref{kappa6})).

Of course the detailed expressions of the 2- and 3-point coefficients in the
two approaches are sensitive to the normalization of the chiral and
anti-chiral operators in the quiver theory and to the coupling between these operators and the KK modes in the holographic approach, as one can see from the appearance of the parameters $w_k$ and $\varpi_{\alpha,k}$.
To get rid of these ambiguities we thus consider the structure constants.

\subsection{The holographic structure constants}
\label{subsecn:structureholo}
The structure constants computed using the AdS/CFT correspondence are
easily obtained from (\ref{GUksugra1}), (\ref{GTksugra1}), (\ref{GUklp1}) 
and (\ref{GUTTklp1}). In the untwisted case we have
\begin{equation}
C_{U_k,U_\ell,\overbar{U}_p}
=\frac{1}{\sqrt{M}\,N}\,\sqrt{k\,\ell\,p}~.
\label{CUUUsugra}
\end{equation}
Notice that in this combination all dependence on the arbitrary coefficients
$w_k$ drops out and many other factors cancel as well, leaving us with a result that
matches the one in (\ref{CUUUloc}) obtained using localization.

More importantly, in the case with twisted operators from (\ref{GUksugra1}), (\ref{GTksugra1}) and (\ref{GUTTklp1}), we find
\begin{equation}
C_{U_k,T_{\alpha,\ell},\overbar{T}_{\alpha,p}}
=\frac{1}{\sqrt{M}\,N}\,\sqrt{k\,(\ell-1)\,(p-1)}~.
\label{CUTTsugra}
\end{equation}
Again most of the factors, including the arbitrary coefficients $w_k$
and $\varpi_{\alpha,k}$ and the coupling $\lambda$ that appear in (\ref{GTksugra1}) and
(\ref{GUTTklp1}), cancel between numerator and denominator, and the end result is independent of $\lambda$ and of the twisted sector, and fully matches the localization result
in (\ref{CUTTloc}).

The complete agreement between the two approaches can be interpreted either as a validation of our extrapolation of the localization results at strong coupling or, alternatively, as a check of the AdS/CFT correspondence prescriptions in a model with non-maximal supersymmetry.

\subsection{Effective Witten-like diagrams}
\label{subsecn:effectiveWitten}

\begin{figure}[ht]
\center{\includegraphics[scale=0.45]{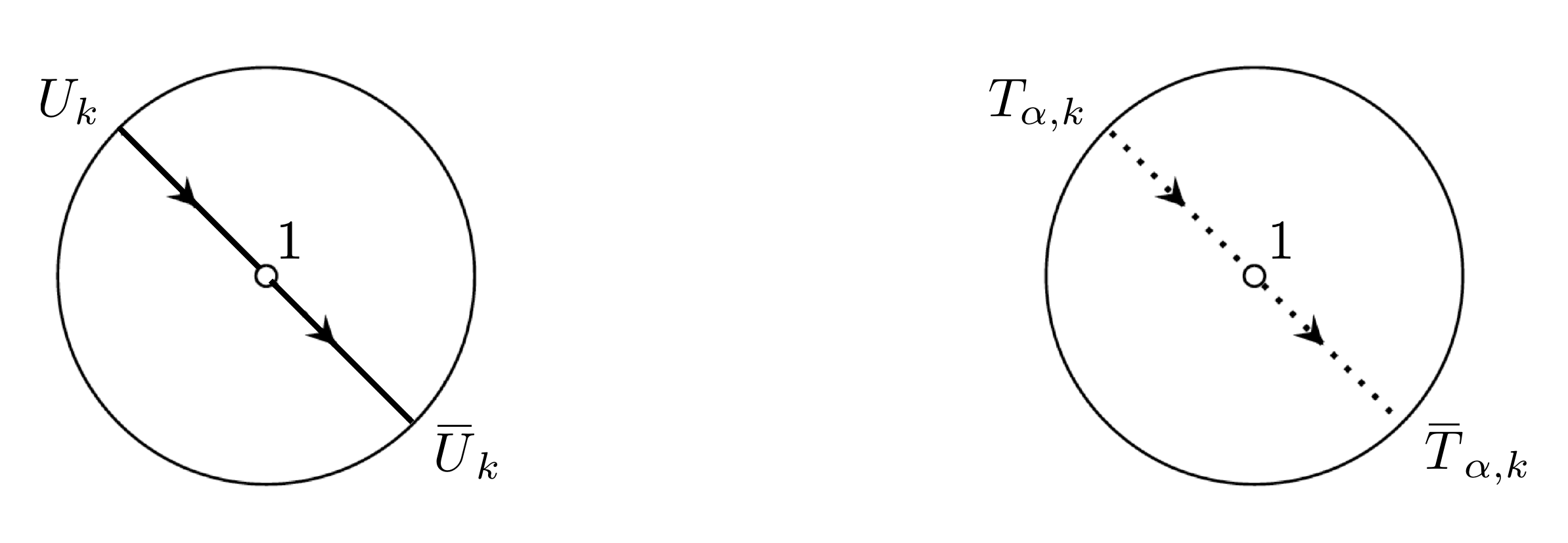}
\caption{The 2-point effective Witten-like diagrams that pictorially represent the untwisted (left)
and twisted (right) correlators in (\ref{effG2}).
The circle represents the space where the gauge theory operators are inserted. Our conventions are such that it yields the factor $M N^2$. 
The full and dotted oriented lines represent
the ``propagators'' of the untwisted and twisted modes, respectively, and the arrows distinguish
between the chiral and anti-chiral ones. Finally, the open dot in the middle represents the quadratic
vertex which, with the coefficients $w_k$ and $\varpi_{\alpha,k}$ we have chosen, is simply 1.
The correlators (\ref{effG2}) are thus given by the product of the global factor $M N^2$ times 1. \label{fig:6.1}} 
    }  
\end{figure}

The remarkable simplicity of the structure constants suggests to exploit the presence of the arbitrary coefficients $w_k$ and $\varpi_{\alpha,k}$ in order to simplify as much as possible the holographic results. In fact, these coefficients can always be chosen in such a way that the untwisted
and twisted 2-point correlators (\ref{GUksugra1}) and (\ref{GTksugra1}) become
\begin{equation}
G_{U_k}=M N^2\quad\mbox{and}\quad
G_{T_{\alpha,k}}=M N^2~.
\label{effG2}
\end{equation}
In this form these correlators can be interpreted as effective Witten-like diagrams in which 
the fields that are dual to the chiral and anti-chiral operators of the
gauge theory are connected with a trivial quadratic vertex as shown in Fig.~\ref{fig:6.1}.

\begin{figure}[ht]
\center{\qquad\quad~\includegraphics[scale=0.45]{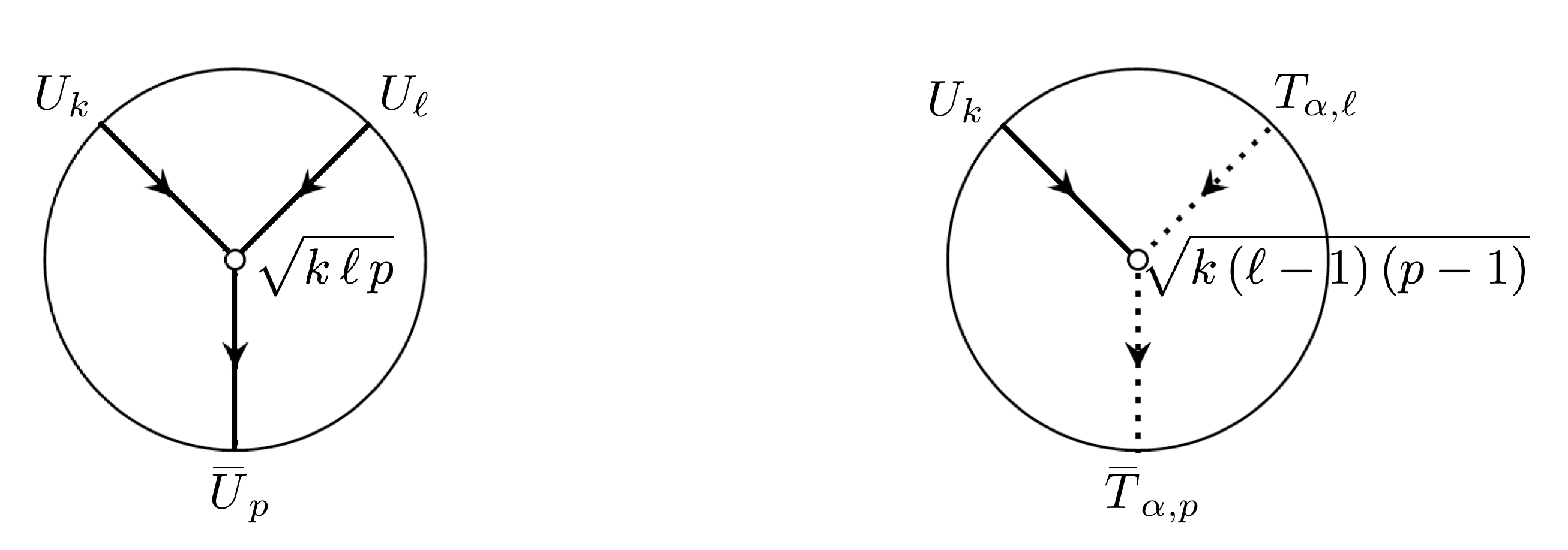}
\caption{The 3-point effective Witten-like diagrams that pictorially represent the correlators (\ref{effG3}). Again the circle, representing the space where the operators are inserted, yields the factor $M N^2$, while the cubic vertices in the middle are, respectively, $\sqrt{k\,\ell\,p}$ (left)
and $\sqrt{k\,(\ell-1)\,(p-1)}$ (right). Notice we have understood the $\delta_{k+\ell-p,0}$ which
enforces charge conservation.
\label{fig:6.2}}
    }  
\end{figure}

The same choice of $w_k$ and $\varpi_{\alpha,k}$ inserted into (\ref{GUklp1}) 
and (\ref{GUTTklp1}), leads to the following 3-point correlators
\begin{equation}
G_{U_k,U_\ell,\overbar{U}_p}=M N^2\,\sqrt{k\,\ell\,p}
\quad\mbox{and}\quad
G_{U_k,T_{\alpha,\ell},\overbar{T}_{\alpha,p}}=M N^2\,\sqrt{k\,(\ell-1)\,(p-1)}
\label{effG3}
\end{equation}
where the $\delta$-function imposing charge conservation is understood.
They too can be represented as effective Witten-like diagrams, this time with a non-trivial cubic vertex as shown in Fig.~\ref{fig:6.2}.

These examples show that all information about the correlators at strong coupling
can be encoded in a cubic vertex which is the product of 
three factors, each one being the square root of the conformal dimension in the case of untwisted operators or the square root of the conformal dimension minus one in the case of twisted operators, independently whether these are chiral or anti-chiral. Moreover, for twisted operators the cubic coupling does not depend on the twisted sector to which they belong. 

\begin{figure}[ht]
\center{\quad\includegraphics[scale=0.455]{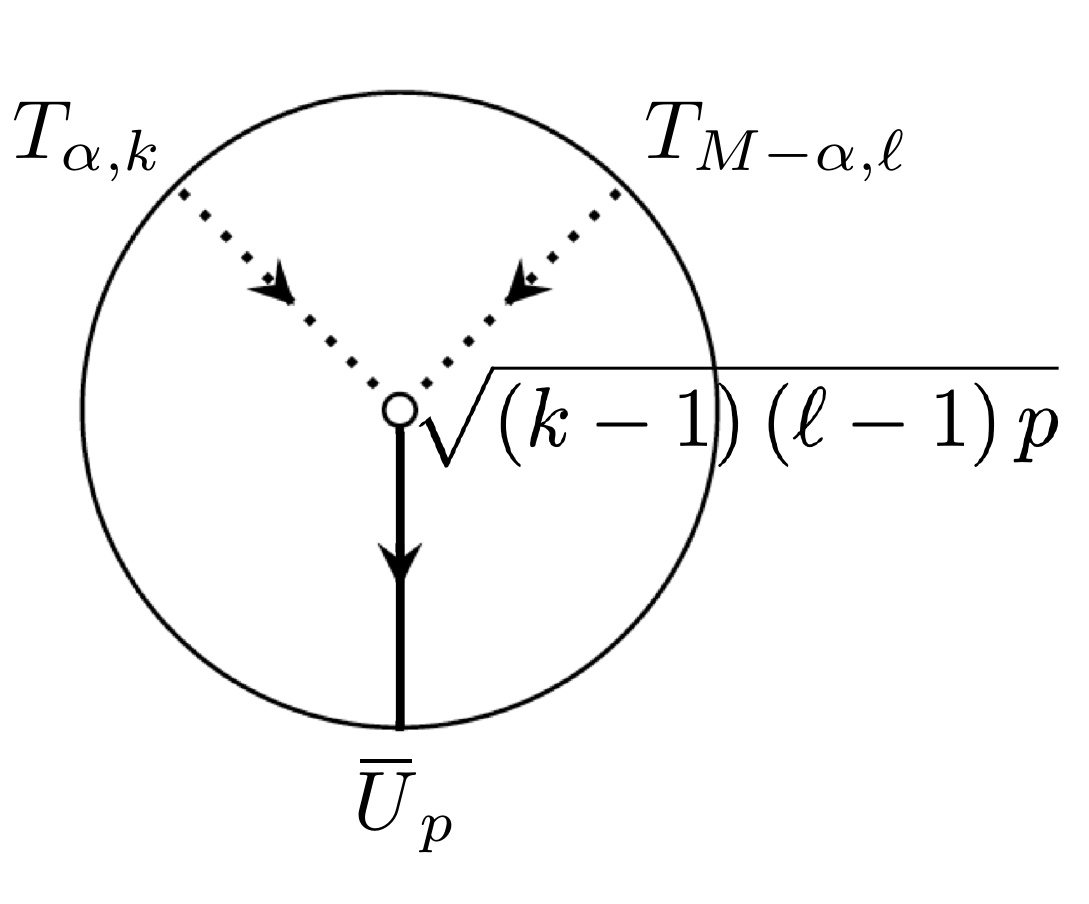}
\caption{The cubic effective Witten-like diagram describing the interaction of two chiral twisted operators belonging to conjugate twisted sectors and one anti-chiral untwisted operator. The cubic vertex is $\sqrt{(k-1)\,(\ell-1)\,p}$ times $\delta_{k+\ell-p,0}$ enforcing charge conservation. As in the previous figures, this $\delta$-function is understood.
\label{fig:6.3}}
    }  
\end{figure}

Using these rules it is natural also to consider the cubic interaction represented in Fig.~\ref{fig:6.3}, in which we have exchanged the orientation of two lines with respect to the right diagram of Fig.~\ref{fig:6.2}, in such a way that the two incoming lines both correspond to twisted chiral operators belonging to conjugate twisted sectors and the outgoing line corresponds to an untwisted anti-chiral operator. 
This cubic interaction, which is not included
in the supergravity derivations considered in Section~\ref{secn:holo}, leads to the following correlator
\begin{equation}
G_{T_{\alpha,k},T_{M-\alpha,\ell},\overbar{U}_{p}}=M N^2\,\sqrt{(k-1)\,(\ell-1)\,p}
\label{G3TTU}
\end{equation}
where the $\delta$-function enforcing the charge conservation is understood, 
and to the structure constants 
\begin{equation}
C_{T_{\alpha,k},T_{M-\alpha,\ell},\overbar{U}_{p}}
=\frac{1}{\sqrt{M}\,N}\,\sqrt{(k-1)\,(\ell-1)\,p}
\label{CTTUsugra}
\end{equation}
which match those in (\ref{CTTUloc}).

\begin{figure}[ht]
\center{\qquad\quad\includegraphics[scale=0.46]{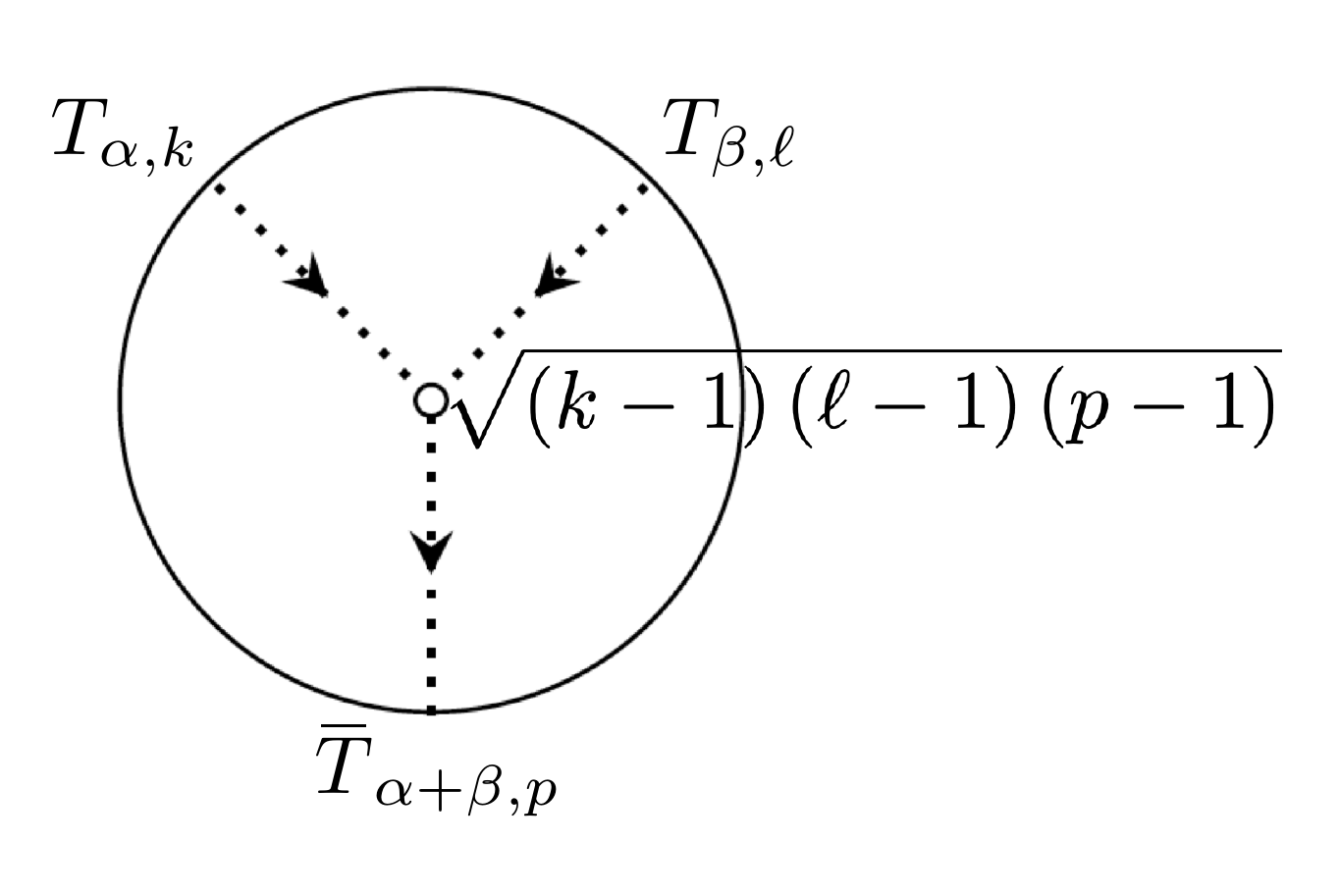}
\caption{A cubic interaction involving three twisted operators whose twist parameters add up to zero modulo $M$. The cubic vertex is $\sqrt{(k-1)\,(\ell-1)\,(p-1)}$ times $\delta_{k+\ell-p,0}$ enforcing charge conservation. Again this $\delta$-function is understood.
\label{fig:6.4}}
    }  
\end{figure}

We finally observe that when $M\geq3$ there is the possibility of having a cubic interaction involving three twisted operators whose twist parameters add up to zero modulo $M$. In this case, following the rules we have introduced, we are naturally led to propose the effective Witten-like diagram represented in Fig.~\ref{fig:6.4}.

This diagram corresponds to the following correlator
\begin{equation}
G_{T_{\alpha,k},T_{\beta,\ell},\overbar{T}_{\alpha+\beta,p}}=M N^2\,\sqrt{(k-1)\,(\ell-1)\,(p-1)}~,
\label{G3TTT}
\end{equation}
and to the structure constants
\begin{equation}
C_{T_{\alpha,k},T_{\beta,\ell},\overbar{T}_{\alpha+\beta,p}}=\frac{1}{\sqrt{M}\,N}\,\sqrt{(k-1)\,(\ell-1)\,(p-1)}
\label{C3TTT}
\end{equation}
which agree with those in (\ref{CTTTloc}) obtained using localization.
Notice that while these structure constants are completely natural, they imply the existence of the 
3-point function $G_{T_{\alpha,k},T_{\beta,\ell},\overbar{T}_{\alpha+\beta,p}}$ which in the holographic normalization should scale as 
$\lambda^{-3/2}$. It would be very interesting to prove this
behavior using the holographic AdS/CFT correspondence. This is particularly challenging since the supergravity action considered 
in Section~\ref{secn:holo} is quadratic in the twisted fields and thus cannot contain a cubic interaction among twisted modes that is needed to produce a 3-point function with all twisted operators. We think that to overcome this problem it is necessary to go beyond the supergravity approximation and consider higher-derivative string corrections which indeed may produce the desired cubic terms in the low-energy effective action of the twisted scalars. Work along this line is in progress.

\vskip 1cm
\noindent {\large {\bf Acknowledgments}}
\vskip 0.2cm
We would like to thank Sujay Ashok, Francesco Galvagno and Igor Pesando for useful discussions.
This research is partially supported by the MUR PRIN contract 2020KR4KN2 ``String Theory as a bridge between Gauge Theories and Quantum Gravity'' and by
the INFN project ST\&FI
``String Theory \& Fundamental Interactions''. The work of A.P. is supported by INFN with a``Borsa di studio post-doctoral per fisici teorici".
\vskip 1cm
\begin{appendix}


\section{Strong-coupling behavior of \texorpdfstring{$\mathsf{d}^{(\alpha)}_k$}{}}
\label{app:ck}

In this Appendix we provide a derivation of the asymptotic form of the coefficients $\mathsf{d}_{k}^{(\alpha)}$ when $\lambda\to\infty$
given in (\ref{call}) of the main text.
We will do this following two methods.

\subsection{First method}
This first method is somehow heuristic and is suggested by the analysis performed 
in \cite{Beccaria:2021vuc}. It is based on the definition of
the coefficients  $\mathsf{d}_{k}^{(\alpha)}$ as a power series in 
$\mathsf{X}$, namely
\begin{align}
\label{cexp}
\mathsf{d}_{k}^{(\alpha)} = \sum_{k\prime}
\Big(\delta_{k, k^\prime}+s_{\alpha}\,\mathsf{X}_{k, k^\prime}+
s_{\alpha}^2\,(\mathsf{X}^2)_{k, k^\prime}
+s_{\alpha}^3\,(\mathsf{X}^3)_{k, k^\prime}+\cdots\Big)\,\sqrt{k^\prime}
~.
\end{align}
Like $\mathsf{X}$, also its higher powers can be written an
integral representation involving products of Bessel functions
\cite{Beccaria:2021hvt,Billo:2021rdb}. For example,
\begin{subequations}
\begin{align}
& (\mathsf{X}^2)_{k,\ell} = +8(-1)^{\frac{k+\ell+2k\,\ell}{2}}
\sqrt{k\,\ell}\int\! 
\mathcal{D}t_1 \,\mathcal{D}t_2 \,
 J_{k}(zt_1)\,\mathcal{G}(zt_1,zt_2)\,J_{\ell}(zt_2)~, \label{X2} \\
& (\mathsf{X}^3)_{k,\ell} = -8(-1)^{\frac{k+\ell+2k\,\ell}{2}}\sqrt{k\,\ell}
\int\! \mathcal{D}t_1 \, \mathcal{D}t_2 \, \mathcal{D}t_3 \, 
J_{k}(zt_1)\,\mathcal{G}(zt_1,zt_2)\,\mathcal{G}(zt_2,zt_3)\,J_{\ell}(zt_3)~, \label{X3}  
\end{align}
\end{subequations}
where
\begin{align}
\mathcal{D}t = \frac{dt}{t}\frac{\textrm{e}^t}{(\textrm{e}^t-1)^2}~,
\qquad z=\frac{\sqrt{\lambda}}{2\pi}~,
\label{Dtz}
\end{align}
and the function $\mathcal{G}(t_1,t_2)$ is
\begin{align}
\mathcal{G}(t_1,t_2) = 8 \times \begin{cases}
& \!\!\!\!\displaystyle{\sum_{n=1}^{\infty}(2n+1)\,J_{2n+1}(t_1)\,J_{2n+1}(t_2) \quad \mbox{if} \ \ k,\ell \ \ \mbox{are~odd} }~,\\
& \!\!\!\!\displaystyle{\sum_{n=1}^{\infty} (2n)\,J_{2n}(t_1)\,
J_{2n}(t_2) \quad \mbox{if} \ \ k,\ell \ \ \mbox{are~even} }~.
\end{cases}
\label{GG} 
\end{align}
Using the properties of the Bessel functions, one can show that the above sums lead to
\begin{align}
\mathcal{G}(t_1,t_2) = \begin{cases}
& \!\!\!\!\displaystyle{\frac{4\,t_1\,t_2}{t_1^2-t_2^2}\Big(t_2 J_1(t_2)\,J_2(t_1)
-t_1 \,J_1(t_1)\,J_2(t_2)\Big)\,\equiv\,\mathcal{G}^{\mathrm{odd}}(t_1,t_2)
\quad \mbox{if} \ \ k,\ell \ \ \mbox{are~odd} }~,\\[4mm]
& \!\!\!\!\displaystyle{\frac{4\,t_1\,t_2}{t_1^2-t_2^2}\Big(t_1\, J_1(t_2)\,J_2(t_1)
-t_2 J_1(t_1)\,J_2(t_2)\Big)\,\equiv\,\mathcal{G}^{\mathrm{even}}(t_1,t_2)\quad \mbox{if} \ \ k,\ell \ \ \mbox{are~even} }~.
\end{cases}
\label{GG1} 
\end{align}

To derive how $\mathsf{d}^{(\alpha)}_k$ behaves when $\lambda\to\infty$, in each power of $\mathsf{X}$ appearing in \eqref{cexp} we keep only the leading term at large $\lambda$, and then perform the sum over $k^\prime$. 
Let us do this for the linear term in $\mathsf{X}$, whose dominant
contribution is given in (\ref{XtoS})--(\ref{Sres}).  When we sum over $k^\prime$, we find
\begin{align}
\sum_{k^\prime} s_{\alpha}\,\mathsf{X}_{k, k^\prime} \,\sqrt{k^\prime} &
\underset{\lambda\to\infty}{\sim} -\frac{s_{\alpha}\,\lambda }{2\pi^2}\,\sum_{k^\prime} 
\mathsf{S}_{k, k^\prime}\,\sqrt{k^\prime}
\underset{\lambda\to\infty}{\sim} -\frac{s_{\alpha}\,\lambda }{4\pi^2} \times \begin{cases}
&\!\!\!\! \frac{1}{\sqrt{2}} \ \ \ \ \, \textrm{for} \ \ \ k=2~,\\
& \!\!\!\! \frac{1}{2\sqrt{3}} \ \ \ \, \textrm{for} \ \ \ k=3~,\\
& \!\!\!\! \ \ 0 \ \ \ \ \ \textrm{for} \ \ \ k \geq 4~.
\end{cases} 
\end{align}
The leading contributions arising from higher powers of $\mathsf{X}$ can be evaluated in a similar way by using
the integral representations (\ref{X2})--(\ref{X3}) (and their generalizations)
and by exploiting the asymptotic expansion of the inverse Mellin transform of the product of two Bessel functions. For example, at the order
$\mathsf{X}^2$ we obtain
\begin{align}
\sum_{k^\prime} s_{\alpha}^2\,(\mathsf{X}^2)_{k, k^\prime} \,\sqrt{k^\prime} 
\underset{\lambda\to\infty}{\sim} \frac{s_{\alpha}^2\,\lambda^2}{4\pi^4} \times \begin{cases}
&\!\!\!\! \frac{1}{6\sqrt{2}} \ \ \ \ \ \ \textrm{for} \ \ \ k=2~,\\
&\!\!\!\! \frac{1}{32\sqrt{3}} \ \ \ \ \ \textrm{for} \ \ \ k=3~,\\
& \!\!\!\!-\frac{1}{48} \ \ \ \ \ \ \textrm{for} \ \ \ k=4~,\\
&\!\!\!\! -\frac{1}{96\sqrt{5}} \ \ \ \textrm{for} \ \ \ k=5~,\\
& \!\!\!\!\ \ \ 0 \ \ \ \ \ \ \ \textrm{for} \ \ \ k \geq 6 ~.
\end{cases}
\end{align}
In general the leading contribution arising from the term
with $\mathsf{X}^q$, when it is non zero, is proportional to $\lambda^q$.
Collecting all the leading contributions, we get a power series in $\lambda$. For example, for $k=2,3,4,5$
we find
\begin{subequations}
\begin{align}
& \mathsf{d}_{2}^{(\alpha)} = \sqrt{2}-\frac{s_{\alpha}\,
\lambda}{4\sqrt{2} \pi
   ^2}+\frac{s_{\alpha }^2\,\lambda ^2 }{24 \sqrt{2} \pi
   ^4}-\frac{11 s_{\alpha }^3\,\lambda ^3 }{1536 \sqrt{2} \pi
   ^6}+\frac{19 s_{\alpha }^4\,\lambda ^4 }{15360 \sqrt{2}
   \pi ^8}+\ldots~, \label{sc2} \\
& \mathsf{d}_{3}^{(\alpha)}= \sqrt{3}-\frac{s_{\alpha }\,\lambda  }{8\sqrt{3} \pi
   ^2}+\frac{s_{\alpha }^2\,\lambda ^2 }{128 \sqrt{3} \pi
   ^4}-\frac{s_{\alpha }^3\,\lambda ^3 }{1920 \sqrt{3} \pi
   ^6}+\frac{13s_{\alpha }^4\, \lambda ^4 }{368640 \sqrt{3}
   \pi ^8}+\ldots~, \label{sc3}\\
   & \mathsf{d}_{4}^{(\alpha)}= 2-\frac{s_{\alpha }^2\,\lambda ^2 }{192 \pi ^4}+\frac{s_{\alpha }^3\,\lambda ^3 }{960 \pi ^6}
   -\frac{17 s_{\alpha }^4\,\lambda ^4 }{92160 \pi ^8}+\ldots ~, \label{sc4}\\
   & \mathsf{d}_{5}^{(\alpha)}=\sqrt{5}-\frac{s_{\alpha
   }^2 \,\lambda ^2 }{384 \sqrt{5} \pi ^4}+\frac{s_{\alpha }^3\,\lambda ^3 }{4608 \sqrt{5} \pi ^6}
   -\frac{s_{\alpha }^4\,\lambda ^4 }{64512 \sqrt{5} \pi ^8}
   +\ldots~.
\end{align}
\label{scstrong}%
\end{subequations}
Remarkably these series can be resummed in terms of a ratio of modified Bessel functions of the first kind $I_k$. Indeed, we have
\begin{align}
 \mathsf{d}_{2m}^{(\alpha)} &=  \sqrt{2m}\,\bigg[1-\frac{I_{2m}\big(
 \frac{\sqrt{ s_{\alpha}\,\lambda}}{\pi }\big)}{I_{0}\big(\frac{\sqrt{ s_{\alpha}\,\lambda}}{\pi }\big)}\bigg]  \label{keven} ~,\\[2mm]
 \mathsf{d}_{2m+1}^{(\alpha)} &=  \sqrt{2m+1}\,\bigg[1-\frac{I_{2m+1}\big(\frac{\sqrt{ s_{\alpha}\,\lambda}}{\pi }\big)}{I_{1}\big(\frac{\sqrt{ s_{\alpha}\,\lambda}}{\pi }\big)}\bigg]  ~.\label{kodd}
\end{align}
We have checked this result for several values of $m$. In this way, exploiting
the asymptotic properties of the modified Bessel functions $I_k$, we find
that
\begin{align}
\mathsf{d}_{2m}^{(\alpha)}  \underset{\lambda\to \infty}{\sim}
\frac{\pi\,\sqrt{2m}}{\sqrt{s_{\alpha}\,\lambda }} \,(2m^2)~,\qquad
\mathsf{d}_{2m+1}^{(\alpha)}  &\underset{\lambda\to\infty}{\sim}
\frac{\pi\,\sqrt{2m+1}}{\sqrt{s_{\alpha}\,\lambda }}\,(2m^2+2m)
~.\label{devendodd}
\end{align}
We can combine these two expressions and write
\begin{align}
	\label{callstrong}
		\mathsf{d}^{(\alpha)}_k \underset{\lambda\to\infty}{\sim}
		\frac{\pi}{\sqrt{s_\alpha\,\lambda}} \,\Big[
		\frac{\sqrt{k}}{2} \big(k^2 - \delta_{k\!\!\!\!\!\mod 2,1}\big)\Big]
\end{align}
for all $k$, which is Eq.~(\ref{call}) of the main text.

We observe that this same method can be used also to extract the leading asymptotic term of the
propagator $\mathsf{D}^{(\alpha)}_{k,\ell}$ when $\lambda\to\infty$. For example, for $k=\ell=2$
we have
\begin{align}
\mathsf{D}^{(\alpha)}_{2,2}&=1+s_{\alpha}\,\mathsf{X}_{2,2}+
s_{\alpha}^2\,(\mathsf{X}^2)_{2,2}
+s_{\alpha}^3\,(\mathsf{X}^3)_{2,2}+s_{\alpha}^4\,(\mathsf{X}^4)_{2,2}+\cdots\notag\\
&=1-\frac{s_{\alpha}\,
\lambda}{6 \pi
   ^2}+\frac{11s_{\alpha }^2\,\lambda ^2 }{384 \pi
   ^4}-\frac{19 s_{\alpha }^3\,\lambda ^3 }{3840 \pi
   ^6}+\frac{473 s_{\alpha }^4\,\lambda ^4 }{552960
   \pi ^8}+\ldots
\end{align}
where the second line follows from retaining in each power of $\mathsf{X}$ only the leading term in
$\lambda$.
As before, this series can be resummed in terms of modified Bessel functions. In this way we find
\begin{align}
\mathsf{D}^{(\alpha)}_{2,2}=\frac{8\pi^2}{s_\alpha\,\lambda}\,
\frac{I_{2}\big(
 \frac{\sqrt{ s_{\alpha}\,\lambda}}{\pi }\big)}{I_{0}\big(\frac{\sqrt{ s_{\alpha}\,\lambda}}{\pi }\big)}\ 
 \underset{\lambda\to \infty}{\sim} \ \frac{8\pi^2}{s_\alpha\,\lambda}
 \label{D22app}
\end{align}
in full agreement with (\ref{Dllres}). In the same way one can check all other cases and reconstruct the matrix (\ref{Dllrese}) which was firstly obtained in \cite{Billo:2021rdb} with a different more rigorous technique.

\subsection{Second method}

We now provide an alternative method to derive the strong-coupling behavior of $\mathsf{d}^{(\alpha)}_k$ which is based on the following identities
\begin{align}
 \mathsf{d}_{2}^{(\alpha)}\,\mathsf{d}_{2m}^{(\alpha)} &=
\sqrt{2m}\,\sum_{r=1}^{m-1}\sqrt{2r}\ \mathsf{D}_{2,2r}^{(\alpha)}+
\big(m+1+\lambda\partial_{\lambda}\big)\mathsf{D}_{2,2m}^{(\alpha)}~, \label{eveneven} \\
 \mathsf{d}_{3}^{(\alpha)}\,\mathsf{d}_{2m+1}^{(\alpha)}
 &=\sqrt{2m+1}\,\sum_{r=1}^{m-1}\sqrt{2r+1}\ \mathsf{D}_{3,2r+1}^{(\alpha)}+\big(m+2+\lambda\partial_{\lambda}\big)\mathsf{D}_{3,2m+1}^{(\alpha)}~, \label{oddodd}\\
  \mathsf{d}_{2n}^{(\alpha)}\,\mathsf{d}_{2m+1}^{(\alpha)}
  &= \sqrt{2m+1}\,\sum_{r=1}^{m}\sqrt{2r}\ \mathsf{D}_{2n,2r}^{(\alpha)}+\sqrt{2n}\sum_{s=1}^{n-1}\sqrt{2s+1}\ \mathsf{D}_{2m+1,2s+1}^{(\alpha)}~. \label{oddeven} 
\end{align}
These relations can be easily checked perturbatively by expanding both sides for small values of $\lambda$, and we have done this
in a variety of cases up to order $\lambda^{45}$. In the following we present a proof of the third identity based on a different strategy.

\subsubsection{Proof of \eqref{oddeven}}

Using the definitions of $\mathsf{D}^{(\alpha)}$ and
$\mathsf{d}^{(\alpha)}$ in terms of $\mathsf{X}$, we expand
both sides of (\ref{oddeven}) in powers of $\mathsf{X}$ and show that
the identity is satisfied at any order. 

\subsubsection*{Linear order}

At the first order, the identity (\ref{oddeven}) reduces to
\begin{align}
    \sqrt{2m+1}\sum_{r=m+1}^{\infty}\sqrt{2r}\ \mathsf{X}_{2n,2r}+\sqrt{2n}\sum_{s=n}^{\infty}\sqrt{2s+1}\ \mathsf{X}_{2m+1,2s+1}=0 ~.
    \label{id1X}
\end{align}
Using the integral representation of $\mathsf{X}$ given
in (\ref{Xkl}), we see that the left-hand side of (\ref{id1X}) is proportional to
\begin{align}
     \sum_{r=m+1}^{\infty}\!\!(-1)^{r+n}\,(2r)\!
     \int\!\mathcal{D}t\,J_{2n}(zt)\,J_{2r}(zt) +\sum_{s=n}^{\infty} (-1)^{s+m}\,(2s+1)\!\int\!\mathcal{D}t\,J_{2m+1}(zt)\,J_{2s+1}(zt)
     \label{id2X}
\end{align}
where $\mathcal{D}t$ and $z$ are defined in (\ref{Dtz}).
We now exploit the following identities
\begin{subequations}
\begin{align}
J_{2n}(zt)&=\frac{2}{zt}\,\sum_{s=n}^\infty (-1)^{n+s}\,(2s+1)\,J_{2s+1}(zt)~,\label{J2n}\\
J_{2m+1}(zt)&=-\frac{2}{zt}\,\sum_{r=m+1}^\infty (-1)^{m+r}\,(2r)\,J_{2r}(zt)~,\label{J2m+1}
\end{align}
\label{rec}%
\end{subequations}
which can be proven recursively from the recurrence relation
\begin{align}
    J_{q-1}(x)=\frac{2q}{x}J_{q}(x)-J_{q+1}(x)
\end{align}
satisfied by the Bessel functions.
Then, inserting (\ref{rec}) into (\ref{id2X}), we find zero, so that the relation
(\ref{id1X}) is satisfied.

\subsubsection*{Quadratic order}

At the quadratic order, the identity \eqref{oddeven} yields the following relation
\begin{align}
 &\sqrt{2m+1}\!\!\sum_{r=m+1}^{\infty}\!\!\sqrt{2r}\ (\mathsf{X}^{2})_{2n,2r}
 +\sum_{r,s=1}^{\infty}\!\!\sqrt{(2r)(2s+1)}\ \mathsf{X}_{2n,2r}\,\mathsf{X}_{2m+1,2s+1}
 \notag\\ &\hspace{2cm}
 +\sqrt{2n}\,\sum_{s=n}^{\infty}\sqrt{2s+1}\ (\mathsf{X}^{2})_{2m+1,2s+1} 
 =0 ~.
\label{quadratic}
\end{align}
Using the integral representation for $\mathsf{X}$ and $\mathsf{X}^2$, one can show that the left-hand side of (\ref{quadratic}) is proportional to
\begin{align}
&\sum_{r=m+1}^{\infty}\!\!(-1)^{r+n}\,(2r)\!
     \int\!\mathcal{D}t_1\mathcal{D}t_2
     \,J_{2n}(zt_1)\,\mathcal{G}^{\mathrm{even}}(zt_{1},zt_{2})\,J_{2r}(zt_2) \notag \\
     &\ +\sum_{s=n}^{\infty} (-1)^{s+m}\,(2s+1)\!\int\!\mathcal{D}t_1\mathcal{D}t_2\,
     \,J_{2m+1}(zt_2)\,\mathcal{G}^{\mathrm{odd}}(zt_1,zt_2)\,J_{2s+1}(zt_1)\label{intquadratic}\\
     &\ +8\sum_{r,s=1}^{\infty}(-1)^{r+n+s+m}\ (2r)(2s+1)\int
     \mathcal{D}t_{1}\mathcal{D}t_{2}\,J_{2n}(zt_{1})\,J_{2m+1}(zt_{2})\,J_{2r}(zt_{1})\,J_{2s+1}(zt_{2})~,
     \notag
\end{align}
where $\mathcal{G}^{\mathrm{even}}$ and $\mathcal{G}^{\mathrm{odd}}$
are defined in (\ref{GG1}). Exploiting the relations \eqref{rec}, after some algebra we are left with
a term proportional to
\begin{align}
      \int\!\mathcal{D}t_{1}\mathcal{D}t_{2}\,
      J_{2n}(zt_{1})J_{2m+1}(zt_{2})\Big[t_{2}\,\mathcal{G}^{\mathrm{even}}(zt_{1},zt_{2}) 
      -t_{1}\,\mathcal{G}^{\mathrm{odd}}(zt_{1},zt_{2})-4z t_{1} t_{2}\,J_{1}(zt_{1})\,
      J_{2}(zt_{2})\Big]
      \label{GGY}
\end{align}
which vanishes as one can see by using the explicit 
expressions of $\mathcal{G}^{\mathrm{even}}$ and $\mathcal{G}^{\mathrm{odd}}$.
Thus, the identity \eqref{oddeven} is proven also to the quadratic order.

\subsubsection*{Higher orders}
The above calculations can be easily generalized to higher orders. At order $\mathsf{X}^p$
the identity (\ref{oddeven}) implies the following relation
\begin{align}
     \sqrt{2m+1}\!\!\sum_{r=m+1}^{\infty}\!\!&\sqrt{2r}\, (\mathsf{X}^{p})_{2n,2r}+\!
    \sum_{r,s=1}^{\infty}\!\sqrt{(2r)(2s+1)} \Big[(\mathsf{X}^{p-1})_{2n,2r}\,\mathsf{X}_{2m+1,2s+1}
    \label{pthorder}\\
    & \hspace{0.5cm}+\ldots
    +\mathsf{X}_{2n,2r}\,(\mathsf{X}^{p-1})_{2m+1,2s+1}\Big]+\sqrt{2n}\sum_{s=n}^{\infty}\sqrt{2s+1}\ (\mathsf{X}^{p})_{2m+1,2s+1} =0~.
\notag
\end{align}
Proceeding as before, one can show that the left-hand side is proportional to
\begin{align}
     &\int\!\!\mathcal{D}t_{1}\ldots\mathcal{D}t_{n}\, J_{2n}(zt_{1})J_{2m+1}(zt_{p})
     \,\Big[ \,\mathcal{G}^{\mathrm{even}}(zt_{1},zt_{2})\ldots
     \mathcal{G}^{\mathrm{even}}(zt_{p-2},zt_{p-1})\,Y(zt_{p-1},zt_{p})
     \notag\\
    &\hspace{1cm}+\mathcal{G}^{\mathrm{even}}(zt_{1},zt_{2})\ldots 
    \mathcal{G}^{\mathrm{even}}(zt_{p-3},zt_{p-2}) \,Y(zt_{p-2},zt_{p-1})\,
    \mathcal{G}^{\mathrm{odd}}(zt_{p-1},zt_{p})+\dots+ \label{pth1}\\
    & \hspace{2cm}
    +Y(zt_{1},zt_{2})\,\mathcal{G}^{\mathrm{odd}}(zt_{2},zt_{3})
    \ldots\mathcal{G}^{\mathrm{odd}}(zt_{p-1},zt_{p})\Big]\notag
\end{align}
where $Y$ is the function inside the square brackets of (\ref{GGY}) which vanishes. Since each term
in (\ref{pth1}) contains this vanishing combination, we have shown that (\ref{pthorder})
is true for a generic $p$. This concludes the proof of the identity (\ref{oddeven}).

\subsubsection{$\mathsf{d}^{(\alpha)}_{k}$ for $\lambda\rightarrow\infty$}
The identity (\ref{eveneven}) for $m=1$ reads
\begin{align}
    \big(\mathsf{d}_{2}^{(\alpha)}\big)^{2}
    =\big(2+\lambda\partial_{\lambda}\big)\mathsf{D}_{2,2}^{(\alpha)} ~.
    \label{d2d2}
\end{align}
From the strong-coupling behavior of $\mathsf{D}_{2,2}^{(\alpha)}$ (see (\ref{D22app})), we easily
deduce that
\begin{align}
    \mathsf{d}_{2}^{(\alpha)}
    \underset{\lambda\to \infty}{\sim} \ \frac{2\sqrt{2}\pi}{\sqrt{s_\alpha\,\lambda}}
\label{d2app}
\end{align}
which agrees with what we have found with the first method (see (\ref{callstrong}) for $k=2$).
Then, exploiting the identity (\ref{oddeven}) for $n=1$ and the previous result, we deduce that
\begin{align}
    \mathsf{d}^{(\alpha)}_{2m+1}\underset{\lambda\to \infty}{\sim} \ 
    \frac{\pi\sqrt{2m+1}\,(2m^{2}+2m)}{\sqrt{s_{\alpha}\,\lambda }}
    \label{doddapp}
\end{align}
in agreement with (\ref{devendodd}). Finally, using this behavior in the identity
(\ref{oddeven}) we obtain 
\begin{align}
    \mathsf{d}^{(\alpha)}_{2m}=\frac{\pi\sqrt{2m}(2m^{2})}{\sqrt{s_{\alpha}\,\lambda }}
    \label{devenapp}
\end{align}
which coincides with (\ref{devendodd}). These results show that the two methods we have presented
lead to the same strong-coupling behavior for $\mathsf{d}^{(\alpha)}_{k}$.


\section{Results in the basis associated to the quiver nodes}
\label{app:node_basis}

In the sections of the main text we have computed the structure constants for the single-trace scalar operators of the quiver theory utilizing the untwisted and twisted operators defined in (\ref{operators}). In this Appendix we rephrase these results in the basis of the operators $\tr\phi_I^k(x)$ associated to the nodes of the quiver.  

\subsubsection*{Free theory} 
When $\lambda=0$, the matrices $a_I$ are decoupled 
and one can work out the large-$N$ behavior of the correlators of the operators $\tr a_I^k$ exploiting the recursion relations independently in each factor. To diagonalize the 2-point functions, we introduce operators $\mathcal{R}_{I,k}$ (with $I=0,\ldots,M-1$ and $k\geq 2$) which are expressed as Chebyshev polynomials in terms of the traces $\tr a_I^k$ and obey 
\begin{align}
	\label{cR2}
		\big\langle 
		\mathcal{R}_{I,k}\,\mathcal{R}_{J,\ell}\big\rangle_0 = \delta_{k,\ell}\delta_{I,J}~.
\end{align}
Their 3-point functions are given by 
\begin{align}
	\label{cR3}
		\big\langle
		\mathcal{R}_{I,k}\, \mathcal{R}_{J,\ell}\,  \mathcal{R}_{K,p} 
		\big\rangle_0 = \sqrt{M}\, C_{k,\ell,p}\,\delta_{I,J}\, \delta_{J,K}
\end{align}
and thus vanish unless $I=J=K$. 
The set of operators $\boldsymbol{\mathcal{R}}_I$  are the analogue of the operators $\boldsymbol{\mathcal{P}}_{\widehat\alpha}$ in the twisted basis
defined in (\ref{calP}), and are related to them by the (inverse) discrete Fourier transform on $\mathbb{Z}_M$:
\begin{align}
	\label{cRidft}
		\boldsymbol{\mathcal{R}}_I = 
		\frac{1}{\sqrt{M}}\sum_{\widehat\alpha=0}^{M-1} \rho^{I\widehat\alpha}\,
		\boldsymbol{\mathcal{P}}_{\widehat{\alpha}}~. 
\end{align}	
Substituting this relation into (\ref{cPcP}) and (\ref{cPcPcP}),
one retrieves\,%
\footnote{Recall that $\rho = \exp(2\pi\ii/M)$ and therefore
	\begin{align}
		\label{deltaft}
			\frac 1M \sum_{\widehat\alpha=0}^{M-1} \rho^{\widehat\alpha(I-J)} = \delta_{I,J}~.
	\end{align}
}, consistently, the 2- and 3-point functions (\ref{cR2}) and (\ref{cR3}). One can also introduce the rescaled operators 
\begin{align}
	\label{Rdef}
		R_{I,k}(0) = \sqrt{\mathcal{G}_k} \,\mathcal{R}_{I,k}
\end{align}	
which are the matrix-model representatives of the chiral scalar operators $\tr\phi_I^n(x)$ of the gauge theory at $\lambda=0$. They satisfy
\begin{align}
	\label{R2}
		\big\langle R_{I,k}(0)\,R_{J,\ell}(0)\big\rangle_0 = \mathcal{G}_k\,\delta_{k,\ell}\,\delta_{I,J}
\end{align}
and
\begin{align}
	\label{R3}
		\big\langle R_{I,k}(0)\, R_{J,\ell}(0)\,  R_{K,p}(0) \big\rangle_0 
		= \sqrt{M}\,\mathcal{G}_{k,\ell,p}\,\delta_{I,J} \,\delta_{J,K}~.
\end{align}
Even if the spectrum of the gauge theory is described by the quiver of Fig.~\ref{fig:1_quiver}, at zero coupling the correlators of the gauge-invariant operators $\tr\phi_I^k(x)$ in each node are not influenced by the presence of the hypermultiplets corresponding to the links of the quiver, which should run inside loops, and are therefore diagonal in the space of the indices $I$. Within each node, at $\lambda=0$ the results coincide with those of the $\mathcal{N}=4$ SU$(N)$ theory, a part from the $\sqrt{M}$ factor in (\ref{R3}).

\subsubsection*{Interacting theory}
At zero coupling, both the operators $\boldsymbol{\mathcal{P}}_{\widehat\alpha}$ and the operators $\boldsymbol{\mathcal{R}}_I$ we just introduced have canonical 2-point functions. In the interacting case, 
the operators $\boldsymbol{\mathcal{P}}_{\widehat\alpha}$ remain diagonal in the space of the indices $\widehat\alpha$. The 2-point functions of the $\boldsymbol{\mathcal{R}}_I$ operators, instead, are no longer diagonal in the space of the indices $I$. In fact, using (\ref{cRidft}) and then (\ref{intprop}), one has
\begin{align}
	\label{intRprop}
		\big\langle \mathcal{R}_{I,k}\, \mathcal{R}_{J,\ell}\big\rangle 
		= \frac{1}{M}\sum_{\widehat\alpha,\widehat\beta=0}^{M-1}
		\rho^{\widehat\alpha I + \widehat\beta J}\,
		\big\langle \mathcal{P}_{\widehat\alpha,k}\, \mathcal{P}_{\widehat\beta,\ell}\big\rangle
		= \frac{1}{M}\sum_{\widehat\alpha=0}^{M-1}\rho^{\widehat\alpha (I - J)} \,
		\mathsf{D}^{(\widehat\alpha)}_{k,\ell}~, 
\end{align}
which is not proportional to $\delta_{I,J}$ because $\mathsf{D}^{(\alpha)}_{k,\ell}$ depends 
on $\widehat\alpha$ and we cannot use (\ref{deltaft}).

The normal-ordered operators $R_{I,k}(\lambda)$, which correspond for generic $\lambda$ to the operators $\tr\phi_I^k(x)$, are obtained via inverse discrete Fourier transform from the operators $P_{\widehat\alpha,k}(\lambda)$ defined in (\ref{PktocPk}):
\begin{align}
	\label{RIl}
		R_{I,k}(\lambda) & = \frac{1}{\sqrt{M}} \sum_{\widehat\alpha=0}^{M-1} \rho^{I \widehat\alpha}\, P_{\widehat\alpha,k}(\lambda)
		= \frac{\sqrt{\mathcal{G}_k}}{\sqrt{M}} \,\sum_{\widehat\alpha=0}^{M-1} \rho^{I \widehat\alpha} 
		\Big(\mathcal{P}_{\widehat\alpha,k} - \sum_{\ell<k} \mathsf{Q}^{(\widehat\alpha)}_{k,\ell}(\lambda) \mathcal{P}_{\widehat\alpha,\ell}\Big)\notag\\[1mm]
		& = \sqrt{\mathcal{G}_k} \Big(\mathcal{R}_{I,k} - \sum_{J=0}^{M-1}\sum_{\ell<k} \mathsf{U}^{(I,J)}_{k,\ell}(\lambda)\, \mathcal{R}_{J,\ell} \Big)~,		
\end{align}
where the coefficients 
\begin{align}
	\label{Ucoff}
		\mathsf{U}^{(I,J)}_{k,\ell}(\lambda)= \frac 1M \sum_{\widehat{\alpha}=0}^{M-1} \rho^{\widehat\alpha(I-J)}
		\mathsf{Q}^{(\widehat\alpha)}_{k,\ell}(\lambda) 
\end{align}		
are not diagonal in $I,J$. Thus, for a generic value of $\lambda$ 
the matrix-model operators that represent the gauge theory operators 
$\tr\phi_I^k(x)$ are linear combinations constructed with the matrices at different nodes. Moreover, their 2-point functions are not diagonal and have a complicated, $\lambda$-dependent expression. These features of the normal ordering in the matrix model and of the 2-point functions were studied at the perturbative level in \cite{Galvagno:2020cgq}, where in fact it was shown that at 1-loop the operators in a node of the quiver start mixing with those of the adjacent nodes and that by increasing the loop order they mix with farther nodes.

\subsubsection*{Strong coupling behavior}
When $\lambda\to\infty$ the map (\ref{RIl}) simplifies because the expression of the normal-ordered operators in the twisted basis reduces to that in (\ref{inftyno}). Therefore, we get
\begin{align}
	\label{RIlinf}
		R_{I,k}(\infty) & = \frac{1}{\sqrt{M}} \sum_{\widehat\alpha=0}^{M-1} \rho^{I \widehat\alpha}\, P_{\widehat\alpha,k}(\infty)
		= \sqrt{\mathcal{G}_k}\, \frac{1}{\sqrt{M}} \sum_{\widehat\alpha} \rho^{I \widehat\alpha} 
		\Big(\mathcal{P}_{\alpha,k} - \sqrt{\frac{k}{k-2}}\,\mathcal{P}_{\alpha,k-2}\Big)
		\notag\\[1mm]
		& = \sqrt{\mathcal{G}_k} \Big(\mathcal{R}_{I,k} - \,\sqrt{\frac{k}{k-2}}\, \mathcal{R}_{I,\ell} 
		\Big)~,		
\end{align}
where in the last step we used (\ref{cRidft}). Thus, at strong coupling, just as it happened in the free theory, the normal-ordered procedure is carried out independently in each node. To study the 2-point correlators it is convenient to separate the untwisted ($\widehat{\alpha}=0$) and twisted ($\widehat\alpha = \alpha\not= 0$) component of these operators by writing
\begin{align}
	\label{RIhat}
		R_{I,k}(\infty) = \frac{1}{\sqrt{M}} P_{0,k}(\infty) + \widehat R_{I,k}(\infty)~.
\end{align}
The operators  
\begin{align}
	\label{defRIhat}
		\widehat R_{I,k}(\infty)
		=  \frac{1}{\sqrt{M}} \sum_{\alpha=1}^{ M-1} \rho^{I \alpha}\, P_{\alpha,k}(\infty)~,
\end{align}		
which are the twisted operators in the node basis, are not independent since
they satisfy the relation
\begin{align}
	\label{sRh0}
		\sum_{I=0}^{M-1} \widehat R_{I,k}(\infty) = 0~.		
\end{align}
Their 2-point functions are given by 		
\begin{align}
	\label{intRpropstrong}
		\big\langle
		\widehat R_{I,k}(\infty)\, \widehat R_{J,\ell}(\infty)\big\rangle & = \frac{1}{M}\sum_{\alpha,\beta=1}^{M-1}
		\rho^{\alpha I + \beta J}
		\vvev{P_{\alpha,k}(\infty)\, P_{\beta,\ell}(\infty)} 
		= \frac{4\pi^2}{\lambda}k(k-1)\delta_{k,l}\, g_{I,J}~,
\end{align}
where in the second step we used (\ref{PPin}) and introduced the $M\times M$ matrix
\begin{align}	
	\label{defgIJ}
		g_ {I,J}
		 = \frac{1}{M}\sum_{\alpha=1}^{M-1}\frac{\rho^{\alpha (I - J)}}{s_\alpha}~.
\end{align}
This matrix arises from the strong-coupling behavior of the quiver gauge theory but has a geometrical meaning related to the resolution of the orbifold space $\mathbb{C}^2/\mathbb{Z}_M$ involved in the string-theory embedding of the $\mathcal{N}=2$ quiver gauge, as discussed at the beginning of Section~\ref{secn:holo}. Recall that the exceptional 2-cycles $e_i$, with $i=1,\ldots M-1$, of the resolved space have an intersection matrix $C_{i,j}$ which is the Cartan matrix of the Lie algebra $\mathfrak{su}_{M-1}$. Introducing the extra cycle $e_0 = -\sum_i e_i$, the intersection matrix of the $e_I =(e_0,e_i)$ cycles is the affine Cartan matrix $\widehat C_{I,J}$ whose Dynkin diagram is exactly a circular quiver with $M$ nodes.  The quantities $4 s_\alpha = 4\sin^2 \frac{\pi\alpha}{M}$
are exactly the $M-1$ non-zero eigenvalues of $\widehat C_{I,J}$, and one has \cite{Billo:2000yb}
\begin{align}
	\label{Ctos}
		 \frac 1M \sum_{\alpha=1}^{M-1} \rho^{\alpha(I-J)}\, s_\alpha = \frac 14 \widehat C_{I,J}~.
\end{align}		
In the expression of the matrix $g_{I,J}$ the reciprocal of the eigenvalues $s_\alpha$ appear. However, the extended Cartan matrix $\widehat C_{I,J}$ has a zero eigenvalue, so $g_{I,J}$ cannot directly be proportional to its (ill-defined) inverse. Rather, we have to restrict ourselves to a set of $M-1$ independent operators among the $\boldsymbol{\widehat R}_{J}(\infty))$ ones, for instance setting $\boldsymbol{\widehat R}_{0}(\infty)= - \sum_{i=1}^{M-1}\boldsymbol{\widehat R}_i(\infty)$ and focusing on the $\boldsymbol{\widehat R}_{i}(\infty)$ ones. Then, the 
symmetric matrix $g_{I,J}$ represents a scalar product and defines a bilinear form according to
\begin{align}
	\label{bilg}
		\sum_{I,J=0}^{M-1} g_{I,J}\, \boldsymbol{\widehat R}_I(\infty) \boldsymbol{\widehat R}_J(\infty) 
		& = \sum_{i,j=1}^{M-1} 
		\left(g_{i,j} - g_{i,0} - g_{0,j} + g_{0,0}\right) \boldsymbol{\widehat R}_i(\infty) \boldsymbol{\widehat R}_j(\infty)\notag\\
		& = 4 \sum_{i,j=1}^{M-1} 
		\left(C^{-1}\right)^{i,j}\, \boldsymbol{\widehat R}_i(\infty) \boldsymbol{\widehat R}_j(\infty)~.
\end{align}	
This argument shows that the geometry of the resolution of the orbifold
$\mathbb{C}^2/\mathbb{Z}_M$, which is the crucial ingredient of the holographic description of the theory, emerges directly from the large-$N$, strong-coupling behavior of the correlators in the quiver gauge theory
as obtained from localization.

\section{The scalar spherical harmonics on \texorpdfstring{$S^5$}{} and \texorpdfstring{$S^5/\mathbb{Z}_M$}{} }
\label{app:harmonics}

\subsection{\texorpdfstring{$S^5$}{}}
A 5-sphere of unit radius $S^5$ can be defined through its embedding in $\mathbb{R}^6$ given by
\begin{equation}
y_1^2+y_2^2+y_3^2+y_4^2+y_5^2+y_6^2=1~.
\label{5sphere}
\end{equation}
This equation, which is clearly invariant under SO(6) rotations, can be conveniently parametrized by
\begin{equation}
\begin{aligned}
y_1&=\cos\phi \,\cos \theta~,\,\,\,\,~~\quad\quad\,
y_2=\cos\phi \,\sin \theta~,\\
y_3&=\sin\phi \,\cos\phi^\prime \,\cos\theta^\prime~,~~
y_4=\sin\phi\,\cos\phi^\prime \,\sin\theta^\prime~,\\
y_5&=\sin\phi\, \sin\phi^\prime \,\cos\theta^{\prime\prime}~,~~\,
y_6=\sin\phi \,\sin\phi^\prime\, \sin\theta^{\prime\prime}~,
\end{aligned}
\label{x123456}
\end{equation}
where $\phi,\phi^\prime\in[0,{\pi}/{2}]$ and $\theta,\theta^\prime,\theta^{\prime\prime}\in[0,2\pi]$. Using these
angular coordinates, the metric of $S^5$ becomes
\begin{equation}
ds^2_{S^5}=d\phi^2+\sin^2\phi\,d\phi^{\prime\,2}+
\cos^2\phi\,d\theta^2+\sin^2\phi\,\cos^2\phi^\prime\,d\theta^{\prime\,2}
+\sin^2\phi\,\sin^2\phi^\prime\,d\theta^{\prime\prime\,2}
\label{S5metric}
\end{equation}
and its volume is
\begin{equation}
\mathrm{vol}(S^5)=\int_{0}^{\frac{\pi}{2}} \!\!d\phi \int_{0}^{\frac{\pi}{2}}\!\!d\phi^\prime
\int_{0}^{2\pi}\!\!d\theta\int_{0}^{2\pi}\!\!d\theta^\prime\int_{0}^{2\pi}\!\!d\theta^{\prime\prime}\,
\big(\sin^3\phi\,\cos\phi\,\sin\phi^\prime\,\cos\phi^\prime\big)=\pi^3
\label{volS5}
\end{equation}

Let us now consider the totally symmetric traceless rank $n$ tensors of SO(6) which we denote by $\mathcal{C}^{\mathcal{A}}_{i_1\ldots i_n}$. Here the integer index $\mathcal{A}$ labels the different such tensors and its range is
\begin{equation}
1\leq \mathcal{A}\leq \frac{(n+1)(n+2)^2(n+3)}{12}~.
\label{range}
\end{equation}
Thus, there are 6 different tensors of rank $1$, 20 different tensors of rank $2$,
50 different tensors of rank $3$ and so on. 
We normalize all these tensors in such a way that
\begin{equation}
\mathcal{C}^{\mathcal{A}}_{i_1\ldots i_n}\,\mathcal{C}^{\mathcal{B}\,,\,i_1\ldots i_n}=\delta^{\mathcal{A}\mathcal{B}}~.
\label{norm}
\end{equation}
The combination
\begin{equation}
\mathcal{Y}^\mathcal{A}=\mathcal{C}^\mathcal{A}_{i_1\ldots i_n}\,y^{i_1}\ldots\,y^{i_n}~,
\end{equation}
which through the map (\ref{x123456}) is a function of the five angular coordinates
of $S^5$, is a scalar spherical harmonic of rank $n$.
One can check that these harmonics satisfy the following relations
\begin{equation}
\int_{S^5} \mathcal{Y}^{\mathcal{A}_1}\,\mathcal{Y}^{\mathcal{A}_2}=\frac{\pi^3}{2^{n-1}(n+1)(n+2)}\,\delta^{\mathcal{A}_1\mathcal{A}_2}
\label{YY}
\end{equation}
where $n$ is the rank of the two harmonics, and
\begin{equation}
\int_{S^5} \mathcal{Y}^{\mathcal{A}_1}\,\mathcal{Y}^{\mathcal{A}_2}\,\mathcal{Y}^{\mathcal{A}_3}=\frac{n_1!\,n_2!\,n_3!\,\pi^3}{2^{\frac{n_1+n_2+n_3}{2}-1}
\,(\frac{n_1+n_2+n_3+4}{2})!\,
(\frac{n_1+n_2-n_3}{2})! \,(\frac{n_2+n_3-n_1}{2})!\,(\frac{n_3+n_1-n_2}{2})!}\,\big\langle
\mathcal{C}^{\mathcal{A}_1}\mathcal{C}^{\mathcal{A}_2}\mathcal{C}^{\mathcal{A}_3}\big\rangle
\label{YYY}
\end{equation}
where $n_1$, $n_2$ and $n_3$ are the ranks of three harmonics and 
$\big\langle \mathcal{C}^{\mathcal{A}_1}\mathcal{C}^{\mathcal{A}_2}\mathcal{C}^{\mathcal{A}_3}\big\rangle$ is
the unique SO(6) invariant that can be formed out of the three tensors $\mathcal{C}^{\mathcal{A}_1}$,
$\mathcal{C}^{\mathcal{A}_2}$ and $\mathcal{C}^{\mathcal{A}_3}$.

For any integer $n\geq0$, we consider in particular the two following harmonics of rank $n$:
\begin{equation}
Y^{\pm n}=\frac{1}{2^{\frac{n}{2}}}\,(y_1\pm\ii\,y_2)^n=\frac{1}{2^{\frac{n}{2}}}\,\cos^n\!\phi\,\mathrm{e}^{\pm\,\ii\,n\,\theta}~,
\label{Y+-n}
\end{equation}
which satisfy the following relations
\begin{align}
\int_{S^5} Y^n\,Y^{-m}&=\frac{\pi^3}{2^{n-1}(n+1)(n+2)}\,\delta_{n,m}~,\label{YYbis}\\[2mm]
\int_{S^5} Y^n\,Y^{m}\,Y^{-p}&=\int_{S^5} Y^{-n}\,Y^{-m}\,Y^{p}=\frac{\pi^3}{2^{\frac{n+m+p}{2}-1}(\frac{n+m+p}{2}+1)(\frac{n+m+p}{2}+2)}\,\delta_{n+m,p}~.\label{YYYbis}
\end{align}
The fact that the overlap coefficient in (\ref{YYbis}) is the same as the one in (\ref{YY}) shows that
$Y^{\pm n}$ are normalized as the general harmonics $\mathcal{Y}^\mathcal{A}$. The reason to consider these particular harmonics is that they are the ones that remain non-trivial at the fixed point of a
$\mathbb{Z}_M$ orbifold, as we are going to see in the next subsection.

\subsection{\texorpdfstring{$S^5/\mathbb{Z}_M$}{}}
\label{subapp:ZM}
Let us consider a $\mathbb{Z}_M$ orbifold in $\mathbb{R}^6$ corresponding to a rotation
of an angle $(2\pi/M)$ in the $(y^3,y^4)$-plane accompanied by a rotation of an angle 
$(-2\pi/M)$ in the $(y^5,y^6)$-plane (see also (\ref{orbifold})). 
In the parametrization (\ref{x123456}), this action simply corresponds to
\begin{equation}
\theta^\prime~\to~\theta^\prime+\frac{2\pi}{M}~,\qquad
\theta^{\prime\prime}~\to~\theta^{\prime\prime}-\frac{2\pi}{M}~,
\end{equation}
with all other angles unchanged. This action in $\mathbb{R}^6$ is inherited by $S^5$ and 
leads to the orbifold $S^5/\mathbb{Z}_M$. The invariant locus under the orbifold is the sub-space
$y_3=y_4=y_5=y_6=0$ corresponding to $\phi=0$. This locus is the $(y_1,y_2)$-plane which in
$S^5$ describes a circle $S^1$ parametrized by $\theta$.

The orbifold space $S^5/\mathbb{Z}_M$ has the same metric (\ref{S5metric}) of $S^5$, but with a modified range for the angles. A possible choice is to reduce the range of $\theta^\prime$ 
to $[0,2\pi/M]$, instead of $[0,2\pi]$ as it was in $S^5$. 
In this way one gets
\begin{equation}
\mathrm{vol}(S^5/\mathbb{Z}_M)=\frac{\pi^3}{M}~.
\label{volS5M}
\end{equation}
Then, in $S^5/\mathbb{Z}_M$
it is easy to see that the harmonics $Y^{\pm n}$ in (\ref{Y+-n}) satisfy the relations
(\ref{YYbis}) and (\ref{YYYbis}) with $\pi^3$ replaced by $\pi^3/M$. Furthermore, at the fixed point
locus $\phi=0$, they simply reduce to
\begin{equation}
Y^{\pm n}=\frac{1}{2^{\frac{n}{2}}}\,\mathrm{e}^{\pm\,\ii\,n\,\theta}~.
\label{Y+-n0}
\end{equation}

\section{Useful identities}
\label{app:identities}

In this Appendix, following the analysis of \cite{DHoker:1999jke},  we prove some identities that are useful in deriving the cubic vertex between one untwisted and two twisted KK modes that we presented in Section~\ref{subsecn:3pointholo}.

Let us first consider the bulk fields $s_k^*$, $\eta_{\alpha,\ell}^*$ and
$\eta_{\alpha,p}$, with $k,\ell,p>0$. In the supergravity Lagrangian
(\ref{cubic2}), we find three types of terms. The first one is proportional to
$s_k^*\,\nabla_\mu \eta_{\alpha,\ell}^*\nabla^\mu \eta_{\alpha,p}$. 
Using the Leibniz rule for the covariant derivatives, it is very easy to show that
\begin{align}
s_k^*\,\nabla_\mu \eta_{\alpha,\ell}^*\nabla^\mu \eta_{\alpha,p}\,
&=\frac{1}{2}\,(\nabla_\mu \nabla^\mu s_k^*)\,\eta_{\alpha,\ell}^*\,\eta_{\alpha,p}-\frac{1}{2}\,s_k^*\,(\nabla_\mu \nabla^\mu \eta_{\alpha,\ell}^*)\,\eta_{\alpha,p}
-\frac{1}{2}\,s_k^*\,\eta_{\alpha,\ell}^*\,(\nabla_\mu \nabla^\mu\eta_{\alpha,p})\notag\\
&\qquad+\frac{1}{2}\,\nabla_\mu\big[s_k^*\,\nabla^\mu
(\eta_{\alpha,\ell}^*\,\eta_{\alpha,p})
-\nabla^\mu s_k^*\,\eta_{\alpha,\ell}^*\,\eta_{\alpha,p}
\big]\notag\\
&=
\frac{1}{2}\,\big[k(k-4)-\ell(\ell-4)-p(p-4)\big]
s_k^*\,\eta_{\alpha,\ell}^*\,\eta_{\alpha,p}+\mbox{total derivative}~,
\label{id1a}
\end{align}
where in the last step we used the on-shell conditions in AdS$_5$ given in (\ref{masssk}) and (\ref{massgammak}) for the untwisted and twisted modes
respectively.
In a very similar way, one can show that
\begin{equation}
\begin{aligned}
\nabla_\mu s_k^*\,\eta_{\alpha,\ell}^*\,\nabla^\mu \eta_{\alpha,p}
&=\frac{1}{2}\,\big[\ell(\ell-4)-k(k-4)-p(p-4)\big]
s_k^*\,\eta_{\alpha,\ell}^*\,\eta_{\alpha,p}+\mbox{total derivative}~,\\
\nabla_\mu s_k^*\,\nabla^\mu \eta_{\alpha,\ell}^*\,\eta_{\alpha,p}
&=
\frac{1}{2}\,\big[p(p-4)-\ell(\ell-4)-k(k-4)\big]
s_k^*\,\eta_{\alpha,\ell}^*\,\eta_{\alpha,p}+\mbox{total derivative}~.
\end{aligned}
\label{id2a}
\end{equation}

The second term in the Lagrangian (\ref{cubic2}) is proportional to
$\nabla_{(\mu}\nabla_{\nu)}s_k^*
\nabla^\mu \eta_{\alpha,\ell}^* \,\nabla^\nu \eta_{\alpha,p}$. Again by repeatedly using the Leibniz rule for the covariant derivatives, we can show that
\begin{align}
\nabla_{(\mu}\nabla_{\nu)}s_k^*
\nabla^\mu \eta_{\alpha,\ell}^* \,\nabla^\nu \eta_{\alpha,p}&=
\Big(\nabla_{\mu}\nabla_{\nu}s_k^*-\frac{1}{5}\,g_{\mu\nu}
\nabla_{\lambda}\nabla^{\lambda}s_k^*\Big)\nabla^\mu 
\eta_{\alpha,\ell}^* \,\nabla^\nu \eta_{\alpha,p}\notag\\
&=\frac{1}{2}\,\nabla_{\mu}\nabla_{\nu}s_k^*
\Big(\nabla^\mu\eta_{\alpha,\ell}^* \,\nabla^\nu\eta_{\alpha,p}
+\nabla^\nu\eta_{\alpha,\ell}^* \,\nabla^\mu\eta_{\alpha,p}
-g^{\mu\nu} \nabla_\lambda\eta_{\alpha,\ell}^* \,\nabla^\lambda\eta_{\alpha,p}\Big)\notag\\
&\qquad+\frac{1}{2}\,\nabla_\lambda \nabla^{\lambda}s_k^* \nabla^\mu\eta_{\alpha,\ell}^* \,\nabla_\mu\eta_{\alpha,p}-\frac{1}{5}\,
\nabla_{\lambda}\nabla^{\lambda}s_k^* \nabla^\mu \eta_{\alpha,\ell}^* \,\nabla_\mu \eta_{\alpha,p}\notag\\
&=-\frac{1}{2}\,\nabla_{\nu}s_k^*\Big(\nabla_{\mu}\nabla^\mu
\eta_{\alpha,\ell}^* 
\,\nabla^\nu\eta_{\alpha,p}+\nabla^\nu\eta_{\alpha,\ell}^* \,
\nabla_\mu 
\nabla^\mu\eta_{\alpha,p}\Big)\notag\\
&\qquad
+\frac{3}{10}\,\nabla_{\lambda}\nabla^{\lambda}s_k^* \nabla^\mu \eta_{\alpha,\ell}^* \,\nabla_\mu \eta_{\alpha,p}
+\mbox{total derivative}\notag\\
&=-\frac{1}{2}\,\ell(\ell-4)\,\nabla_{\nu}s_k^*\,\eta_{\alpha,\ell}^*
\,\nabla^\nu\eta_{\alpha,p}
-\frac{1}{2}\,p(p-4)\,
\nabla_{\nu}s_k^*\,\nabla^\nu\eta_{\alpha,\ell}^* \, \eta_{\alpha,p}
\notag\\
&\qquad
+\frac{3}{10}\,k(k-4)\,s_k^* \nabla^\mu\eta_{\alpha,\ell}^* 
\,\nabla_\mu\eta_{\alpha,p}
+\mbox{total derivative}~,
\label{id3a}
\end{align}
where in the last step we have inserted the mass-shell conditions (\ref{masssk}) and (\ref{massgammak}).
If we now use the identities (\ref{id1a}) and (\ref{id2a}), we can rewrite (\ref{id3a}) as
\begin{align}
\nabla_{(\mu}\nabla_{\nu)}s_k^*
\nabla^\mu \eta_{\alpha,\ell}^* \,\nabla^\nu \eta_{\alpha,p}
&=-\frac{1}{4}\,\ell(\ell-4)\,\Big[\ell(\ell-4)-k(k-4)-
p(p-4)\Big]\,s_k^*\,\eta_{\alpha,\ell}^*\,\eta_{\alpha,p}\notag\\
&\quad-\frac{1}{4}\,p(p-4)\,\Big[p(p-4)-k(k-4)-\ell(\ell-4)\Big]\,s_k^*\,\eta_{\alpha,\ell}^*\,\eta_{\alpha,p}\notag\\
&\quad+\frac{3}{20}\,k(k-4)\,\Big[k(k-4)-\ell(\ell-4)-p(p-4)\Big]\,s_k^*\,\eta_{\alpha,\ell}^*\,\eta_{\alpha,p}\notag\\[2mm]
&\quad+\mbox{total derivative}~.
\label{id4a}
\end{align}
Exploiting (\ref{id1a}) and (\ref{id4a}), we see that the combination appearing in the Lagrangian (\ref{cubic2}) can be written as follows
\begin{align}
-\frac{4}{5}&\,k\,s_k^* \nabla_\mu \eta_{\alpha,\ell}^* \,\nabla^\mu \eta_{\alpha,p}
-\frac{4}{k+1}\,\nabla_{(\mu}\nabla_{\nu)}s_k^*
\nabla^\mu \eta_{\alpha,\ell}^* \,\nabla^\nu \eta_{\alpha,p}
-4\,k\,\ell\,p\,s_k^* \,\eta_{\alpha,\ell}^* \, \eta_{\alpha,p}
\notag\\
&=\bigg\{\!-\frac{2\,k}{5}\,\big[k(k-4)-\ell(\ell-4)-p(p-4)\big]
\notag\\
&\qquad
+\frac{1}{(k+1)}\Big[\ell(\ell-4)\big(\ell(\ell-4)-k(k-4)-p(p-4)\big)\label{id5a}\\
&\qquad\qquad\qquad+p(p-4)\big(p(p-4)-k(k-4)-\ell(\ell-4)\big)\notag\\
&\qquad\qquad\qquad-\frac{3}{5}\,k(k-4)\big(k(k-4)-\ell(\ell-4)-p(p-4)\big)\Big]-4\,k\,\ell\,p
\bigg\}\,s_k^* \,\eta_{\alpha,\ell}^* \, \eta_{\alpha,p}~,\notag
\end{align}
which leads to the cubic coupling given in (\ref{Lklp}) of the main text.

\end{appendix}

%
%

\providecommand{\href}[2]{#2}\begingroup\raggedright\endgroup

\end{document}